\documentclass[aps,preprint,amsmath,amssymb,nofootinbib]{revtex4-2}
\usepackage{xcolor}
\usepackage{graphicx} %including figures
\usepackage[caption=false]{subfig}

\usepackage{gensymb} 
\usepackage{amsmath} 
\usepackage{braket} 
\usepackage{grffile}
\usepackage{booktabs} 
\usepackage[colorlinks=true]{hyperref} %color in references to fig,bib,etc
\hypersetup{colorlinks,
	citecolor=blue
}
\usepackage[capitalize]{cleveref}

\newcommand{\ba}{\begin{eqnarray}}
\newcommand{\ea}{\end{eqnarray}}

\newcommand{\nne}{{\bf     e}}

\newcommand{\nk}{{\bf      k}}
\newcommand{\np}{{\bf      p}}
\newcommand{\nq}{{\bf      q}}

\begin{document}

%\preprint{} 

\title{Semi-inclusive charged-current neutrino-nucleus cross sections in the relativistic plane wave impulse approximation}


\author{J.M. Franco-Patino}
\author{J. Gonzalez-Rosa}
\author{J.A. Caballero}
%\email[]{}
%\thanks{}
\affiliation{Departamento de F\'isica at\'omica, molecular y nuclear, Universidad de Sevilla, 41080 Sevilla, Spain}
\author{M.B. Barbaro}
\affiliation{Dipartimento di Fisica, Universit\`a di Torino and INFN,
Sezione di Torino, Via P. Giuria 1, 10125 Torino, Italy}


\date{\today}

\begin{abstract}
	Neutrino-nucleus quasielastic scattering  is studied in the plane wave impulse approximation for three nuclear models: the relativistic Fermi gas (RFG), the independent-particle shell model (IPSM) and the natural orbitals (NO) model with Lorentzian dependence of the excitation energy. A complete study of the kinematics of the semi-inclusive process and the associated cross sections are presented and discussed for $^{40}$Ar and $^{12}$C. Inclusive cross sections are also obtained by integrating the semi-inclusive expressions over the outgoing hadron. Results are consistent with previous studies restricted to the inclusive channel. In particular, a comparison with the analytical results for the RFG model is performed. Explicit expressions for the hadronic tensor and the 10 semi-inclusive nuclear responses are given. Theoretical predictions are compared with semi-inclusive experimental data from T2K experiment.  
\end{abstract}

%\keywords{}

\maketitle

\section{\label{Introduccion}Introduction}

In the last years Neutrino Physics has become one of the most flourishing fields in nuclear and particle physics. In particular, physicists have devoted a great effort in pursuing the physics responsible for neutrino masses. As stated in the NuSTEC White Paper~\cite{Alvarez-Ruso17}, accelerator based neutrino scattering experiments have been identified as the highest priority intermediate-future effort by the world physics community. This explains the high interest in international projects like the Deep Underground Neutrino Experiment (DUNE)~\cite{Acciarri:2016crz}, to be hosted by Fermilab, and the Tokai-to-HyperKamioKande (T2HK) in Japan~\cite{HyperK15}. The main goal of these experiments is focused on the analysis of the oscillations that neutrinos undergo in travelling from a near to a far detector. 
The aim is  not only to improve our present knowledge on the oscillation mixing angles, but also to explore the CP-violating phase, that is related to the matter-antimatter asymmetry in the Universe, assess the neutrino mass hierarchy, and investigate possible physics beyond the standard model~\cite{Alvarez-Ruso17,Katori17,electron-vs-neutrino}.

The analysis of neutrino properties, due to the smallness of the weak cross sections, requires the use of large amounts of target material. In fact, most of the presently running (T2K, NoVA) and planned (T2HK, DUNE) neutrino oscillation experiments use different complex nuclei as targets, such as carbon, oxygen, argon and iron. Thus, a precise enough description of the interaction between neutrinos and nuclei is needed. Only by having an excellent control of the nuclear effects in the weak scattering process, will it be possible to access without ambiguity to the real nature of neutrinos and their properties. This clearly shows that only a close collaboration between theoretical and experimental groups from both the nuclear and high energy physics communities will make it possible to overcome the challenges we face.
 
In past years a great effort has been devoted to the description of neutrino-nucleus observables with high accuracy. Very different models, initially designed to describe electron scattering reactions where there exist a large amount of data to compare with, have been extended to neutrino processes. As a general constraint any nuclear model aiming to describe neutrino-nucleus interaction should be first tested against electron scattering data. Starting with the simple relativistic Fermi Gas (RFG), still widely used in the analysis of neutrino oscillation data, models with different levels of complexity have been applied to weak interaction processes: nuclear spectral function~\cite{Rocco:2018mwt,Vagnoni:2017hll,Benhar:2015xga,Ankowski:2007uy}, 
Relativistic Mean Field~\cite{Caballero:2005sj,Martinez06,Gonzalez-Jimenez:2013plb,Caballero:2007tz,PhysRevC.100.045501,Maieron03}, Relativistic Green Function (RGF)~\cite{Meucci15,Meucci:2004,Meucci:2011pi,Meucci:2011vd}, Random Phase Approximation (RPA)~\cite{Martini:ee,Pandey15,PhysRevC.94.024611}, scaling-based approaches~\cite{Caballero:2005sj,Caballero:2007tz,Amaro:2005dn,Amaro05,Caballero05,Gonzalez-Jimenez:2014eqa,PhysRevD.99.113002,Megias:2018oxygen} and ab-initio Green's function Monte Carlo (GFMC)~\cite{Lovato16,Rocco:2016ejr,Lovato:2017cux}. 

A basic difference between electron and neutrino scattering processes, in addition to the weak versus electromagnetic interaction, concerns the beam energy. Whereas for electrons the energy is perfectly known, the situation is clearly different for neutrinos where the energy is distributed along the neutrino flux that can be extended from a few MeV up to several GeV. This makes a crucial difference between the two scattering reactions that affects not only the particular description of the interaction between the lepton and the nuclear target, but also the analysis of oscillation experiments and the specific information on the inner properties of neutrinos.

At present most of the studies related to neutrino-nucleus scattering have concerned inclusive measurements where only the scattered lepton is detected in the final state. This corresponds to $(\nu_\ell, \ell)$ processes where $\ell$ refers to the lepton in the final state. Note that this process is kinematically equivalent to inclusive electron scattering $(e,e')$. However,  contrary to $(e,e')$ processes where the momentum ($q$) and energy ($\omega$) transferred to the nucleus are perfectly known, for $(\nu_\ell, \ell)$ this is not so because of the neutrino beam energy distribution (flux). As a consequence, in the analysis of $(\nu_\ell, \ell)$ reactions very different reaction mechanisms can contribute significantly to the cross sections for a given kinematics of the final lepton. Not only the quasielastic (QE) region should be carefully evaluated but also nucleon resonances, two-particle two-hole (2p-2h) effects, deep inelastic scattering and even the region at very low momentum/energy transfer where the impulse approximation is not applicable. This makes the theoretical description of neutrino-nucleus scattering processes more demanding than the one corresponding to electrons. An illustrative example of this came with the first neutrino-carbon cross sections published by the MiniBooNE collaboration~\cite{PhysRevD.81.092005}. It was found that data were largely underestimated by all theoretical predictions based on the impulse approximation unless the value of the axial mass was increased by more than $\sim 30\%$ compared with the standard value, $M_A\simeq 1$ GeV. This was known as the {\it ``$M_A$-puzzle"}. It was soon realized that nuclear effects beyond the impulse approximation, in particular, the excitation of 2p-2h states, could remedy significantly the discrepancy between theory and data without need to modify the value of $M_A$ (see Refs.~\cite{Martini:2009uj,Martini:2010ex,Amaro:2010sd,Amaro:2011aa,Nieves:2011pp} for details).

The determination of oscillation mixing angles, CP-violating phase and neutrino mass ordering from neutrino-nucleus scattering data requires to know precisely the neutrino energy. Since the neutrino beams are not monochromatic, the incoming neutrino energy should be reconstructed from the final states in the reaction. Up to present this determination has been based on the analysis of inclusive $(\nu_\ell, \ell)$ reactions assuming that the neutrino interacts with a neutron in the nucleus at rest (likewise for antineutrinos and protons). This is a very crude approximation that allows to determine the neutrino energy from the variables of the lepton in the final state, but with high uncertainty due to the effects associated to the nuclear dynamics and the different channels involved in the reaction. The reader interested in a study of the impact
of nuclear effects on the neutrino energy reconstruction can go to~\cite{Alvarez-Ruso17} and references therein. 

A way to improve significantly the analysis, constraining the incident neutrino energy much better, is to consider events in which not only the final-state charged lepton is detected but  some hadron as well. In fact, a proper description of the hadrons and mesons in the final-state will be essential for the next-generation of neutrino experiments. This is connected to the fact that modern experimental studies of neutrino-nucleus reactions rely on the use of data simulations to determine the behavior of the detectors involved. This requires having a reasonable control on the reconstruction of the energy neutrino which can be achieved more precisely by analyzing the kinematics of the final particles. This clearly shows the importance to have realistic theoretical predictions corresponding to more exclusive processes where, in addition to the final lepton, other particles are detected. Although the theoretical prediction of semi-inclusive reactions is much harder than modelling inclusive processes, the richer structure of the cross sections allows one to better discriminate among different models. The extension of nuclear models to semi-inclusive reactions is one of the main challenges to be faced by nuclear theorists working in the field.  

In this work we present a detailed study of semi-inclusive charged current (CC) processes: $(\nu_\ell, \ell^- \, p)$ and $(\overline{\nu}_\ell, \ell^+ \, n)$. We follow the seminal works developed in \cite{Donnelly:2014,Cenni_1997,Donnelly:2017,Donnelly:2019} and restrict ourselves to the plane wave impulse approximation (PWIA) in which the semi-inclusive cross section factorizes in a term dealing with the neutrino-nucleon interaction, namely, the single-nucleon cross section, and the spectral function that incorporates the whole dependence on the nuclear dynamics. The whole formalism is presented in the paper by writing the explicit expressions of all the semi-inclusive weak responses entering in the cross section. Moreover, the consistency of the calculations is proved by comparing the inclusive cross section obtained by integrating the corresponding semi-inclusive one over the emitted nucleon variables with the inclusive results already presented in the literature~\cite{Amaro05,Caballero05,electron-vs-neutrino,Caballero_2006}. This is shown for different nuclear models. Although being aware of the oversimplified description of the scattering process provided by PWIA, a comparison with some semi-inclusive data recently measured by the T2K collaboration is performed. In forthcoming work we will extend our study to all available semi-inclusive data, and will include in our analysis the role played by the final state interactions (FSI) treated within the framework of the Relativistic Distorted Wave Impulse Approximation (RDWIA), exploiting our past experience on the description of semi-inclusive ($e,e'N)$ reactions within a fully relativistic microscopic approach.

To conclude, in our present investigation we have restricted our interest to the kinematics corresponding to T2K (using carbon in the near detector) and DUNE (argon). Semi-inclusive cross sections for both cases have been shown for selected kinematics as functions of the ejected nucleon variables (momentum and angle) for the different nuclear models. The development and implementation in experimental event generators of a complete semi-inclusive formalism for neutrino reactions will have a huge impact in the analysis of new experiments more sensitive to hadron detection. %All this will be crucial for the reduction of systematics and experimental costs, and the precise determination of oscillation parameters, thus shedding light on relevant cosmological questions. 

The paper is organized as follows: In Sect.~\ref{General formalism} we present the general formalism for semi-inclusive neutrino-nucleus scattering reactions. We discuss the general kinematics and evaluate the cross section in both the semi-inclusive and inclusive regimes. The discussion on the nuclear models is presented in Sect.~\ref{spectral function}: here we show the cross sections obtained for the Relativistic Fermi Gas (RFG), Independent Particle Shell Model (IPSM) and Natural Orbitals (NO). Explicit expressions for the flux-averaged semi-inclusive cross sections are provided for the three cases. In Sect.~\ref{Results} we present our results for different kinematics corresponding to T2K and DUNE experiments. Sect.~\ref{semi-inclusive cross sections} contains a detailed discussion on the semi-inclusive cross sections, while in Sect.~\ref{Semi-inclusive experimental} we compare our predictions with some data taken by the T2K 
collaboration~\cite{PhysRevD.98.032003}. The case of inclusive responses is considered in Sect.~\ref{inclusive cross sections} proving the consistency of the calculations. Finally, in Sect.~\ref{off-shell effects} we discuss the effects associated to the use of different descriptions of the weak current operator, {\it i.e.}, off-shell effects. This is discussed for the semi-inclusive cross sections as well as for the inclusive ones considering the RFG and IPSM. In Sect.~\ref{Conclusions} we summarize our conclusions.

\section{\label{General formalism} General formalism}

In this section we define the kinematics of the quasi-elastic neutrino-nucleus reaction and set up the general formalism for the corresponding differential cross section in both the semi-inclusive and inclusive channels. 

\subsection{\label{Kinematics}Kinematics}

\begin{figure}[!htbp]
	\includegraphics{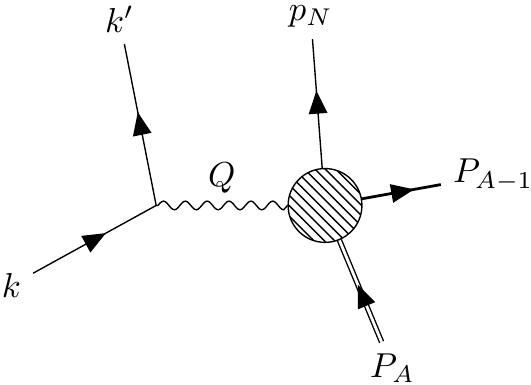}
	\caption{\label{feynmanneut} Schematic representation of the process analyzed in this work in the Born approximation.}
\end{figure}

We start by defining the kinematic variables entering into the reaction represented in Fig.~\ref{feynmanneut}.
The  four-momenta of the initial neutrino and final lepton are 
\begin{equation}
K^{\mu}=(\varepsilon, \textbf{k}), \ \ \ K'^{\mu}=(\varepsilon',\textbf{k}') \, ,
\end{equation}
where
\begin{equation}
\varepsilon=\sqrt{k^{2} + m_{\nu}^{2}} \approx k , \ \ \ \varepsilon'=\sqrt{k'^{2} + m_{l}^{2}}
\end{equation}
with $m_l$ the mass of the final lepton.

\begin{figure}[!htbp]
		\centering
		\includegraphics[width=0.48\textwidth]{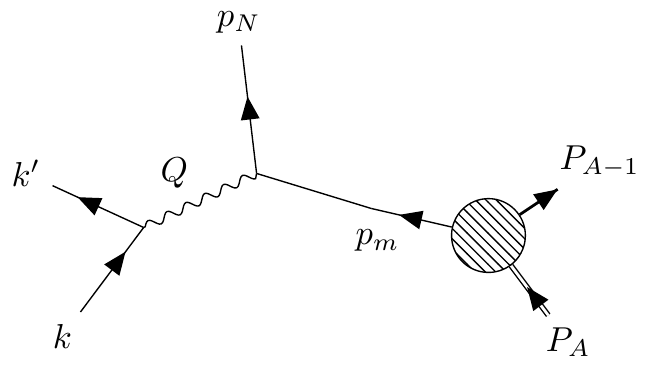}
		\caption{\label{neutrino-nuc} Feynman diagram for neutrino-nucleus quasielastic scattering in the plane wave impulse approximation.}
\end{figure}

The four-momentum transfer is defined as
\begin{equation}
Q^{\mu}=(\omega,\textbf{q}) = (\varepsilon-\varepsilon', \textbf{k}-\textbf{k}')\, .
\end{equation}
We work in the laboratory system, where the four-momentum of the target nucleus  is
\begin{equation}
    P^{\mu}_{A}=(M_{A}, \bf{0})
\end{equation}
with $M_{A}$ the rest mass of the nucleus. The four-momentum of the outgoing nucleon is
\begin{equation}
P^{\mu}_{N}=(E_N, \textbf{p}_{N})\, ,
\end{equation}
where $E_N=\sqrt{p^{2}_{N} + m^{2}_{N}}$ is the on-shell energy, and the four momentum of the residual nucleus, having invariant mass $W_{A-1}$ and momentum $-\textbf{p}_m$, is
\begin{equation}
P^{\mu}_{A-1}= \left(\sqrt{p^{2}_{m} + W^{2}_{A-1}}, -\textbf{p}_{m}\right)\, .
\end{equation}
In the above, we have introduced the missing momentum
\begin{equation}
\textbf{p}_m = \textbf{k}'+\textbf{p}_N-\textbf{k} =\textbf{p}_N-\textbf{q}\, ,
\label{eq:pcons}
\end{equation}
which, in the PWIA approximation represented in Fig.~\ref{neutrino-nuc}, is simply the momentum of the hit nucleon.
We also introduce the missing energy 
\begin{equation}
E_m = \sqrt{p^{2}_{m} + W^{2}_{A-1}} + m_N - M_A \, .
\end{equation}
For fixed values of $p_N$ and $q$, the allowed values of the missing momentum are
\begin{equation}
p_m^- \leq p_m \leq p_m^+ 
\end{equation}
with
\begin{equation}
p_m^\pm = |p_N\pm q|\, .
\label{eq:pmlim}
\end{equation}

\begin{figure*}[!hptb]
	\centering
	\subfloat[ $q=0.5$ GeV and $\omega=0.06$ GeV  (below quasielastic peak).]{
		\includegraphics[width=0.33\textwidth]{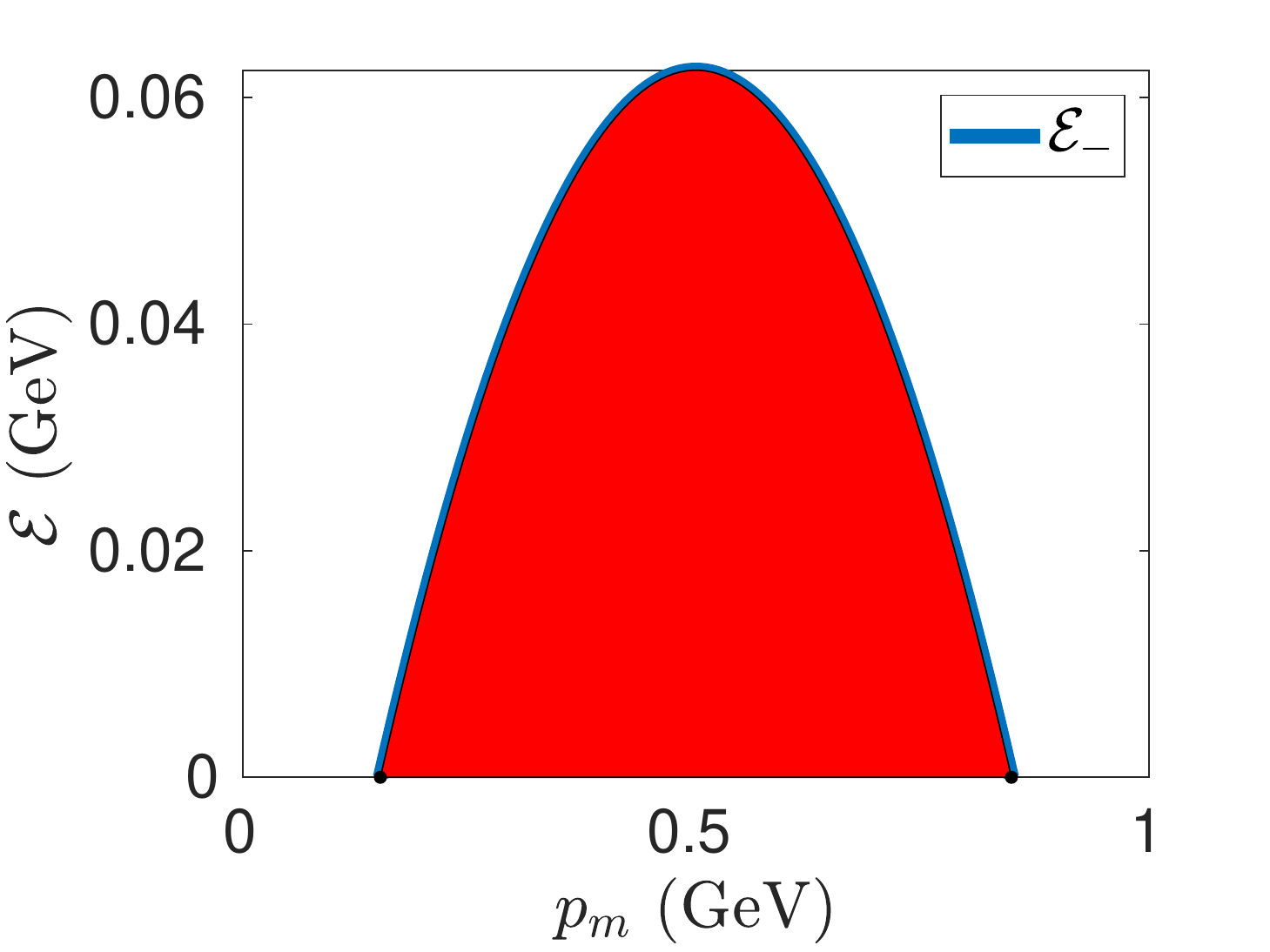}
			}
	\subfloat[ $q=0.5$ GeV and $\omega=0.12$ GeV (at the quasielastic peak).]{
		\includegraphics[width=0.33\textwidth]{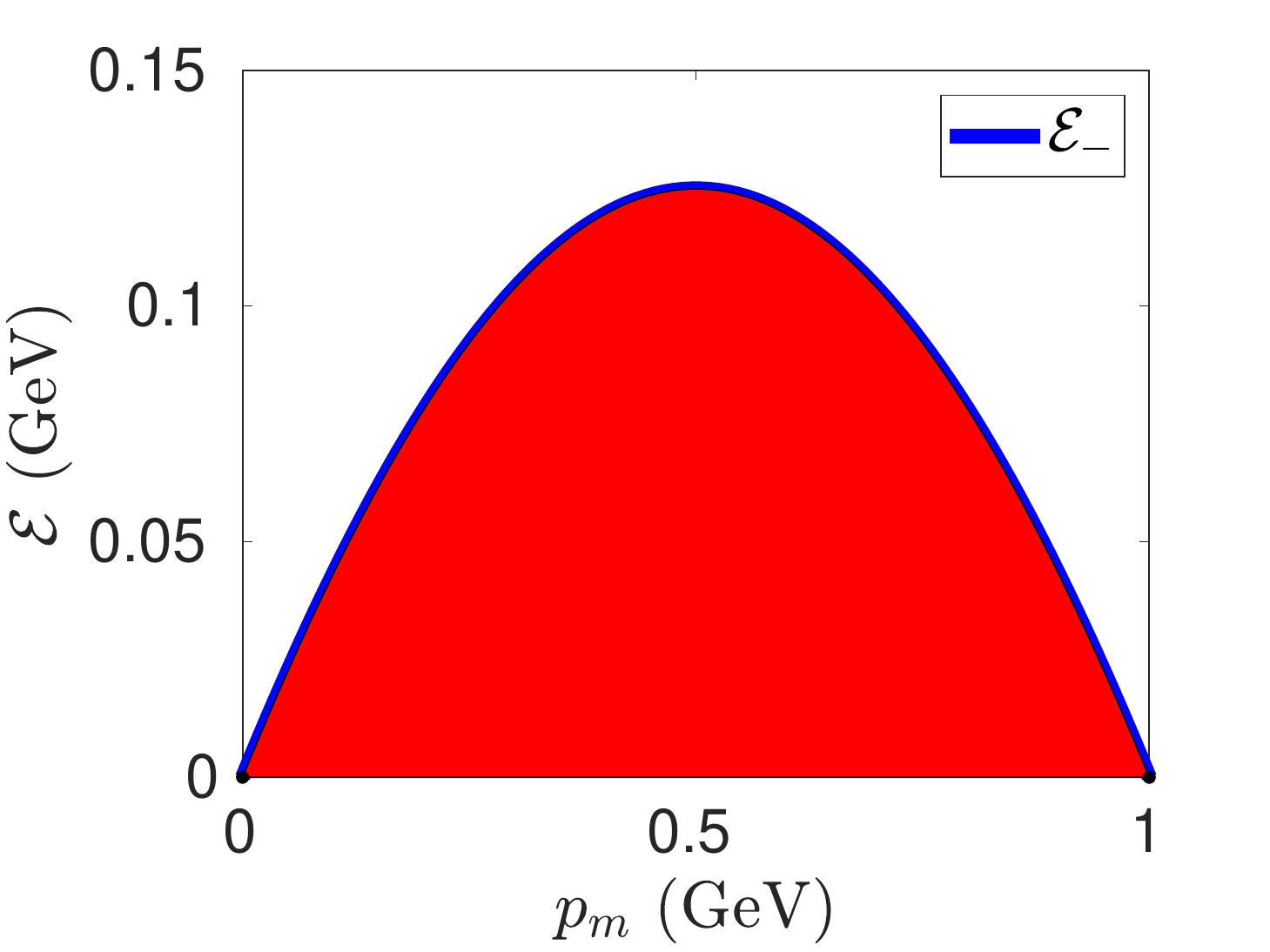}
			}
	\subfloat[ $q=0.5$ GeV and $\omega=0.19$ GeV  (above quasielastic peak).]{ 
		\centering
		\includegraphics[width=0.33\textwidth]{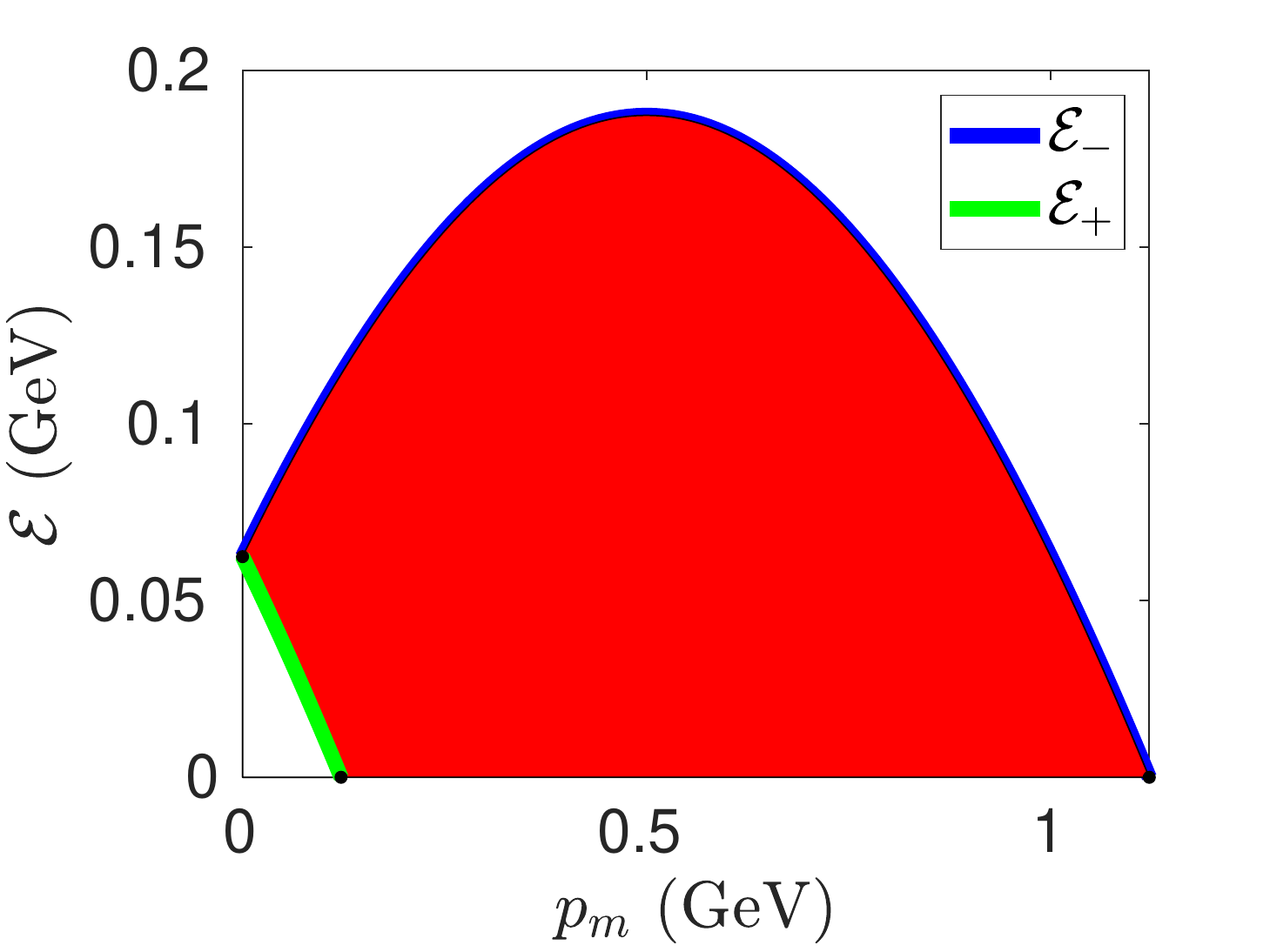}
		}
	\caption{\label{FigKine} Planes defined by the excitation energy $\mathcal{E}$ and the missing momentum $p_m$. 
	The allowed region for the quasielastic reaction is the red area between the curves $\mathcal{E}_{\pm}$ defined in Eqs.~\eqref{eq:limE}-\eqref{eq:limE-}. In these plots we take $E_{s}=0$.}
\end{figure*}
Next, it is convenient to introduce the variable 
\begin{equation}
	\mathcal{E}=\sqrt{p^{2}_{m} + W^{2}_{A-1}}-\sqrt{p^{2}_{m} + M^{2}_{A-1}} \geq 0\, ,
\end{equation}
 the excitation energy of the residual nucleus. It is related to the missing energy and momentum by the expression
\begin{equation}
\mathcal{E} = E_m-E_s -\left(\sqrt{p_m^2+M^2_{A-1}} -M_{A-1}\right)\, ,
\label{eq:cale}
\end{equation}
where we have introduced the nuclear separation energy
\begin{equation}
E_s=M_{A-1} + m_N - M_A \, ,
\end{equation}
namely  the minimum energy necessary to remove a nucleon from a nucleus of mass $A$.
The last term in Eq.~\eqref{eq:cale} represents the nuclear recoil energy and can be neglected
for medium-heavy nuclei ($p_{m}<<M_{A-1}$). In this case we can write
\begin{eqnarray}
\mathcal{E} &\simeq& \omega - E_{s} + m_{N} -  E_N
\label{eq:Esimp}
\\
&=& \omega - E_{s} + m_{N} -\sqrt{q^{2} + p^{2}_{m} + 2p_{m}q\cos{\theta_{m}} + m^{2}_{N}} \, ,
\nonumber\end{eqnarray}
where $\theta_{m}$
is the angle between $\textbf{p}_m$ and $\textbf{q}$. At given values of $\omega$, $q$ and $p_m$, the boundary limits of the variable $\mathcal{E}$ are obtained when $\cos\theta_m=\pm 1$, namely
\begin{equation}
\max(\mathcal{E}_+,0) \leq \mathcal{E} \leq \mathcal{E}_-
\label{eq:limE}
\end{equation}
with
\begin{eqnarray}
\label{eq:limE+}
\mathcal{E}_+ &=& \omega - E_{s} + m_{N} -\sqrt{q^{2} + p^{2}_{m} + 2p_{m}q + m^{2}_{N}} \\
 \mathcal{E}_- &=& \omega - E_{s} + m_{N} -\sqrt{q^{2} + p^{2}_{m} - 2p_{m}q + m^{2}_{N}}\,  .
\label{eq:limE-}
\end{eqnarray}
By exploiting Eq.~\eqref{eq:Esimp}, the limits~\eqref{eq:pmlim} can be written in terms of $\mathcal{E}$ as
\begin{align}
p^{+}_{m}= \sqrt{(\omega - E_{s} - \mathcal{E})(\omega - E_{s}  -\mathcal{E} + 2m_{N})} + q \label{uplimit-general}\\
p^{-}_{m}= \left|\sqrt{(\omega - E_{s} -  \mathcal{E})(\omega - E_{s} -  \mathcal{E} + 2m_{N})} - q\right| \, .
\label{downlimit-general}
\end{align}
The region of the plane $(\mathcal{E},p_m)$ kinematically allowed for the quasielastic reaction  is represented in Fig.~\ref{FigKine} for fixed $q$ and for three values of $\omega$ around the quasielastic peak value $\omega_{QE}=\sqrt{q^{2} + m^{2}_{N}} -m_{N} + E_{s}$.

\begin{figure}[!htb]
		\centering
		\includegraphics{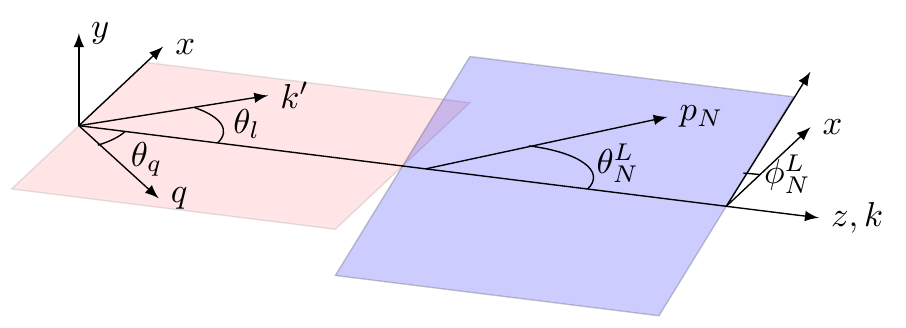}
		\caption{\label{Fig2} Kinematical variables in the $k$-system where the beam direction is chosen as the $z$-axis. 
		%The plane of the outgoing nucleon (reaction plane) is represented in blue and the plane of the neutrino beam is represented in pink (scattering plane). {\red I have some questions/remarks on this figure. First, 1',2',3' are not defined: we can just call them x,y,z. Moreover, a vector does not define a plane! So we can write: "}
		The plane formed by the neutrino beam and the outgoing nucleon (reaction plane, $(\nk,\np_N)$) is represented in blue and the plane identified by the incident neutrino and scattered lepton (scattering plane, $(\nk,\nk')$) is represented in pink. 
		%The angle between both planes is given by $\phi_N^L$ whereas $\theta_N^L$ determines the outgoing nucleon momentum, $\np_N$ measured with respect to $\nk$ (defining the $z$-axis). See text for more details.
		The unit vectors defining the axes $x$ and $y$ are given by $\textbf{e}_y=(\textbf{k} \times \textbf{k}')/|\textbf{k}\times \textbf{k}'|$ and $\textbf{e}_x=\textbf{e}_y\times \textbf{e}_z$ with $\textbf{e}_z = \textbf{k}/|\textbf{k}|$. 
		%{\ttred Juan: Remove 1',2' and 3' and introduce $x,y,z$-axes. Should we modify a little bit the graph to make easier the identification of the two planes? Take a look to the papers by Bill and Wally. Finally, $\theta_N^L$ is the angle between $p_N$ and $k$ with positive sign, while here it is represented as the angle between $p_N$ and $-k$ (that is, $\pi-\theta_N^L$).}
		}
\end{figure}

Finally, let us fix the axes direction. We consider that the three-momentum $\textbf{k}$ defines the direction of the $z$-axis. We choose this frame - referred to as the $k$-system - because the direction of the neutrino beam is known in experiments, so we can directly compare our results with experimental data.
%~\footnote{In Sect. \ref{From semi-inclusive to inclusive} and Appendix \ref{apendixa} we will use the "q-system" where the $z$ axis is chosen to be along $\textbf{q}$. Notice that the two systems are simply related by a rotation of an angle $\theta_q$ within the scattering plane. The $q-system$, used in the analysis of semi-inclusive electron scattering processes, has special symmetries that simplify the calculation of the response functions.}.  
All the kinematic variables are represented in Fig.~\ref{Fig2}. To make clear the discussion we distinguish between the scattering plane and the reaction one. The former, represented in pink, is
defined by the neutrino beam momentum $\textbf{k}$ ($z$-axis) and the ejected lepton momentum $\textbf{k}'$. 
%The unit vectors defining the axes $x$ and $y$ are given by $\textbf{e}_y=(\textbf{k} \times \textbf{k}')/|\textbf{k}\times \textbf{k}'|$ and $\textbf{e}_x=\textbf{e}_y\times \textbf{e}_z$ with $\textbf{e}_z\equiv \textbf{k}/|\textbf{k}|$. 
The reaction plane, represented in blue, contains $\textbf{k}$ and the ejected nucleon momentum $\textbf{p}_N$.  Thus, the three-momenta defined in the $x,y,z$ frame (see Fig.~\ref{Fig2}) are
\begin{align}
\textbf{k}&= k \textbf{e}_{z}, \nonumber\\
\textbf{k}'&=k'(\sin{\theta_{l}}\textbf{e}_{x} + \cos{\theta_{l}}\textbf{e}_{z}), \nonumber\\
\textbf{p}_{N}&=p_{N}(\cos{\phi^{L}_{N}}\sin{\theta^{L}_{N}}\textbf{e}_{x} + \sin{\phi^{L}_{N}}\sin{\theta^{L}_{N}}\textbf{e}_{y} + \cos{\theta^{L}_{N}}\textbf{e}_{z}) \, ,
\end{align}
where $\phi_N^L$ is the angle formed by the two planes. Note that the transferred momentum, $\textbf{q}$, is contained in the scattering plane ($xz$) and $\theta_q$ represents the angle between $\textbf{q}$ and $\textbf{k}$ whereas $\theta_l$ is the scattering angle, {\it i.e.,} the angle between $\textbf{k}$ and $\textbf{k'}$. Finally, $\theta_N^L$ represents the polar angle that defines the direction of the ejected nucleon momentum $\textbf{p}_N$ with respect to the $z$-axis ($\textbf{k}$-direction). 

It is important to distinguish between the $k$-system defined above and the $q$-system usually considered in the analysis of semi-inclusive electron scattering processes. In the latter the $z$-axis is chosen to be along the momentum transfer $\nq$. Hence the two systems are simply related by a rotation of an angle $\theta_q$ within the scattering plane. In Sect.~\ref{From semi-inclusive to inclusive} and Appendix~\ref{apendixa} we will use the $q$-system to evaluate the response functions because of its special symmetries.

\subsection{\label{Semi-inclusive} Semi-inclusive cross section}

In this work we restrict our attention to the Plane Wave Impulse Approximation (PWIA) where, neglecting the contribution of the lower components in the relativistic bound nucleon wave function, the cross section factorizes into a term dealing with the weak interaction of a single nucleon in the nucleus and the nuclear spectral function that embodies the nuclear dynamics in the process. The ``factorized" ansatz has been shown to work properly in the case of inclusive processes providing good agreement with experiment. In the future we will extend our study by including the role of final state interactions (FSI) that break in general the factorizable form. 
The sixth-differential semi-inclusive cross section with respect to the momenta $k'$ and $p_N$ and the solid angles $\Omega_{k'}=(\theta_l,\phi_l)$ and $\Omega_N^L=(\theta_N^L,\phi_N^L)$ in the factorization approximation is given by~\cite{Donnelly:2014} 
\begin{widetext}
\begin{align}\label{general-semics-i}
\frac{d\sigma}{dk'd\Omega_{k^{'}}dp_{N}d\Omega^{L}_{N}}=\frac{(G_{F}\cos{\theta_{c}} k'p_{N})^{2}m_{N}}{8k\varepsilon'E_{N}(2\pi)^{6}}  \int_{0}^{\infty}d\mathcal{E}
\int d^{3}p_{m} \upsilon_0\mathcal{F}^2_\chi S\bigl(p_{m},E_{m}(\mathcal{E}, p_{m})\bigr)\nonumber\\
\times\delta(M_{A} + k - \varepsilon' - E_{N} - \sqrt{p^{2}_{m} + M^{2}_{A-1}} - \mathcal{E}) 
\delta (\textbf{k}- \textbf{k}' -\textbf{p}_{N} + \textbf{p}_{m})\, ,
\end{align}
\end{widetext}
where $G_F$ is the Fermi constant, $\theta_c$ is the Cabibbo angle, $\mathcal{F}^2_\chi$ ($\chi=+1$ for neutrinos and $\chi=-1$ for antineutrinos) is a reduced single nucleon cross section and $\upsilon_{0}$ is a kinematic factor. Those are defined in Appendix~\ref{apendixa}. The spectral function $S\bigl(p_{m},E_{m}(\mathcal{E}, p_{m})\bigr)$, which describes the possibility to find a nucleon in a nucleus with given momentum and excitation energy of the residual nuclear system, embodies the nuclear model dependence. It will be discussed in 
Sect.~\ref{spectral function} in different models. 
In the case of relativistic nuclear models, as the relativistic Fermi gas, an extra factor $m_N/\sqrt{p_m^2+m_N^2}$ must be inserted inside the integral, according to the Feynman rules~\cite{Bjorken:100769}.

The integrals over $\mathcal{E}$ and $\textbf{p}_{m}$ can be performed using the delta functions, and the following analytical expression for the cross section results:
\begin{widetext}
\begin{equation}
\label{analytic-semics-i}
\frac{d\sigma}{dk'd\Omega_{k^{'}}dp_{N}d\Omega^{L}_{N}} = \frac{(G_{F}\cos{\theta_{c}} k'p_{N})^{2}m_{N}}{8k\varepsilon'E_{N}(2\pi)^{6}}  
 \upsilon_0\mathcal{F}^2_\chi S\bigl(p_{m},E_{m}(\mathcal{E}, p_{m})\bigr) \,\theta(\mathcal{E})\, ,
\end{equation}
\end{widetext}
where the missing momentum and excitation energy in the previous expression are fixed by the conditions:
\begin{eqnarray}
p_m &=& \left|  \textbf{k}' +\textbf{p}_{N} - \textbf{k} \right| ,
\label{eq:pmbar}
\\
{\mathcal{E}} &=& M_{A} + k - \varepsilon' - E_{N} - \sqrt{p^{2}_{m} + M^{2}_{A-1}}
\nonumber\\
 &\simeq& k - \varepsilon' - E_{N} + m_N-E_s 
 \label{eq:calebar}
\end{eqnarray}
and $\upsilon_0\mathcal{F}^2_\chi$ is meant to be evaluated at the values of $p_m$ and $\mathcal{E}$ given by Eqs.~(\ref{eq:pmbar}) and (\ref{eq:calebar}).

%\subsection{\label{neutrino flux}Neutrino flux}

\begin{figure}[!htb] %fluxes normalized
	\centering
	\includegraphics[width=0.5\textwidth]{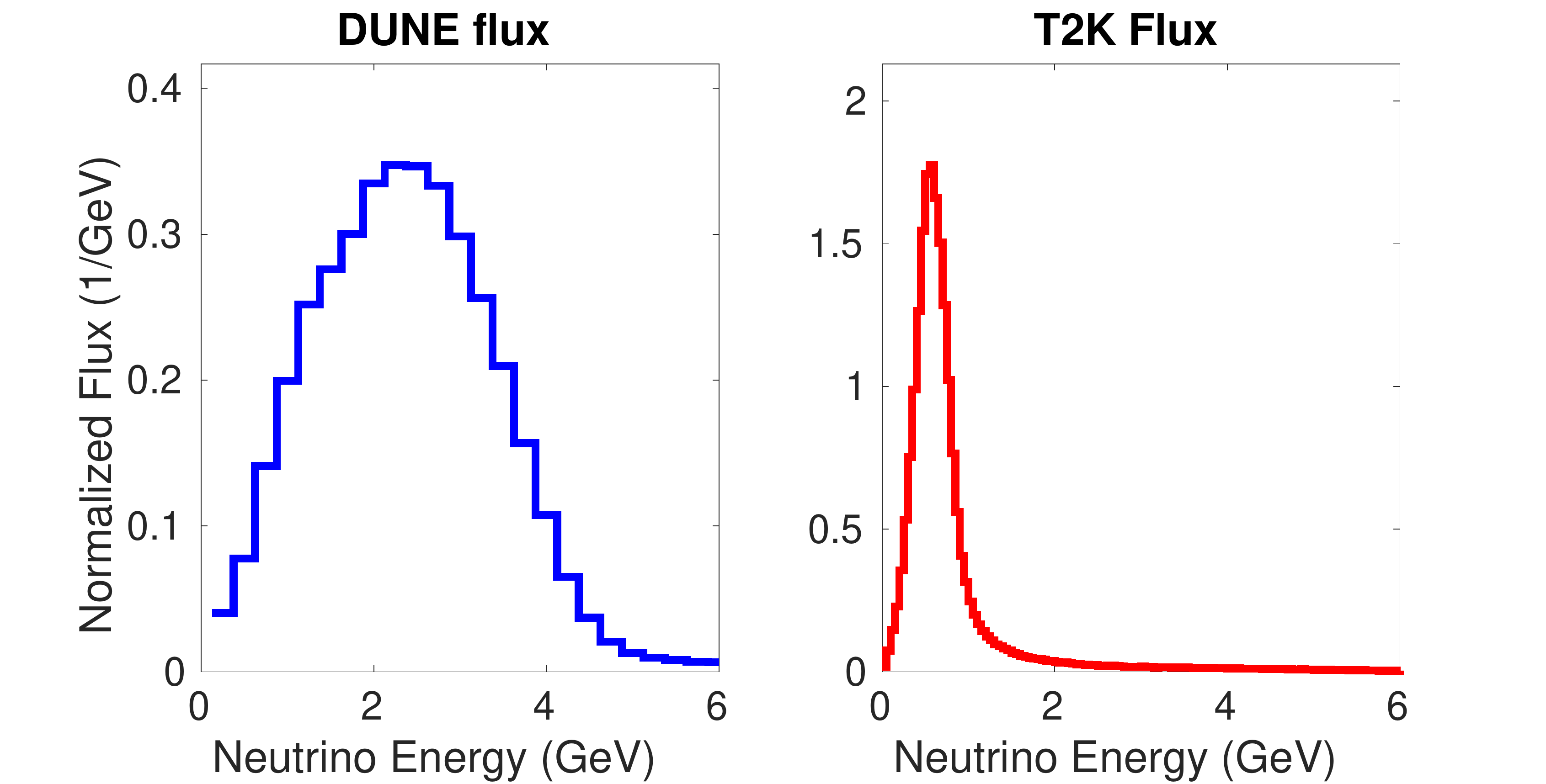}
	\caption{\label{fluxes} Muonic neutrino flux with total area normalized to 1 for the DUNE and T2K experiments.}
\end{figure}

The expressions \eqref{general-semics-i} and \eqref{analytic-semics-i} depend on the variables of the final lepton and the outgoing nucleon and assume that the neutrino energy $k$, and therefore the transferred four-momentum $(\omega,\textbf{q})$, are fixed. However, in comparing the results obtained using this equation with experimental data it is necessary to take into account that in long-baseline oscillation experiments the neutrino beam does not have a well-defined energy: a particle accelerator boosts protons which collide with a target, for instance graphite or beryllium, producing charged pions and kaons. Then, these positive (negative) hadrons decay to produce a flux usually highly dominated by $\nu_\mu$ ($\Bar{\nu}_\mu$): depending of the specific experiment, there is a more or less  extended range of initial neutrino energies that participate in the reaction. As a consequence one needs to average over all the possible energies in order to compare with the experimental data. 
As already mentioned, in this work we will concentrate our attention on two of these experiments, namely T2K~\cite{PhysRevLett.113.241803} 
and DUNE~\cite{Acciarri:2016crz}. 
The corresponding fluxes for the muonic neutrinos are presented in Fig.~\ref{fluxes} \cite{DUNEFlux, T2KFlux}.

%On each experiment, we know the flux which represents the weighting of the contributions of possible momenta from the neutrino associated  with the momentum profile of the neutrino beam. The experiment whose data are we using in this article are  DUNE . The flux allow us to make an average over all possible neutrino energies.

After including an integration over the initial neutrino energy in   %Eq. \ref{general-semics-i} 
Eq.~\eqref{analytic-semics-i} we get  the following flux-averaged semi-inclusive cross section 
%\begin{widetext}
%\begin{align}\label{general-semics}
%< \frac{d\sigma}{dk'd\Omega_{k^{'}}dp_{N}d\Omega^{L}_{N}} > &=\frac{(G_{F}\cos{\theta_{c}} k'p_{N})^{2}m_{N}}{8\varepsilon'E_{N}(2\pi)^{6}}  \int_{0}^{\infty}d\mathcal{E}\int_{0}^{\infty}dk \frac{P(k)}{k}
%\int d^{3}p_{m} \upsilon_0\mathcal{F}^2_\chi S\bigl(p_{m},E_{m}(\mathcal{E}, p_{m})\bigr)\nonumber\\
%&\times\delta(M_{A} + k - \varepsilon' - E_{N} - \sqrt{p^{2}_{m} + M^{2}_{A-1}} - \mathcal{E}) 
%\delta (\textbf{k}- \textbf{k}' -\textbf{p}^{L}_{N} + \textbf{p}_{m}) ,
%\end{align}
%\end{widetext}
\begin{widetext}
\begin{align}\label{general-semics}
\left < \frac{d\sigma}{dk'd\Omega_{k^{'}}dp_{N}d\Omega^{L}_{N}} \right > &=\frac{(G_{F}\cos{\theta_{c}} k'p_{N})^{2}m_{N}}{8\varepsilon'E_{N}(2\pi)^{6}}  \int_{0}^{\infty}dk \frac{P(k)}{k} \upsilon_0\mathcal{F}^2_\chi S\bigl(p_{m},E_{m}(\mathcal{E}, p_{m})\bigr)\theta(\mathcal{E})\, ,
\end{align}
\end{widetext}
where $P(k)$ is the normalized experimental neutrino flux.

\subsection{\label{From semi-inclusive to inclusive} From semi-inclusive to inclusive}

Starting from the above expression one can recover the inclusive cross section by integrating over the variables of the outgoing nucleon. 
In this case we use the $q$-system (see Fig.~\ref{qframe}) where the transfer momentum determines the $z$-axis. This frame presents some special symmetries that simplify significantly the calculation of the various response functions that enter in the scattering process. In Appendix~\ref{apendixa} we present in detail the connection between the variables defined in the  $k$- and $q$-systems and show the explicit calculation of all the weak hadronic responses.
In the $q$-system the outgoing nucleon momentum is given by
\begin{equation}
\textbf{p}_{N}=p_{N}(\cos{\phi_{N}}\sin{\theta_{N}}\textbf{e}_{1} + \sin{\phi_{N}}\sin{\theta_{N}}\textbf{e}_{2} + \cos{\theta_{N}}\textbf{e}_{3}) \, ,
\end{equation}
where we have introduced the unit vectors $\nne_1$, $\nne_2$ and $\nne_3$ that define the reference frame (see Fig.~\ref{qframe}). Note that 
the connection between these unit vectors and the ones introduced in the $k$-system is simply given by a rotation of the angle $\theta_q$ contained in the scattering plane. The angle between the scattering and reaction planes is given by $\phi_N$ while $\theta_N$ determines the direction of the ejected nucleon momentum $\np_N$ with respect to $\nq$.

Obviously the use of the $k$ or $q$-systems does not affect the result for the inclusive cross section since the differential of the solid angle is the same in all frames connected to each other by a rotation: 
\begin{equation}\label{omega-angle}
d\Omega^{L}_{N} = d\Omega^{q}_{N} \equiv d\Omega_N\, .
\end{equation}

By integrating Eq.~\eqref{analytic-semics-i} over $\textbf{p}_N$ we get
\begin{widetext}
\begin{eqnarray}
\frac{d\sigma}{dk' d\Omega_{k'}} &=& \frac{(G_{F}\cos{\theta_{c}} k')^{2}m_{N}}{8k\varepsilon'(2\pi)^{6}}  
 \int_0^\infty dp_N \int d\Omega_N \frac{p_N^2}{E_N}
 \upsilon_0\mathcal{F}^2_\chi S\bigl({p}_{m},E_{m}({\mathcal{E}}, {p}_{m})\bigr) \,\theta({\mathcal{E}}) \, .
\label{inclcs}
\end{eqnarray}
\end{widetext}
Since the only dependence upon the angle $\phi_N$ occurs in the single-nucleon function $\upsilon_0\mathcal{F}^2_\chi$, we define
the $\phi_N$-averaged quantity
\begin{equation}
\int_{0}^{2\pi}d\phi_{N} \upsilon_{0} \mathcal{F}^{2}_{\chi} = 2\pi \upsilon_{0} \overline{\mathcal{F}_{\chi}^2}\, .
\label{eq:Fchi2-av}
\end{equation}
Then we change the integral over $\cos\theta_N$ into an integral over $\mathcal{E}$. The energy conservation relation
\begin{equation}
M_{A} + \omega=\mathcal{E}+E_N+
\sqrt{p^{2}_{N} + q^{2} + m^{2}_{A-1} -2p_{N}q\cos{\theta_{N}}}
\end{equation}
implies
\begin{equation}
d\cos\theta_N = \frac{M_A+\omega-E_N-\mathcal{E}}{p_N q}\,d\mathcal{E} \, .
\end{equation}
Then
\begin{widetext}
\begin{equation}
\label{inclcs2}
\frac{d\sigma}{dk' d\Omega_{k'}} = \frac{(G_{F}\cos{\theta_{c}} k')^{2}m_{N}}{8k\varepsilon' q(2\pi)^5}  
\int_0^\infty dp_N \int_0^\infty d\mathcal{E} \frac{p_N}{E_N} \left(M_A+\omega-E_N-\mathcal{E}\right)
 \upsilon_0\overline{\mathcal{F}^2_\chi} S\bigl(p_m,E_{m}(\mathcal{E}, p_m)\bigr)\, .
\end{equation}
\end{widetext}
Next we change the integral over $p_N$ into an integral over the missing momentum $p_m$ using again the energy conservation written as
\begin{equation}
\sqrt{p_N^2+m_N^2} = \omega+M_A-\mathcal{E}-\sqrt{p_m^2+M_{A-1}^2} \, ,
\end{equation}
which entails
\begin{equation}
\frac{p_{N}}{E_{N}}dp_{N}=\frac{p_{m}}{\sqrt{p^{2}_{m}  +  M_{A-1}^2}} dp_{m}
= \frac{p_{m}}{M_A+\omega-E_N-\mathcal{E}} dp_{m} \, .
\end{equation}
This yields the inclusive cross section as an integral over the $(\mathcal{E},p_m)$ plane previously introduced:
\begin{widetext}
\begin{equation}\label{general-inclusive-cross-section}
\frac{d\sigma}{dk'd\Omega_{k'}} =\frac{(G_{F}\cos{\theta_{c}}k')^{2}m_{N}}{8k\varepsilon'(2\pi)^{5}}\int_{0}^{\infty}d\mathcal{E}
\int_{p_m^-}^{p_m^+} dp_{m}  \frac{p_{m}}{q}\upsilon_{0}\overline{\mathcal{F}^{2}_{\chi}}S\bigl(p_{m},E_{m}(\mathcal{E},p_{m})\bigr)\, ,
\end{equation}
\end{widetext}
where $p^{+}_{m}$ and $p^{-}_{m}$ are the kinematic limits given in Eqs.~\eqref{uplimit-general} and \eqref{downlimit-general}.
Note that the same expression is obtained by integrating Eq.~\eqref{general-semics-i} by exploiting the $\delta$-function in order to integrate over $\textbf{p}_N$.

The inclusive cross section can be also expressed in terms of nuclear responses~\cite{Amaro05}  
\begin{widetext}
\begin{equation}\label{analytical-inclusive-RFG}
\frac{d\sigma}{d\varepsilon'd\cos{\theta_l}} = \sigma_0\bigl( V_{CC}R_{CC} + 2V_{CL}R_{CL} + V_{LL}R_{LL} + 2\chi V_{T'}R_{T'}\bigr)\, ,
\end{equation}
\end{widetext}
where
\begin{equation}
\sigma_0 = \frac{G_F^2\cos{\theta_c}^2}{4\pi}\frac{k'}{\varepsilon'}\upsilon_0\, ,
\end{equation}
$V_K$ are the inclusive leptonic responses given in Appendix~\ref{apendixa} and $R_K$ are the weak nuclear response functions. These embody the whole dependence on the nuclear model and are given by taking the appropriate, charge ($C$), longitudinal ($L$) and transverse ($T$), components of the weak nuclear tensor~\cite{Amaro05}.

As for the semi-inclusive case, before comparing the theoretical predictions with experimental data an average over the neutrino flux must be performed:
\begin{equation}\label{general-inclusive-cross-section-aver}
\left< \frac{d\sigma}{dk'd\Omega_{k'}} \right > =  \int_0^\infty dk P(k) \frac{d\sigma}{dk'd\Omega_{k'}} \, .
\end{equation}

It is worth mentioning that, although this work only deals with the charged-current reaction, by integrating the semi-inclusive cross section over the final lepton variables one  obtains the `$u$-inclusive" cross section~\cite{Barbaro:1996vd} that only depends on the variables of the outgoing nucleon:
\begin{align}
\frac{d\sigma}{dp_{N}d\Omega_{N}}&=\int_{0}^{\infty}dk'\int d\Omega_{k'}
\frac{d\sigma}{dk'd\Omega_{k'}dp_{N}d\Omega_{N}}\, .
\end{align}
This is the measured cross section in neutral current reactions, where the outgoing neutrino cannot be detected.

In the next Section we describe the spectral function $S\bigl(p_{m},E_{m}\bigr)$ within different nuclear models.

\section{\label{spectral function} Nuclear models: the spectral function}
The lepton-hadron cross section is proportional to the contraction of the leptonic and hadronic tensors. However, the hadronic tensor, $W^{\mu\nu}$, for a complex nucleus is in general a complicated object and can be evaluated only under some approximations.
The expression~\eqref{general-semics-i} for the cross section is based on the assumption that it can be factorized as~\cite{Donnelly:2017}
%\begin{widetext}
\begin{align}
W^{\mu\nu}=
&\frac{1}{8\pi}\mathcal{W}^{\mu\nu}(P_{A}-P_{A-1},Q)S(p_{m},E_{m})\, ,
\end{align}
%\end{widetext}
where $\mathcal{W}^{\mu\nu}(P_{A}-P_{A-1},Q)$ is the off-shell single-nucleon response tensor~\cite{DeForest:1983ahx}, corresponding to the scattering with a moving off-shell nucleon, and 
$S(p_m,E_m)$ is the nuclear spectral function, which describes the joint probability of finding a nucleon with given momentum $p_m$ in a nucleus  and of reaching a final state with excitation energy (or, equivalently, missing energy $E_m$) of the residual nuclear system~\cite{Donnelly:2017aaa}. 
The off-shell single-nucleon tensor will be analyzed in Appendix~\ref{apendixa} for the case of charged current reactions.

The spectral function is normalized as~\cite{Donnelly:2019}
\begin{align}
n(p_{m})=\int_{0}^{\infty}dE_{m}S(p_{m},E_{m})\, ,
\end{align}
where $n(p_m)$ is the proton or neutron  momentum distribution.
Therefore $S$ is correctly normalized if the relation
\begin{equation}\label{momentum-normaliz}
\mathcal{N}=\frac{1}{(2\pi)^{3}}\int_{0}^{\infty}dp_{m}p^{2}_{m}n(p_{m})
\end{equation}  
is fulfilled. 
Here, $\mathcal{N}$ is the number of nucleons that are active in the scattering, {\it i.e.} the number of neutrons, $N$, for the case of neutrino scattering  ($CC_{\nu}$) and the number of 
protons, $Z$, for antineutrinos ($CC_{\bar{\nu}}$).

We shall now provide the explicit expression for the spectral function in three simple nuclear models: the Independent-Particle Shell Model (IPSM), the Natural Orbitals Shell Model (NO) and the Relativistic Fermi Gas (RFG).

\subsection{\label{IPSM} Independent-Particle Shell Model (IPSM)}

\begin{figure}[!htbp]
	\begin{center}
		\includegraphics{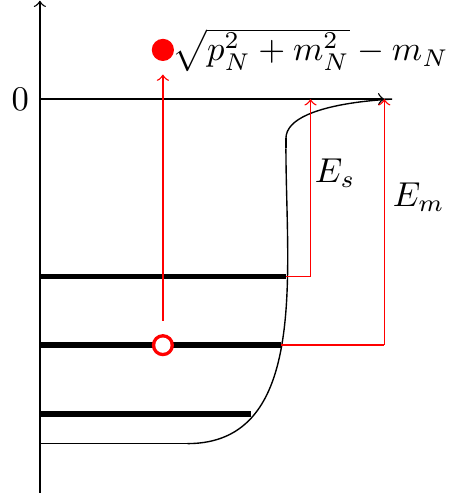}
		\caption{Schematic representation of an electroweak reaction within the Independent-Particle Shell Model.}
		\label{Fig4}
	\end{center}
\end{figure}

In the IPSM  the nucleons are bound by a potential and occupy discrete energy levels $-E_{nlj}$. The scattering process for this model is represented in Fig.~\ref{Fig4}: a nucleon absorbs energy from the probe and produces an on-shell nucleon with relativistic kinetic energy $\sqrt{p^{2}_{N} + m^{2}_{N}} - m_{N}$, leaving a hole in the residual nucleus. The spectral function of this model is~\cite{Donnelly:2019}
\begin{equation}\label{ipsm-spectralfunction}
S_{IPSM}(p_{m},\mathcal{E})=\sum_{n,l,j}(2j +1)n_{nlj}(p_{m})\delta(\mathcal{E} + E_{s} - E_{nlj})\, ,
\end{equation} 
where $n_{nlj}(p_{m})$ is the momentum distribution of a single nucleon in the $nlj$ shell. Energy conservation implies that 
\begin{equation}
\mathcal{E}= E_{nlj} - E_{s}\, .
\label{eq:enconsSM}
\end{equation}
Since $\mathcal{E}\geq 0$, the separation energy in the IPSM model is the energy of the highest shell, as shown in Fig.~\ref{Fig4}.

By inserting Eq.~\eqref{ipsm-spectralfunction} into Eq.~\eqref{analytic-semics-i} we obtain the semi-inclusive cross section for the IPSM:
\begin{align}\label{semi-inclusive_IPSM-noflux}
\frac{d\sigma}{dk'd\Omega_{k'}dp_{N}d\Omega^{L}_{N}}&=\frac{(G_{F}\cos{\theta_{c}}k'p_{N})^{2}m_{N}}{8(2\pi)^{6}\varepsilon'E_{N}}  \sum_{n,l,j}(2j +1)\nonumber\\
&\times\frac{\upsilon_{0}\mathcal{F}^{2}_{\chi}}{k}n_{nlj}(p_{m})\delta(k-k_{0nlj})\, ,
\end{align}
where 
\begin{equation}
k_{0nlj}=\varepsilon' + E_{N} - m_{N} + E_{nlj}
\label{eq:k0nlj}
\end{equation}
and the missing momentum is given by - see  Eq.~\eqref{eq:pmbar} -
\begin{align}
p^{2}_{m}&= k^{2}_{0nlj} + k'^{2} + p^{2}_{N} -2k_{0nlj}k'\cos{\theta_{l}} -2k_{0nlj}p_{N}\cos{\theta^{L}_{N}} \nonumber \\ &+ 2k'p_{N}(\cos{\theta_{l}}\cos{\theta^{L}_{N}}+ \sin{\theta_{l}}\sin{\theta^{L}_{N}}\cos{\phi^{L}_{N}})\, .
\end{align}
From Eq.~\eqref{eq:k0nlj} we see that a nucleon sitting in the shell $nlj$ can only interact, at given $\varepsilon'$ and $E_N$, with a neutrino of energy $k_{0nlj}$.
Therefore, when we average over the flux (Eq.~\eqref{general-semics}), we get
\begin{align}\label{semi-inclusive_IPSM}
\left <\frac{d\sigma}{dk'd\Omega_{k'}dp_{N}d\Omega^{L}_{N}}\right >&=\frac{(G_{F}\cos{\theta_{c}}k'p_{N})^{2}m_{N}}{8(2\pi)^{6}\varepsilon'E_{N}}  \sum_{n,l,j}(2j +1)\nonumber\\
&\times\frac{P(k_{0nlj})}{k_{0nlj}}\upsilon_{0}\mathcal{F}^{2}_{\chi}n_{nlj}(p_{m})\, .
\end{align}

The inclusive cross section (see Eqs.~\eqref{general-inclusive-cross-section}  and \eqref{general-inclusive-cross-section-aver}) becomes in this case
\begin{widetext}
\begin{align}\label{inclusive-IPSM}
\left <\frac{d\sigma}{dk'd\Omega_{k'}}\right >&=\frac{(G_{F}\cos{\theta_{c}}k')^{2}m_{N}}{8(2\pi)^{5}\varepsilon'}  \sum_{n,l,j}(2j+1)\int_{0}^{\infty}dk\frac{P(k)}{qk}\int_{p^-_m}^{p^+_m}dp_{m}p_{m}\upsilon_{0}\overline{\mathcal{F}^{2}_{\chi}}\,
n_{nlj}(p_m)\theta(\omega-E_{nlj})\, ,
\end{align}
\end{widetext}
where, from Eqs.~(\ref{uplimit-general},~\ref{downlimit-general},~\ref{eq:enconsSM}),
\begin{eqnarray}
p^{+}_{m}&=&\sqrt{(\omega-E_{nlj})(\omega - E_{nlj} + 2m_{N})} + q,  \\
p^{-}_{m}&=&\left|\sqrt{(\omega - E_{nlj})(\omega - E_{nlj} + 2m_{N})} -q\right| \, ,
\end{eqnarray}
and  the last theta-function, $\theta(\omega-E_{nlj})$, corresponds to the condition that the transferred energy must be equal or higher than the selected sub-shell level. 

\subsection{\label{Sofia} Natural Orbitals Shell Model (NO).}

This model takes into account nucleon-nucleon correlations and the smearing of the energy eigenstates. It employs natural orbitals, $\psi_{\alpha}(r)$, which are defined as the complete orthonormal set of single-particle wave functions that diagonalize the one-body density matrix (OBDM)~\cite{PhysRev.97.1474}:
\begin{equation}\label{OBDM}
    \rho(\textbf{r},\textbf{r}' )=\sum_{a} N_{a}\psi_{a}^{*}(\textbf{r})\psi_{a}(\textbf{r}')\, ,
\end{equation}
where the eigenvalues $N_{\alpha} (0\le N_{\alpha} \le 1, \sum_{\alpha}N_{\alpha}=A)$ are the natural occupation numbers.

The NO single-particle wave functions, that include short-range nucleon-nucleon (NN) correlations, are used to obtain the occupation numbers and the wave functions in momentum space, {\it i.e.,} the momentum distributions, and from them the spectral function that is given by~\cite{PhysRevC.89.014607} 
%{\red{I SIMPLIFIED 2/4, PLEASE CHECK}}
\begin{equation}\label{spectral-sofia}
%    S_{NO}(p_{m},\mathcal{E})=\frac{1}{4\pi A}\sum_{i}2(2j_{i} + 1)N_{i}|\psi_{i}(p_{m})|^{2}L_{\Gamma_{i}}(\mathcal{E}-\mathcal{E}_{i}),
    S_{NO}(p_{m},\mathcal{E})=\frac{1}{2\pi A}\sum_{i}(2j_{i} + 1)N_{i}|\psi_{i}(p_{m})|^{2}L_{\Gamma_{i}}(\mathcal{E}-\mathcal{E}_{i})\, ,
\end{equation}
where  $A$ is the mass number and the dependence upon the energy is given by the Lorentzian function:
\begin{equation}\label{lorentzian}
    L_{\Gamma_{i}}(\mathcal{E}-\mathcal{E}_{i})=\frac{1}{2\pi}\frac{\Gamma_{i}}{(\mathcal{E} - \mathcal{E}_{i})^{2} + (\Gamma_{i}/2)^{2}}\, ,
\end{equation}
where $\Gamma_{i}$ is the width for a given single-particle state and $\mathcal{E}_{i}$ is the energy eigenvalue of the state. 

The semi-inclusive cross section in this model is given by 
%{\red{I SIMPLIFIED 2/16, PLEASE CHECK}}:
\begin{align}\label{seminclusive-sofia}
%<\frac{d\sigma}{dk'd\Omega_{k'}dp_{N}d\Omega^{L}_{N}}>=\int d\mathcal{E} \frac{(G_{F}\cos{\theta_{c}}k' p_{N})^{2}m_{N}P(k_{s})}{16k_{s}\varepsilon'E_{N}(2\pi)^{7}A} \times \nonumber \\
%\sum_{i}2(2j_{i} +1)N_{i}|\psi_{i}(p_{m})|^{2}L_{\Gamma_{i}}(\mathcal{E} - \mathcal{E}_{i})\upsilon_{0}\mathcal{F}^{2}_{\chi},
\left <\frac{d\sigma}{dk'd\Omega_{k'}dp_{N}d\Omega^{L}_{N}}\right >=\int d\mathcal{E} \frac{(G_{F}\cos{\theta_{c}}k' p_{N})^{2}m_{N}P(k)}{8k\varepsilon'E_{N}(2\pi)^{7}A} \nonumber \\
\times\sum_{i}(2j_{i} +1)N_{i}|\psi_{i}(p_{m})|^{2}L_{\Gamma_{i}}(\mathcal{E} - \mathcal{E}_{i})\upsilon_{0}\mathcal{F}^{2}_{\chi}\, ,
\end{align}
where the neutrino momentum is 
\begin{equation}\label{neutrino-momentum-sofia}
    k=E_{s} + E_{N} + \varepsilon' - m_{N} + \mathcal{E}
\end{equation}
and
\begin{align}\label{missing-momentum-sofia}
p^{2}_{m}&= k^{2} + k'^{2} + p^{2}_{N} -2kk'\cos{\theta_{l}} -2kp_{N}\cos{\theta^{L}_{N}} \nonumber \\ &+ 2k'p_{N}(\cos{\theta_{l}}\cos{\theta^{L}_{N}}+ \sin{\theta_{l}}\sin{\theta^{L}_{N}}\cos{\phi^{L}_{N}})\, .
\end{align}
Note that in this case the integral over $\mathcal{E}$ has to be performed numerically because, unlike in the IPSM model, the single-particle energies are not discrete.

The inclusive cross-section is 
\begin{align}\label{inclusive-sofia}
%<\frac{d\sigma}{dk'd\Omega_{k'}}>=\frac{(G_{F}\cos{\theta_{c}}k')^{2}m_{N}}{16A\varepsilon'(2\pi)^{6}}\int_{0}^{\infty}d\mathcal{E}\int_{0}^{\infty}dk \int_{p^-_m}^{p^+_m} dp_{m} \nonumber \\  \times  \frac{P(k)p_{m}}{kq}\upsilon_{0}\overline{\mathcal{F}^{2}_{\chi}}\sum_{i}2(2j_{i} +1)N_{i}|\psi_{i}(p_{m})|^{2}L_{\Gamma_{i}}(\mathcal{E} - \mathcal{E}_{i}) ,
\left <\frac{d\sigma}{dk'd\Omega_{k'}}\right >=\frac{(G_{F}\cos{\theta_{c}}k')^{2}m_{N}}{8A\varepsilon'(2\pi)^{6}}\int_{0}^{\infty}d\mathcal{E}\int_{0}^{\infty}dk \int_{p^-_m}^{p^+_m} dp_{m} \nonumber \\  \times  \frac{P(k)p_{m}}{kq}\upsilon_{0}\overline{\mathcal{F}^{2}_{\chi}}\sum_{i}(2j_{i} +1)N_{i}|\psi_{i}(p_{m})|^{2}L_{\Gamma_{i}}(\mathcal{E} - \mathcal{E}_{i}) \, ,
\end{align}
where the limits in the missing momentum are given by Eqs.~\eqref{uplimit-general} and \eqref{downlimit-general}.

\subsection{\label{RFG} Relativistic Fermi Gas (RFG)}

\begin{figure}[!tb]
	\begin{center}
		\includegraphics{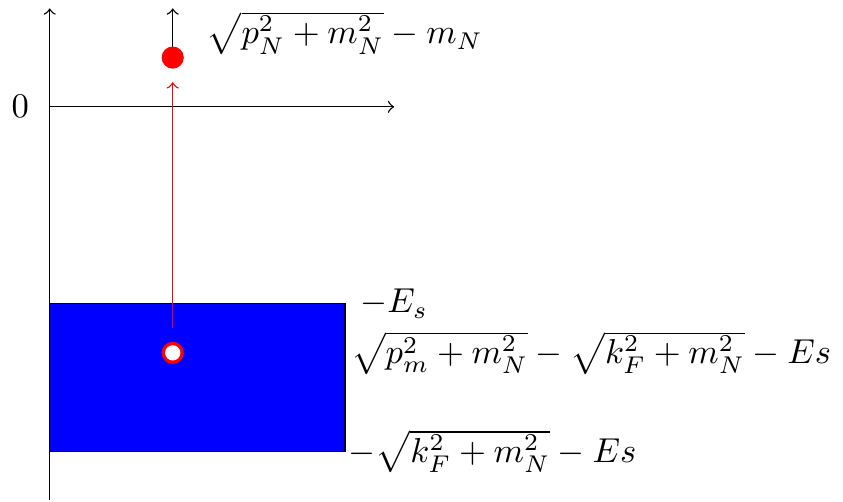}
		\caption{Schematic representation of an electroweak reaction within the relativistic Fermi gas model.}
		\label{Fig3}
	\end{center}
\end{figure} %Fig.~\ref{fig:epsart}

This model consists in describing the nucleus as an infinite gas of free relativistic nucleons that, in the nuclear ground state, occupy all the levels up to the Fermi momentum $k_{F}$ while the levels above that are empty. The Fermi momentum is the only free parameter of the model. It is usually fitted to the width of the quasielastic peak in electron scattering data~\cite{Maieron:2001it} and varies with the nucleus.
Since in the pure RFG the nucleons are unbound, the separation energy in this model is negative~\cite{Cenni_1997}: 
\begin{equation}
E_s^{RFG}=-T_F\equiv -E_F+m_N \, , 
\label{EsRFG}
\end{equation}
being $E_F=\sqrt{k_F^2+m_N^2}$ the Fermi energy and $T_F$ the corresponding kinetic energy. In order to cure this problem and to be more consistent with the other models considered in this work, we adopt the prescription of Ref.~\cite{Donnelly:2019}: we shift the RFG energies by a constant in such a way that the last occupied level in the Fermi sea coincides with -$E_s$, as shown in Fig.~\ref{Fig3}. This amounts to putting the nucleons off-shell by changing their free energy as
\begin{equation}
E=\sqrt{p^2+m_N^2} \ \ \longrightarrow \ \ E-\left(E_F + E_s\right) \, .
\end{equation}
In the scattering process, illustrated in Fig.~\ref{Fig3}, a nucleon with momentum $p_m$ absorbs enough energy to leave the Fermi sea, $E_F + E_s - \sqrt{p^{2}_{m} + m^{2}_{N}}$, and to be knocked out  with positive kinetic energy $\sqrt{p^{2}_{N} + m^{2}_{N}}-m_N$, namely
\begin{equation}\label{omega-fermi}
\omega=T_F + E_{s} + \sqrt{p^{2}_{N} + m^{2}_{N}}  -\sqrt{p_{m}^{2} + m^{2}_{N}}\, .
\end{equation}
The missing energy is then
\begin{equation}
E_m = E_F+E_s-\sqrt{p^{2}_{m} + m^{2}_{N}}
\end{equation}
and the excitation energy of the residual nucleus
\begin{equation}
\mathcal{E} = E_m-E_s = E_F-\sqrt{p^{2}_{m} + m^{2}_{N}}\, .
\end{equation}
The normalized spectral function is~\cite{Donnelly:2019}
\begin{equation}
S_{RFG}(p_m,\mathcal{E})=\frac{3(2\pi)^{3}\mathcal{N}}{k^{3}_{F}}\theta(k_{F} - p_{m}) \delta\left(\mathcal{E} - E_F + \sqrt{p^{2}_{m} + m^{2}_{N}}\right) 
\end{equation}
with $\mathcal{N}$ the number of neutrons (protons) for incoming neutrinos (antineutrinos).

By inserting this spectral function in the general expression~\eqref{analytic-semics-i} we get the semi-inclusive RFG cross section at fixed neutrino energy $k$:
\begin{widetext}
\begin{equation}
\label{semi-inclusive rfgk}
\frac{d\sigma}{dk'd\Omega_{k'}dp_{N}d\Omega^{L}_{N}} = 
\frac{3\mathcal{N} (G_F\cos{\theta_c} k'p_{N} m_N)^2} 
{8k\varepsilon' E_{N}(2\pi k_F)^3 \sqrt{{p}^2_m+m^2_N}}  
 \upsilon_0\mathcal{F}^2_\chi \theta(k_{F} - {p}_{m}) \theta(p_N-k_F) \delta\left({\mathcal{E}} - E_F + \sqrt{{p}^{2}_{m} + m^{2}_{N}}\right) \, ,
\end{equation}
\end{widetext}
where, as mentioned in Sec.~\ref{Semi-inclusive}, the relativistic factor $m_{N}(\sqrt{{p}^{2}_{m} + m^{2}_{N}})^{-1}$ has been introduced, and the values of ${p}_m$ and ${\mathcal{E}}$ entering in the previous expression are given
by Eqs.~(\ref{eq:pmbar}, \ref{eq:calebar}). The theta-function $ \theta(p_N-k_F)$ represents the Pauli-blocking restriction on the momentum of the ejected nucleon, which must be larger than $k_F$.

The flux-averaged  semi-inclusive cross section for the RFG model is then
\begin{widetext}
\begin{align}\label{semi-inclusive rfg}
\left <\frac{d\sigma}{dk'd\Omega_{k'}dp_{N}d\Omega^{L}_{N}}\right >&=\frac{3\mathcal{N}(G_{F}\cos{\theta_{c}}m_{N}k'p_{N})^{2}}{8(2\pi k_{F})^{3}\varepsilon'E_{N}}  \int_{0}^{\infty}dk\frac{P(k)}{k}\frac{\upsilon_{0}\mathcal{F}^{2}_{\chi}}{\sqrt{( \textbf{p}_B-\textbf{k} )^{2} + m^{2}_{N}}} \nonumber \\
&\times \delta\biggl(k - E_B + \sqrt{( \textbf{p}_B-\textbf{k} )^{2} + m^{2}_{N}}\biggr)\theta(k_{F} -\left| \textbf{p}_B-\textbf{k} \right|) \theta(p_N-k_F)\, ,
\end{align}
\end{widetext}
where for brevity  we have defined the  following variables:
\begin{align}
\textbf{p}_{B}= \textbf{k}'& + \textbf{p}_{N}, \\
E_{B}=\varepsilon' + E_{s}& + T_{F} + E_{N}\, .
\end{align}
The delta function can be recast as 
\begin{align}
\delta\biggl(k - E_{B} + \sqrt{(\textbf{p}_{B} - \textbf{k})^{2} + m^{2}_{N}}\biggr) =& \nonumber \\ 
\frac{\sqrt{(\textbf{p}_{B} - \textbf{k})^{2} + m^{2}_{N}}}{E_{B} - p_{B}\cos{\theta_{B}}} \delta(k - k_{0})
\end{align}
with
\begin{eqnarray}
k_{0}&=&\frac{E^{2}_{B} - p^{2}_{B}-m^{2}_{N}}{2\bigl(E_{B} - p_{B}\cos\theta_{B} \bigr)} \\
\cos{\theta_{B}} &=& \frac{k'\cos{\theta_{l}} + p_{N}\cos{\theta^{L}_{N}}}{p_{B}}\, ,
\end{eqnarray}
and used to perform the integral over $k$. 
Finally, the flux-averaged semi-inclusive cross section is
\begin{widetext}
\begin{align}\label{semi-inclusive_RFG}
\left <\frac{d\sigma}{dk'd\Omega_{k'}dp_{N}d\Omega^{L}_{N}}\right >&=\frac{3\mathcal{N}(G_{F}\cos{\theta_{c}}m_{N}k'p_{N})^{2}}{8(2\pi k_{F})^{3}\varepsilon'E_{N}}  \frac{P(k_{0})}{k_{0}}\frac{\upsilon_{0}\mathcal{F}^{2}_{\chi}}{E_{B} - p_{B}\cos{\theta_{B}}}\theta(k_{F} - {p}_{m}) \theta(p_N-k_F) 
\end{align}
\end{widetext}
with the missing momentum given by 
\begin{eqnarray}\label{pmiss_RFG}
&&{p}^{2}_{m}= k^{2}_0 - 2k'k_0\cos{\theta_{l}} + k'^{2} + p^{2}_{N} -2k_{0}p_{N}\cos{\theta^{L}_{N}}  \nonumber\\
&&+ 2k'p_{N}(\cos{\theta_{l}}\cos{\theta^{L}_{N}} 
+ \sin{\theta_{l}}\sin{\theta^{L}_{N}}\cos{\phi^{L}_{N}}).
\end{eqnarray}

If we integrate Eq.~\eqref{semi-inclusive rfg} over $\textbf{p}_N$ and exploit the delta-function to perform the integral over $\theta^L_N$ we obtain the flux-averaged inclusive cross section (\ref{general-inclusive-cross-section}, \ref{general-inclusive-cross-section-aver})
\begin{widetext}
\begin{align}\label{inclusive-fermi}
\left <\frac{d\sigma}{dk'd\Omega_{k'}}\right >=\frac{3\mathcal{N}(G_{F}\cos{\theta_{c}}m_{N}k')^{2}}{8\varepsilon'k_{F}^3 (2\pi)^{2}}\int_{0}^{\infty}dk\frac{P(k)}{qk} 
\int_{p_m^-}^{k_F}dp_{m}\frac{p_{m}}{\sqrt{p^{2}_{m} + m^{2}_{N}}}\upsilon_{0}\overline{\mathcal{F}^{2}_{\chi}}\, \theta(p_{N} - k_{F})\, , 
\end{align}
\end{widetext}
where~\cite{Donnelly:2019}
\begin{equation}\label{pminus}
p_m^{-}=\left|\frac{\sqrt{(\omega - E_{s} - T_{F})^{2}\eta(\eta + 4m^{2}_{N})}}{2\eta} - \frac{q}{2}\right|
\end{equation}
and
\begin{equation}
\eta=q^{2} - (\omega - E_{s} - T_{F})^{2}\, .
\end{equation}
These kinematic limits can be recast in the form
\begin{equation}
p_{m}^{\rm min} \leq p_m \leq  p_{m}^{\rm max} \,,
\end{equation}
with
\begin{eqnarray}
p_{m}^{\rm min} &=&  \left| \frac{q}{2}-\frac{\overline{\omega}}{2} \sqrt{1+\frac{4m_N^2}{\eta}} \right| 
\\
p_{m}^{\rm max}  &=& k_{F},
\end{eqnarray}
where we have introduced the quantity
\begin{equation}
\label{omegabar-definition1}
\overline{\omega} =\omega -E_s - T_F
\end{equation}
and we have used the condition $p_m\leq k_F$. The corresponding limits on the momentum of the ejected nucleon, $p_N=|{\bf p}_m+{\bf q}|$, are
\begin{equation}
{\rm Max}\left\{ k_F, \frac{q}{2}+\frac{\overline{\omega}}{2} \sqrt{1+\frac{4m_N^2}{\eta}} \right\} \leq p_N \leq  k_{F}+q \,,
\end{equation}
where we have imposed the Pauli blocking condition $p_N\geq k_F$.

One advantage of using the RFG model for the description of the neutrino-nucleus inclusive cross section is that the integral over $p_m$ in Eq.~\eqref{inclusive-fermi} can be performed analytically leading to relatively simple 
expressions~\cite{electron-vs-neutrino,PhysRevC.71.065501}. 
In particular, the weak response functions in the RFG can be written as 
%{\ttblue Maria: I have used the shifted scaling variable $\psi'$ since we have already introduced $\overline\omega$}
%
\begin{equation}
R_K = \mathcal{N}\Lambda_0 U_Kf(\psi')\, .
\end{equation}
The  expressions for $U_K$ and the factor $\Lambda_0$ can be found in the Appendix C of~\cite{PhysRevC.71.065501} and $f(\psi')$ is the RFG scaling function
\begin{equation}
f(\psi') = \frac{3}{4}\bigl(1 - \psi'^2\bigr)\theta\bigl(1 - \psi'^2\bigr)\, ,
\end{equation}
where the scaling variable $\psi'$ is the minimum kinetic energy of the bound nucleon in units of the nucleon mass, i.e.~\cite{Alberico:1988bv},
\begin{equation}
\psi'^2 = \frac{1}{\xi_F}\left(\frac{\sqrt{{p_{m}^{-}}^2 + m_N^2}}{m_N}-1\right) \,.
\end{equation}
These expressions allow us to check the reliability of the results presented in the next section. 

%\vspace{0.5cm}

%In the following section we use these three models and compare their predictions for the semi-inclusive and inclusive cross sections .

\section{\label{Results} Results}

In this section we present and discuss semi-inclusive and inclusive results for the IPSM, NO and RFG models illustrated in the previous Section, considering two different  neutrino fluxes (DUNE and T2K) and two nuclear targets,  $^{40}$Ar and $^{12}$C.

Let us briefly summarize the main features of each model.
For the IPSM we describe the bound nucleon states as self-consistent Dirac-Hartree solutions, derived within a relativistic mean field (RMF) approach using a Lagrangian containing $\rho$, $\sigma$ and $\omega$ mesons~\cite{Serot97,Horowitz:1981xw}. These relativistic single-particle wave functions are used to obtain the momentum distribution of each shell for both types of nucleons in the nucleus.
In the NO approach the single-particle wave functions are non-relativistic, but they include short-range NN correlations. 
They are used to generate the momentum distributions for the different shells. In this case the energy delta-functions corresponding to each shell are replaced by  Lorentzian distributions (see~\cite{Ivanov:2013saa,Antonov:2011bi} for details).
The RFG is the simplest among the three models and does not account for the shell structure of the nucleus, because the nucleons are non-interacting. However the model is fully relativistic and is still employed in most event generators used in experiments.

%{\ttred Juan: What about the new sentence following Maria's suggestion?}

\begin{figure}[!htbp] %16o distribution
    \captionsetup[subfigure]{labelformat=empty}
	\centering
	\subfloat[]{{\includegraphics[width=0.49\textwidth]{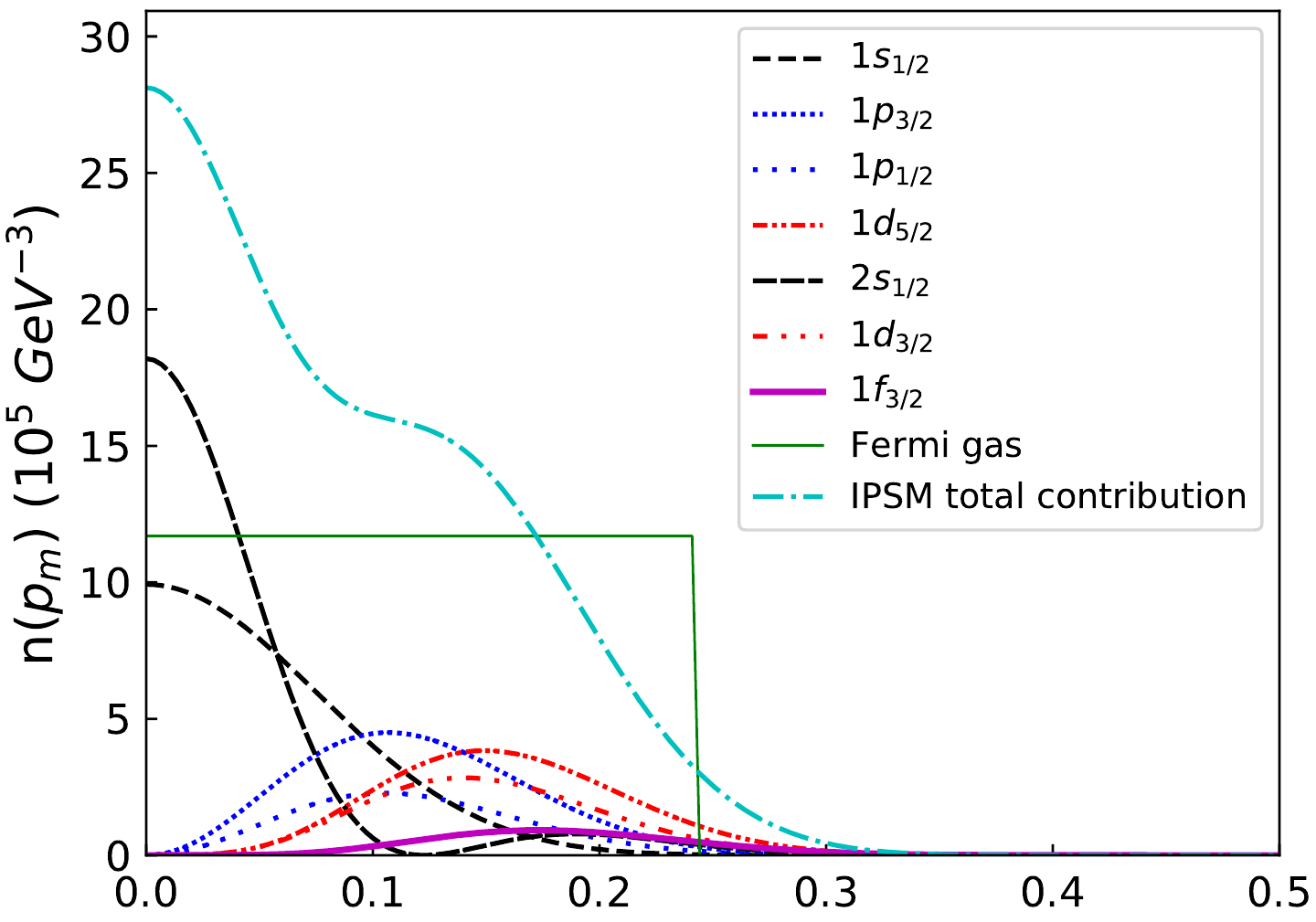} }}%
	\qquad
	\subfloat[]{{\includegraphics[width=0.49\textwidth]{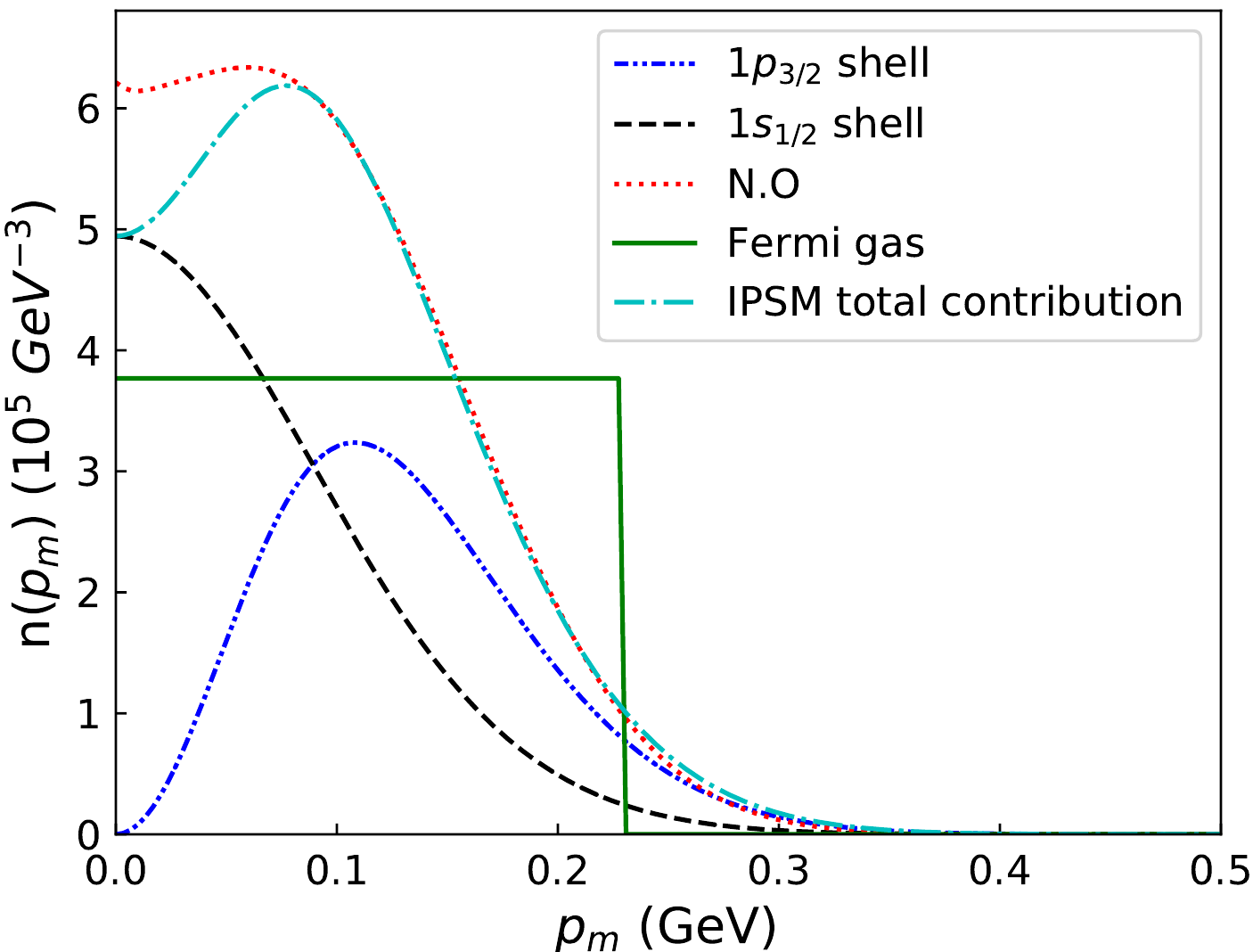} }}%
	\caption{\label{mom_dis_o16} Top panel: IPSM and RFG momentum distributions normalized according to Eq.~(\ref{momentum-normaliz}) for $^{40}$Ar using $k_F=$ 0.241 GeV. Bottom panel: Same as in the top panel but for $^{12}$C and including also the NO momentum distribution. The Fermi momentum in this case is fixed to 0.228 GeV.}
\end{figure}

 In the discussion that follows we focus on the RFG and IPSM in the case of argon (DUNE experiment), whereas for carbon (T2K experiment) we explore in addition the results provided by NO.
Momentum distributions for $^{40}$Ar and $^{12}$C are presented in Fig.~\ref{mom_dis_o16} showing the complex dependence upon the missing momentum for the IPSM and the NO models, whereas the RFG distribution is basically a step function. As we will see later, under some fixed kinematics, the shape and magnitude of the semi-inclusive cross section will be strongly dependent on the momentum distribution, hence the results for the RFG model will be quite different from the other two.

%{\ttred Juan: Remove (a) and (b) from the figure. Also the title in both panels should be removed. Finally, $p_m$ can be shown only for the bottom panel.}

%\begin{figure*}[!htbp] %12C distribution
	%\centering
	%\subfloat[]{{\includegraphics[width=0.49\textwidth]{pictures/Results/distributions_fluxes/mom_disc12_linear.pdf} }}%
	%\qquad
	%\subfloat[]{{\includegraphics[width=0.49\textwidth]{pictures/Results/distributions_fluxes/mom_disc12_log.pdf} }}%
	%\caption{\label{mom_dis_c12}Same as Fig.~\ref{mom_dis_o16} but for $^{12}$C using $k_F =$ 0.228 GeV.}
%\end{figure*}

\subsection{\label{semi-inclusive cross sections} Semi-inclusive cross sections.}

In sections~\ref{IPSM}, \ref{Sofia} and \ref{RFG} we deduced the semi-inclusive cross sections for the IPSM, NO and RFG  models using the spectral function defined for each case. Since we want to shed some light on the discrepancies between the models, in what follows we select some specific kinematics where the dependence of the momentum distribution with the missing momentum is particularly relevant.

\begin{figure*}[!htb] %16O fermi phin:180/165 DUNE
    \captionsetup[subfigure]{labelformat=empty}
	\centering
	\subfloat[]{{\includegraphics[width=0.48\textwidth]{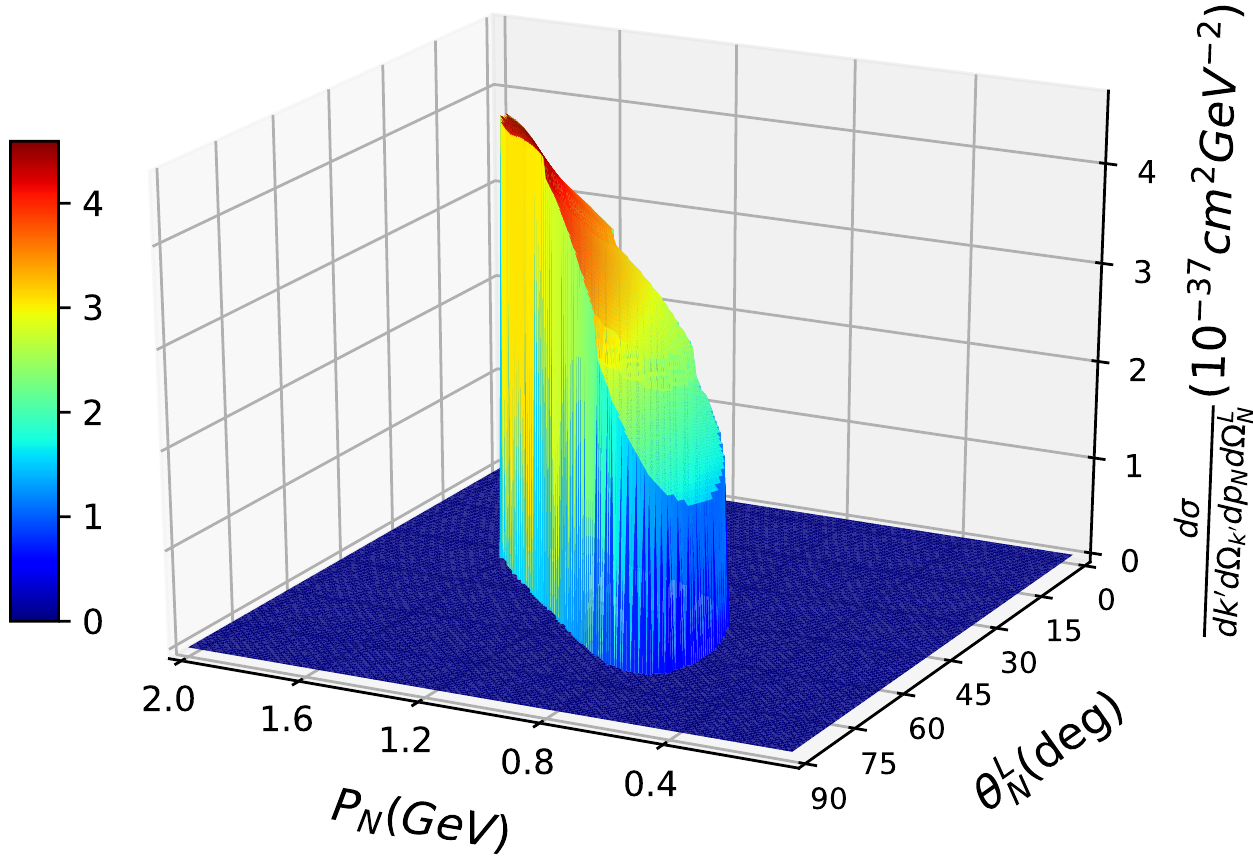} }}%
	\hspace{3.5mm}%
	%\qquad
	\subfloat[]{{\includegraphics[width=0.48\textwidth]{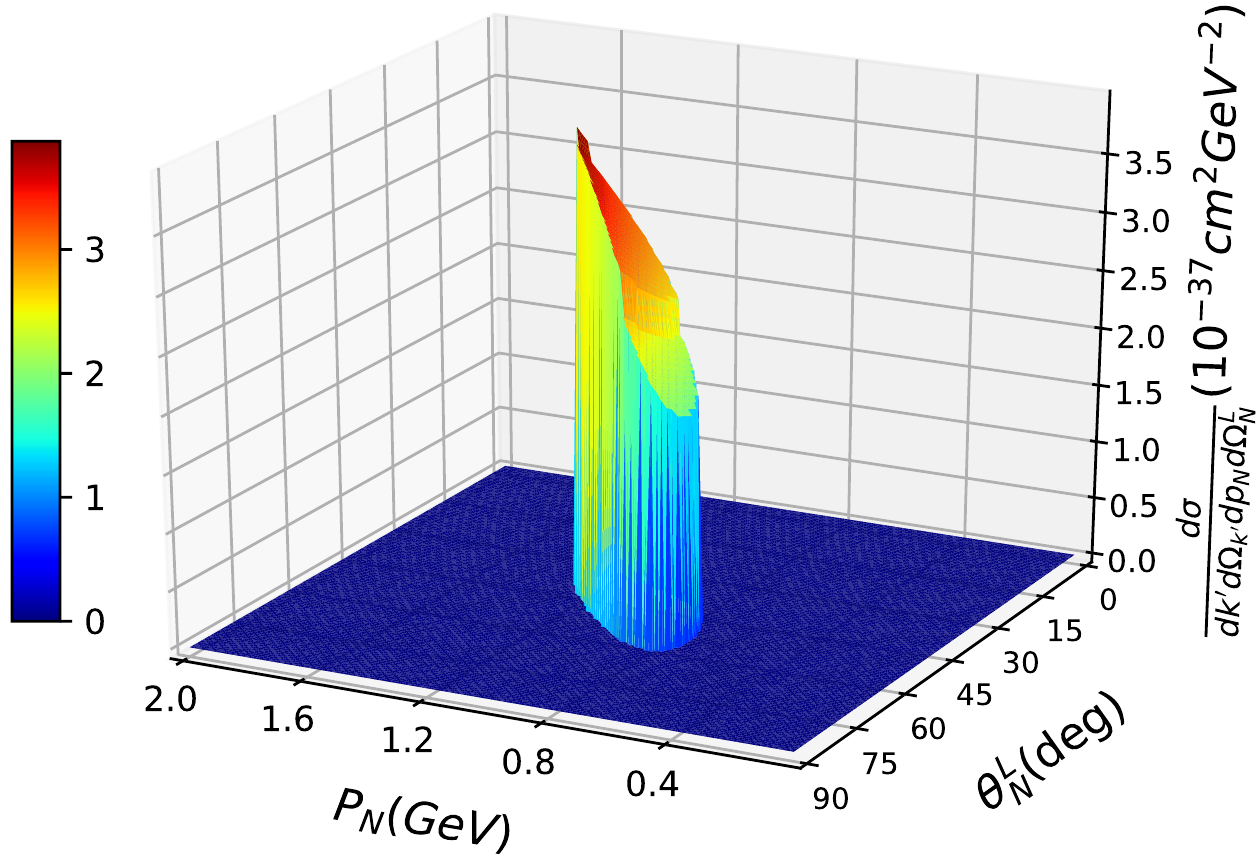} }}%
	\qquad
	{
		\subfloat[]{{\includegraphics[width=0.48\textwidth]{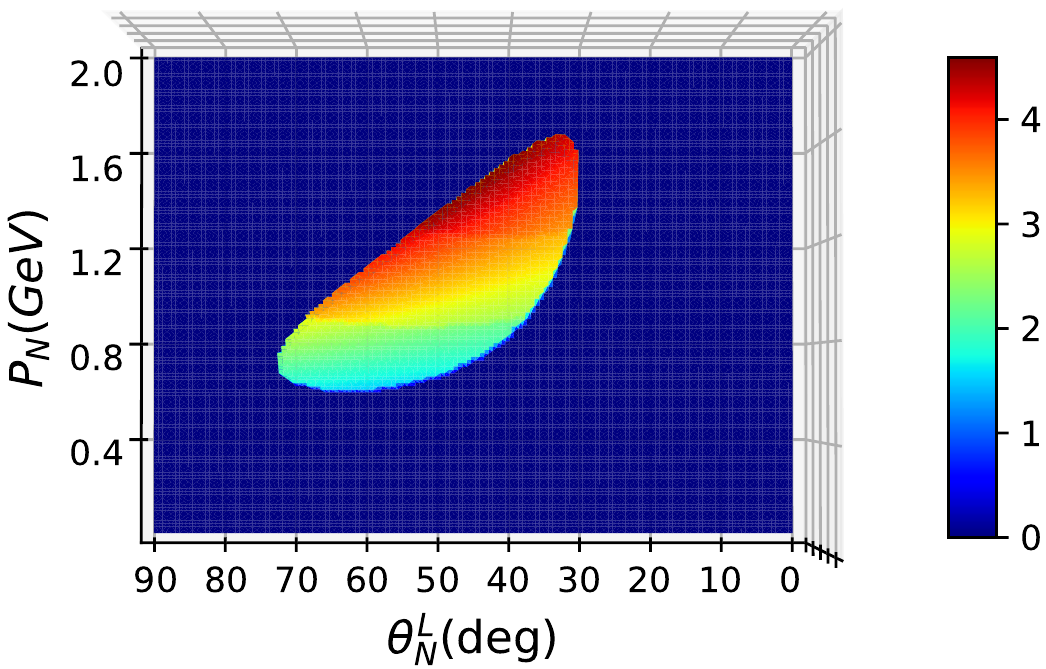} }}%
		\hspace{3.5mm}%
		\subfloat[]{{\includegraphics[width=0.48\textwidth]{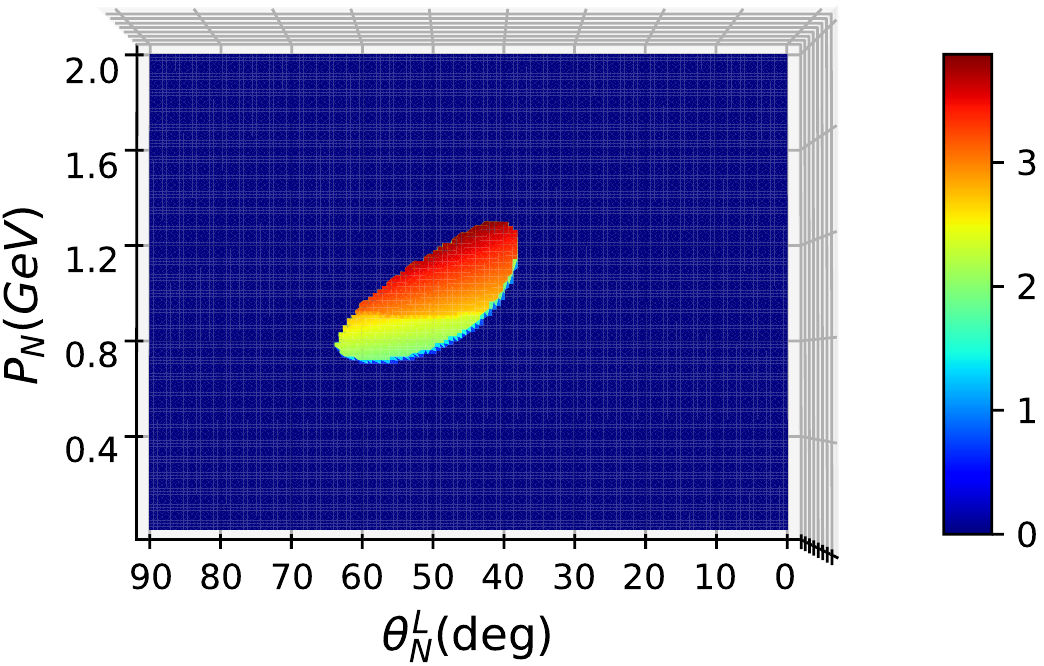} }}%
	}
	\caption{\label{16OfermidonneDUNE}Semi-inclusive cross section for $^{40}$Ar and DUNE flux using $k'$ = 1.5 GeV, $\theta_l$ = 30$\degree$, $\phi_N^L$ = 180$\degree$ (left panels) and $\phi_N^L$ = 165$\degree$ (right panels) for the RFG model.}
	\vspace{0.3cm}
\end{figure*}

We begin considering the semi-inclusive cross sections for IPSM and RFG with muon momentum $k'$ = 1.5 GeV, muon scattering angle $\theta_l$ = 30$\degree$ and two different values for the azimuthal angle defined in the $k$-system, namely $\phi_N^L$ = 180$\degree$ and $\phi_N^L$ = 165$\degree$, as function of the ejected nucleon momentum $p_N$ and the angle $\theta_N^L$. We consider $^{40}$Ar as the target and use the neutrino flux corresponding to DUNE. Results for the RFG model are presented in Fig.~\ref{16OfermidonneDUNE} 
%for $^{40}$Ar 
using $k_F$ = 0.241 GeV for two different points of view, namely ``side" (top panels) and ``hawk" (bottom) views. The graphs on the left correspond to 
$\phi_N^L=180\degree$ while the ones on the right to $\phi_N^L=165\degree$. In both cases the
shape of the cross section is simple being only different from zero in a very well-defined area given by the condition $p_m \leq k_F$. Note that the region where the cross section exists for $\phi_N^L$ = 165$\degree$ is significantly reduced compared with the case at $\phi_N^L$ = 180$\degree$. In the former kinematics, only a few points in the plane ($p_N$, $\theta_N^L$) fulfil the condition that the corresponding missing momentum defined in Eq.~\eqref{pmiss_RFG} is smaller than $k_F$.

\begin{figure*}[!htbp] %16O rmf phin:180/165 DUNE
    \captionsetup[subfigure]{labelformat=empty}
	\centering
	\subfloat[]{{\includegraphics[width=0.48\textwidth]{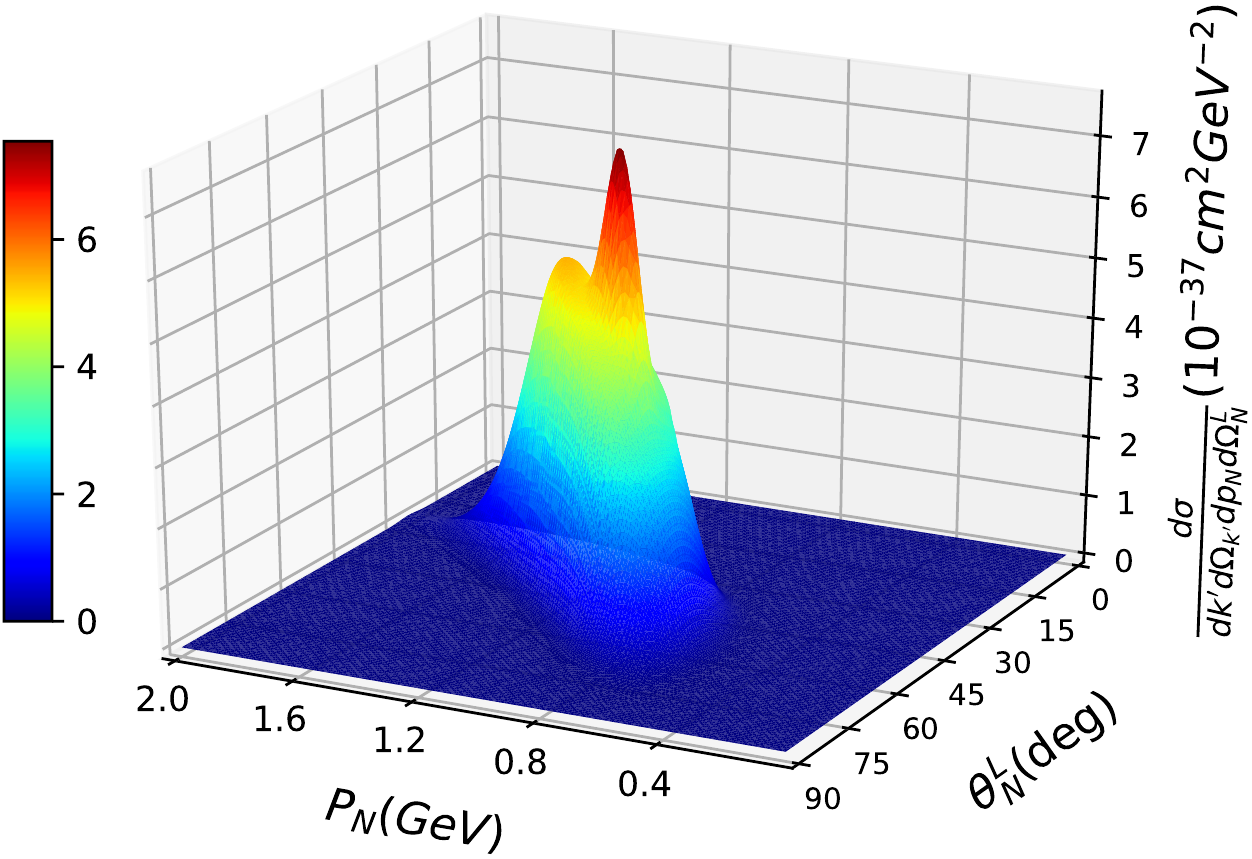} }}%
	\hspace{3.5mm}%
	%\qquad
	\subfloat[]{{\includegraphics[width=0.48\textwidth]{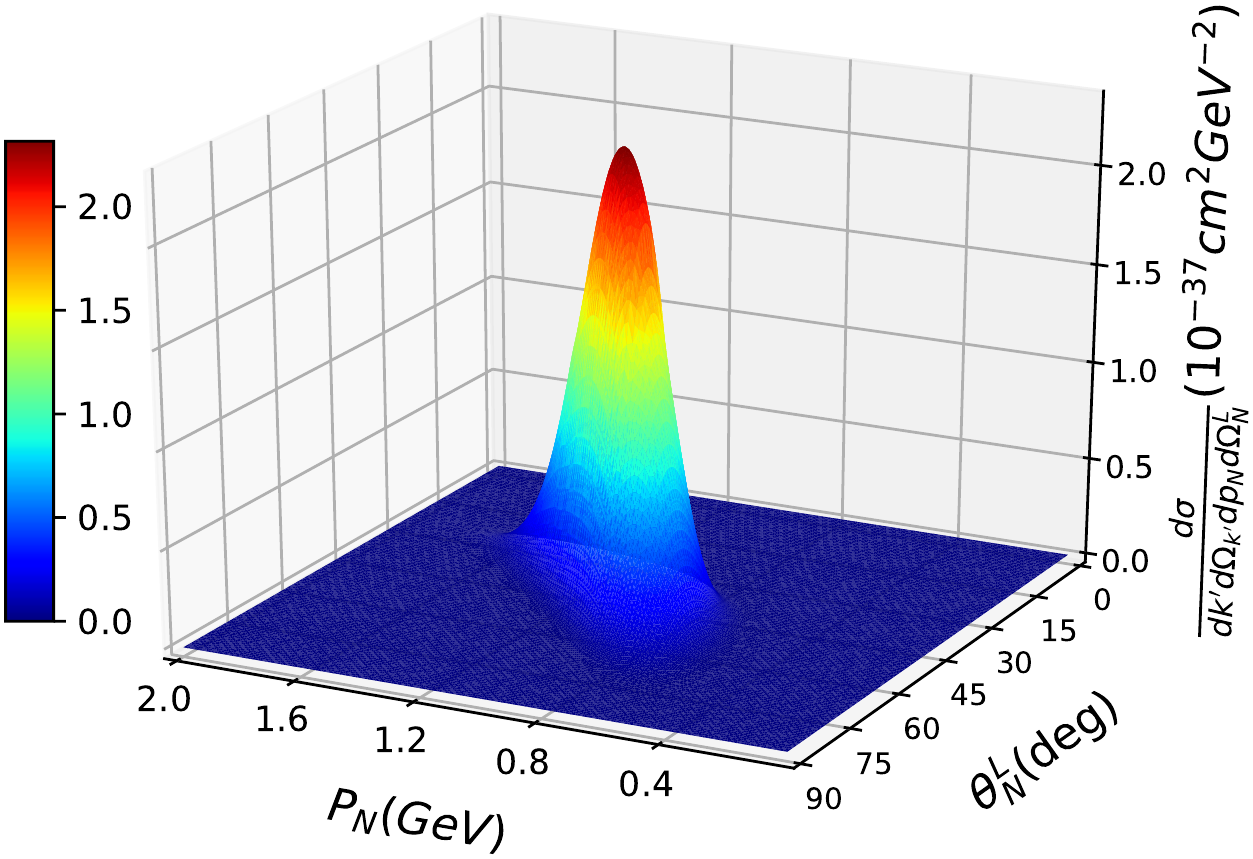} }}%
	\qquad
	{
		\subfloat[]{{\includegraphics[width=0.48\textwidth]{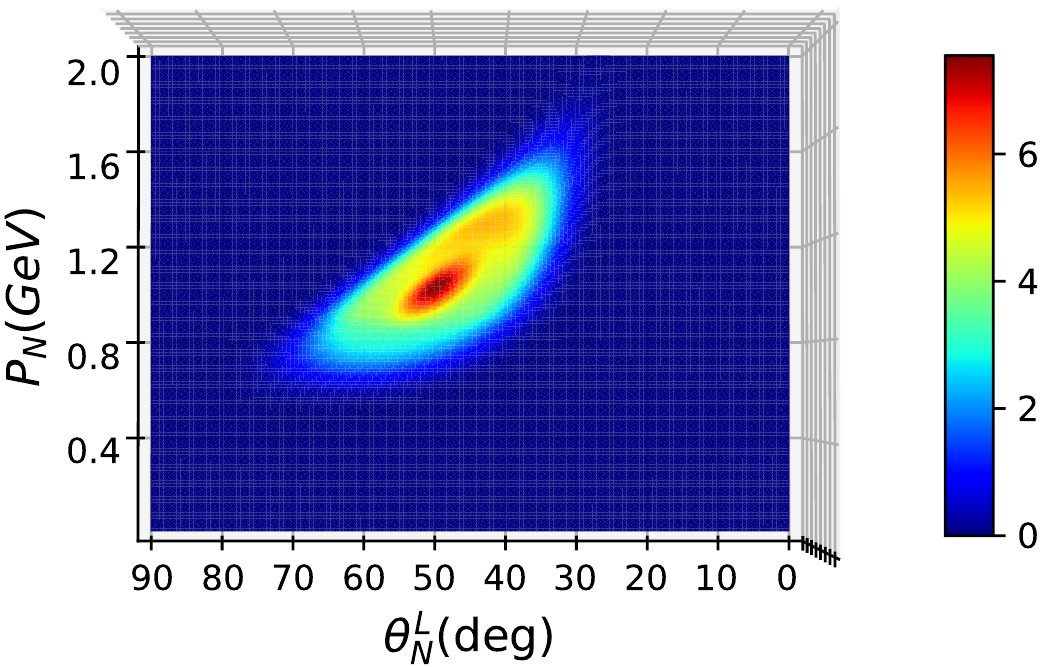} }}%
		\hspace{3.5mm}%
		\subfloat[]{{\includegraphics[width=0.48\textwidth]{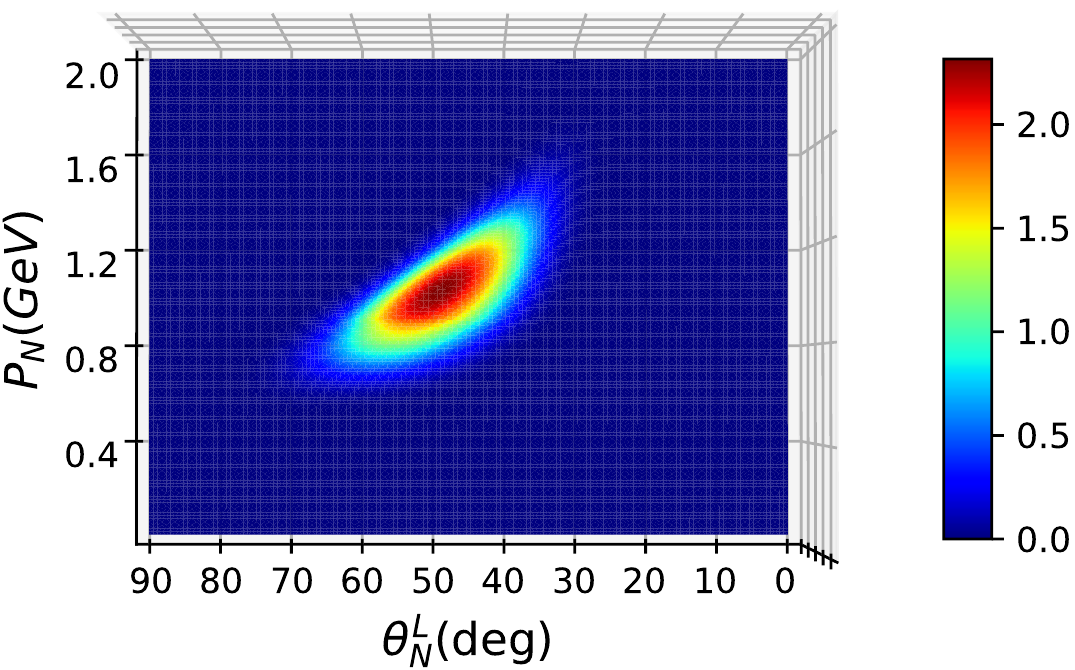} }}%
	}
	\caption{\label{16OrmfdonneDUNE}Semi-inclusive cross section for $^{40}$Ar and DUNE flux using $k'$ = 1.5 GeV, $\theta_l$ = 30$\degree$, $\phi_N^L$ = 180$\degree$ (left panels) and $\phi_N^L$ = 165$\degree$ (right panels) for the IPSM.}
\end{figure*}

Fig.~\ref{16OrmfdonneDUNE} shows the semi-inclusive cross section for the IPSM model.
For $\phi_N^L = 180\degree$ (left panels), the shape is not as simple as for the RFG model because the contour is more diffuse and the maximum is approximately located in the center of the region where the cross section exists, although with a distribution that clearly differs from the results corresponding to $\phi_N^L= 165\degree$ (right panels). Here the cross section shows a more symmetric shape with a very well-defined maximum located in the center of the projected contour and its magnitude decreasing uniformly in all directions as one moves away from the center. It is important to point out that the maximum value of the cross section at $\phi_N^L= 165\degree$ is reduced by $\sim 70$\% compared with the corresponding value at $\phi_N^L = 180\degree$. On the contrary, note that for the RFG model (Fig.~\ref{16OfermidonneDUNE}) the reduction is only $\sim15\%$. This is clearly illustrated in Table~\ref{table} where we present the specific values, denoted by $(\tilde{p}_N, \tilde{\theta}_N^L$), for which the semi-inclusive cross section reaches its maximum (also given in the table) for the two $\phi_N^L$-angles considered and both, IPSM and RFG, models. 
%{\ttred Juan: I don't like too much the notation $p_N^{max}$ because it can be easily thought as the maximum value of $p_N$ in a particular kinematics. I would propose another notation.... $\tilde{p}_N$ and $\tilde{\theta}_N^L$}. {\ttblue Maria: I agree. Since this is the quasielastic peak, we could also use $p_N^{QEP}$ and $\theta_N^{QEP}$, but $\tilde{p}_N$ and $\tilde{\theta}_N^L$ are perfectly fine}. 
The origin of these features is further investigated in the next plots.

\begin{figure}[!htbp]
	\centering
	\includegraphics[width=0.48\textwidth]{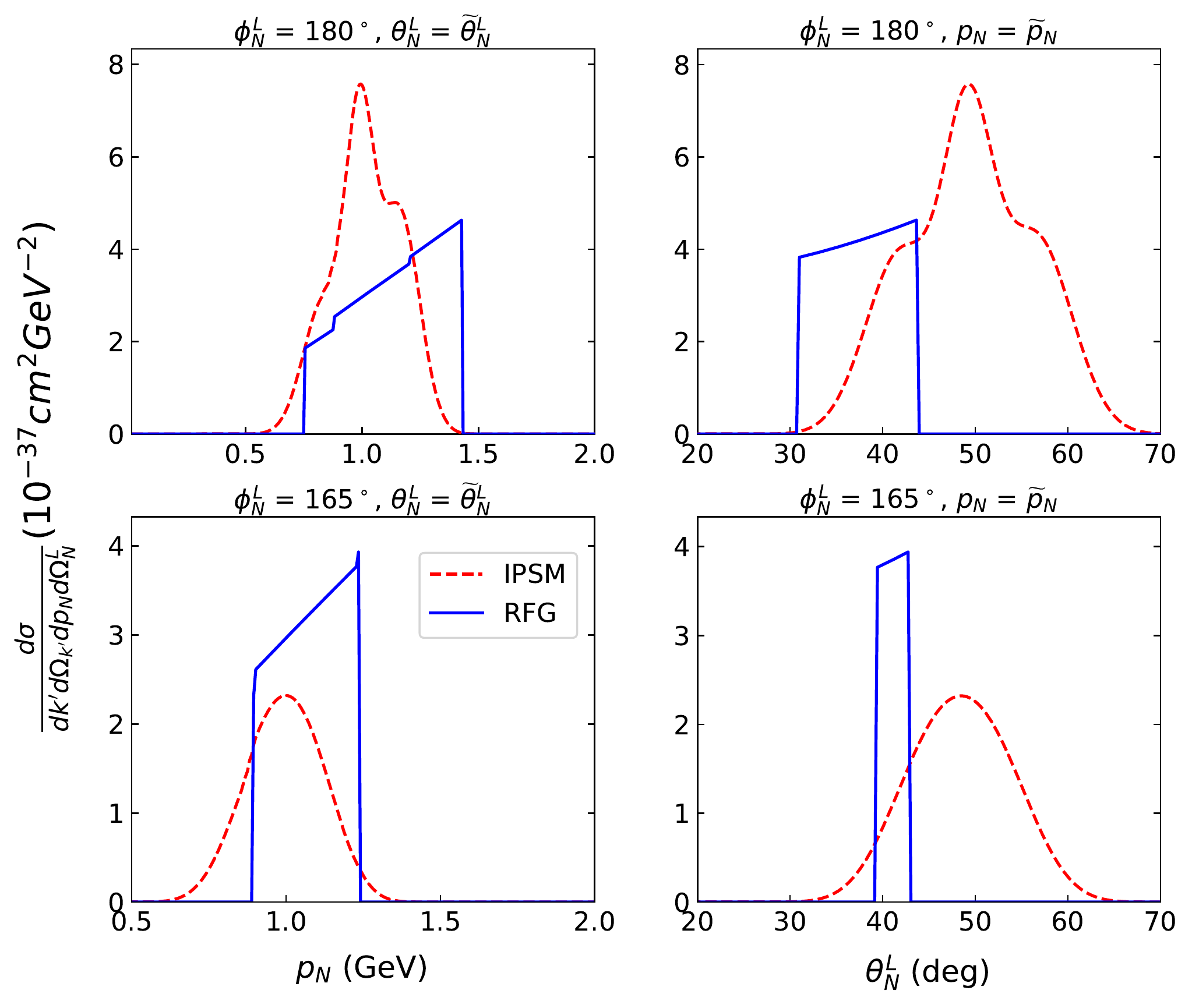}
	\caption{\label{2d-in-the-maximum}Semi-inclusive cross sections as function of $p_N$ ($\theta_N^L$) for the two values considered of the azimuthal angle $\phi_N^L$ (see text for details). In each case the cross section is evaluated at the corresponding values $\tilde{\theta}_N^L$ ($\tilde{p}_N$) that give the maximum cross section in Figs.~\ref{16OfermidonneDUNE} and~\ref{16OrmfdonneDUNE}. The values of $\tilde{p}_N$ and $\tilde{\theta}_N^{L }$ are summarized in Table~\ref{table}.}
\end{figure}

Fig.~\ref{2d-in-the-maximum} shows the semi-inclusive cross sections for 
the IPSM (red dashed) and the RFG (blue solid) models and the two values of $\phi_N^L$. 
%{\ttblue Maria: I suggest to use the same range of $p_N$, (0.5,1.75), for the two left panels.} 
The graphs on the left present the semi-inclusive cross section as a function of the ejected nucleon momentum $p_N$ evaluated at the values of $\tilde{\theta}_N^L$ where the maximum in the cross section occurs in each model. The panels on the right show the corresponding cross sections against $\theta_N^L$ at fixed $\tilde{p}_N$. Not only the shapes in the two models completely disagree but also the region in ($p_N$, $\theta_N^L$) where the cross section is defined differs very significantly.
%The position ($p_N^{max}$, $\theta_N^{L max}$) of the maxima of the semi-inclusive cross sections presented in Fig.~\ref{16OfermidonneDUNE} and Fig.~\ref{16OrmfdonneDUNE} are shown in Table~\ref{table}. The semi-inclusive cross sections in the maximums are also presented in Fig.~\ref{2d-in-the-maximum} for the IPSM and the RFG model and the two values of $\phi_N^L$. 

\begin{figure}[!htbp]
    \captionsetup[subfigure]{labelformat=empty}
	\subfloat[]{\includegraphics[width=0.48\textwidth]{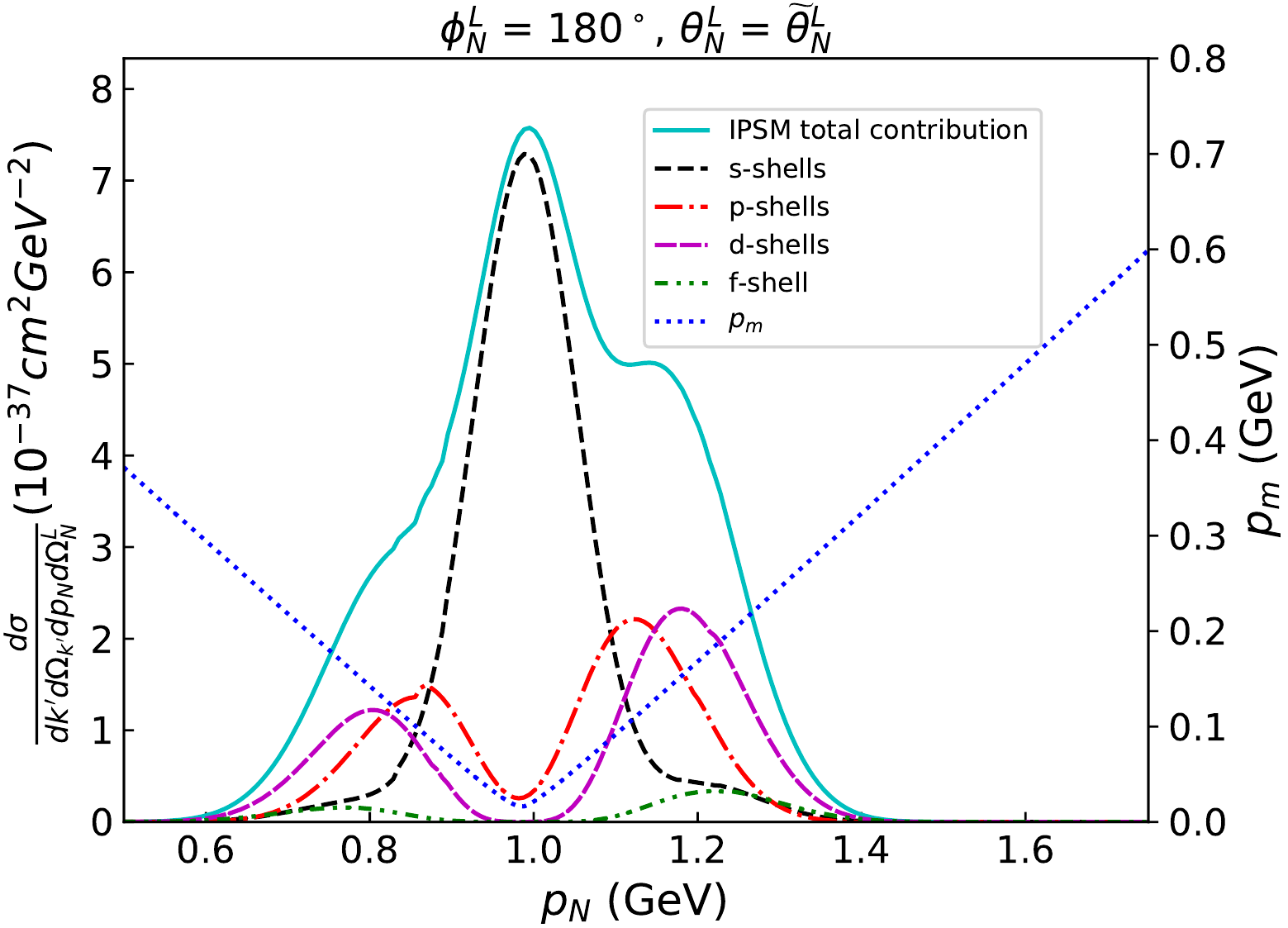}}
	\qquad
	\subfloat[]{\includegraphics[width=0.48\textwidth]{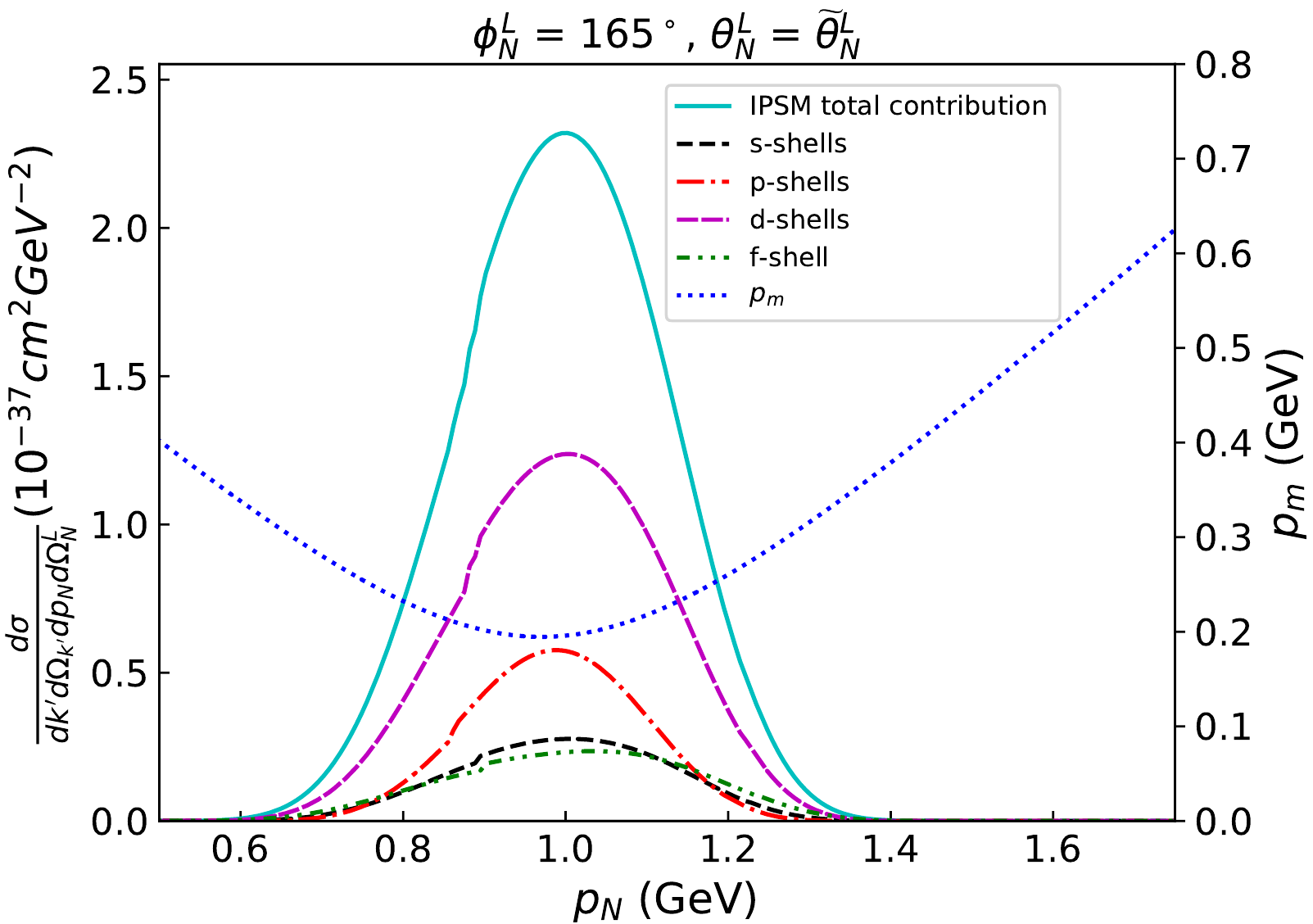}}
	\caption{\label{IPSM-sm-contribu-by-shells}Contributions to the semi-inclusive cross section by each shell in the IPSM for $^{40}$Ar. The value of $\tilde{\theta}_N^{L}$ is given in Table~\ref{table} and the lepton variables are fixed to $k'$ = 1.5 GeV and $\theta_l$ = 30$\degree$.}
\end{figure}

The specific contribution of the various shells in the IPSM model to the semi-inclusive cross section in the case of $^{40}$Ar is shown in Fig.~\ref{IPSM-sm-contribu-by-shells}. Here we present a cut in the plane $\theta_N^L = \tilde{\theta}_N^{L}$ of the semi-inclusive cross section shown in Fig.~\ref{16OrmfdonneDUNE}. Top (bottom) panel in Fig.~\ref{IPSM-sm-contribu-by-shells} corresponds to $\phi_N^L= 180\degree$ ($\phi_N^L=165\degree$). In both graphs we also display the behavior and allowed values of the missing momentum $p_m$ (dotted line).
According to the $^{40}$Ar momentum distribution (see Fig.~\ref{mom_dis_o16}), the $s$-shell contribution is dominant for very low missing momentum. For the kinematics considered in the top panel of Fig.~\ref{IPSM-sm-contribu-by-shells} this very low-$p_m$ region corresponds to values of the ejected nucleon momentum in the vicinity of $p_N$ = 1.0 GeV ($p_m\sim 40-50$ MeV). 
The other shells give a smaller contribution, originating the secondary peaks observed in Fig.~\ref{16OrmfdonneDUNE} at $\phi_N^L= 180\degree$. These peaks disappear at $\phi_N^L= 165\degree$ being also the cross section significantly smaller. 
As shown in the bottom panel of 
Fig.~\ref{IPSM-sm-contribu-by-shells}, the missing momentum gets also its minimum value in the region of $p_N$ close to 1 GeV. However, here $p_m \sim 200$ MeV, {\it i.e.,} much larger than the corresponding value in the previous case. As clearly illustrated by the $^{40}$Ar momentum distribution (Fig.~\ref{mom_dis_o16}), at $p_m\simeq 200$ MeV the shells that contribute the most are the $d$ and $p$-ones. This is consistent with the more symmetric shape of the semi-inclusive cross section shown in Fig.~\ref{16OrmfdonneDUNE} for $\phi_N^L= 165\degree$ (left panels) with only one peak visible. 

%{\ttblue Maria: it would be interesting to show the same also for $\phi_N^L= 165\degree$.} {\ttred Juan: I have completed the previous general discussion with additional figure. Modify as wished.}

\begin{table}[!htbp]
	\resizebox{\linewidth}{!}{%
		\begin{tabular}{ccccccc} \toprule\toprule
			& & $\phi_N^L$ = 180$\degree$ & & & $\phi_N^L$ = 165$\degree$ & \\  \cmidrule(rl){2-4}\cmidrule(l){5-7}
			& $\tilde{p}_N$ & $\tilde{\theta}_N^{L}$ & Cross section & $\tilde{p}_N$ & $\tilde{\theta}_N^{L}$ & Cross section \\\midrule
			RFG & 1.43 GeV & 43.50$\degree$ & 4.62 $\times$ $10^{-37}$ & 1.24 GeV & 42.89$\degree$ & 3.95 $\times$ $10^{-37}$ \\
			IPSM & 1.00 GeV & 49.54$\degree$ & 7.56 $\times$ $10^{-37}$ & 1.00 GeV & 48.33$\degree$ & 2.32 $\times$ $10^{-37}$ \\
			\bottomrule\bottomrule
	\end{tabular}}
	\caption{Values of $p_N$ and $\theta_N^L$ that give the maximum cross sections in Fig.~\ref{16OfermidonneDUNE} and Fig.~\ref{16OrmfdonneDUNE}, {\it i.e.,} ($\tilde{p}_N,\tilde{\theta}_N$). Cross sections in cm$^2$/GeV$^{2}$. (See text for details).}
	\label{table}
\end{table} 

According to the general energy and momentum conservation given by Eqs.~(\ref{eq:Esimp},~\ref{eq:pcons}), it is possible to deduce an expression for $\mathcal{E}$, or equivalently $E_m$, as function of $p_m$ for a selected set of semi-inclusive variables: ($k'$, $\theta_l$, $p_N$, $\theta_N^L$, $\phi_N^L$). This relation is
\begin{equation}
	\mathcal{E} (p_m) = \omega -E_s - E_N + m_N\, ,
\end{equation}
where the neutrino momentum $k$ is the solution of the equation
\begin{align}
	k^2 - 2k(k'\cos{\theta_l} + p_N\cos{\theta_N^L}) + {k'}^2 + p_N^2 + \nonumber\\
	2k'p_N(\cos{\theta_l}\cos{\theta_N^L} + \sin{\theta_l}\sin{\theta_N^L}\cos{\phi_N^L}) - p_m^2 = 0
\,,
\end{align}
and it defines  trajectories in the $(\mathcal{E}, p_m)$ plane  allowed by  energy conservation at each kinematics.

\begin{figure}[!htbp]
	\centering
	\includegraphics[width=0.5\textwidth]{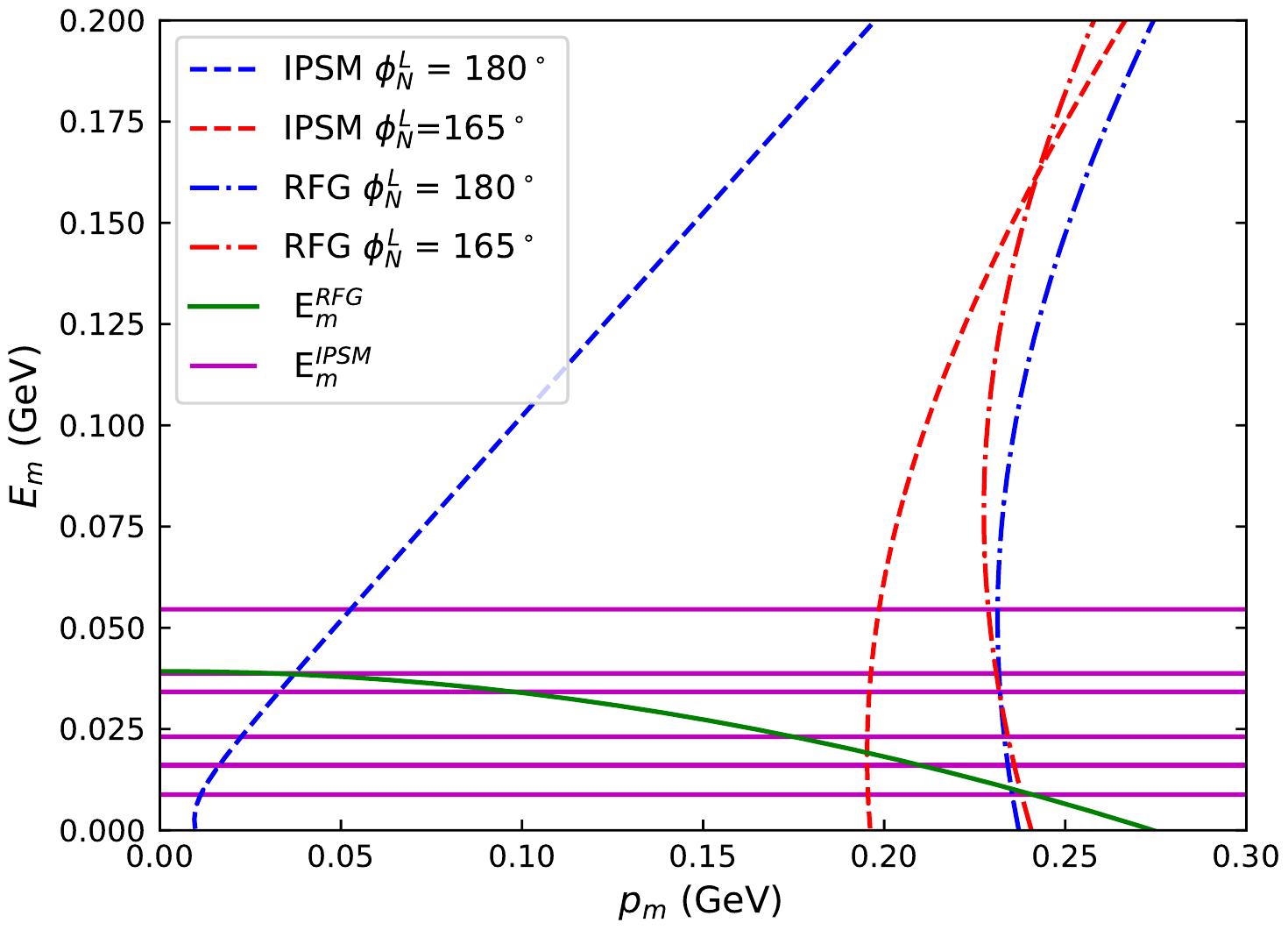}
	\caption{\label{maximum_trajectories}Trajectories for two values of $\phi_N^L$. The outgoing nucleon variables $p_N$ and $\theta_N^L$ are those that give the maximum value of the cross section for each $\phi_N^L$ given in Table~\ref{table}. Also included $E_m(p_m)$ for RFG and IPSM.}
\end{figure}

By plotting the trajectories $\mathcal{E}(p_m)$, likewise $E_m(p_m)$, for a set of semi-inclusive variables, we can observe that the RFG (IPSM) semi-inclusive cross section is different from zero only if the corresponding trajectory crosses the curve $\mathcal{E}_{RFG}(p_m)$ ($\mathcal{E}_{nlj}$), where the RFG (IPSM) spectral function lives. This is illustrated in Fig.~\ref{maximum_trajectories} where we show the trajectories $E_m(p_m)$ for the set of  variables that gives the maximum cross sections in Figs.~\ref{16OfermidonneDUNE} and~\ref{16OrmfdonneDUNE} together with the support of the spectral function for the two models. In the case of the IPSM (dashed curves) the two trajectories corresponding to the two $\phi_N^L$-values cross the specific missing energies for the different shells at very different values of the missing momentum. Whereas for $\phi_N^L=180\degree$ the crossing occurs in the region of low-$p_m$, {\it i.e.,} $p_m\simeq 50$ MeV/c, the situation is clearly different for $\phi_N^L= 165\degree$ where the crossing takes place at larger $p_m$-values ($p_m\simeq 200$ MeV/c), a region where the momentum distribution has dropped very significantly. This explains the great reduction observed in the maximum of the semi-inclusive cross section when going from $\phi_N^L=180\degree$ to $\phi_N^L=165\degree$. 

The situation is clearly different for the RFG model. Here the trajectories for the two $\phi_N^L$-values (dot-dashed lines) are very close to each other and they cross the value of the RFG missing energy at $p_m\simeq 240$ MeV/c, {\it i.e.,} just below the Fermi level. As known, the RFG momentum distribution is constant and different from zero up to $p_m=k_F$. Thus the $15\%$ reduction observed in the maxima of the semi-inclusive cross sections for the two $\phi_N^L$-values cannot be connected with the momentum distribution but with the specific kinematical factors (evaluated at the particular allowed values for the remaining kinematical variables) that enter in the cross section.   
  
\begin{figure}[!htbp]
    \captionsetup[subfigure]{labelformat=empty}
	\centering
	\subfloat[]{\includegraphics[width=0.48\textwidth]{pictures/Results/3d_graphs/40Ar_rmf_180_aguila_kp1.5thl30_DUNE.pdf}}
	\qquad
	\subfloat[]{\includegraphics[width=0.48\textwidth]{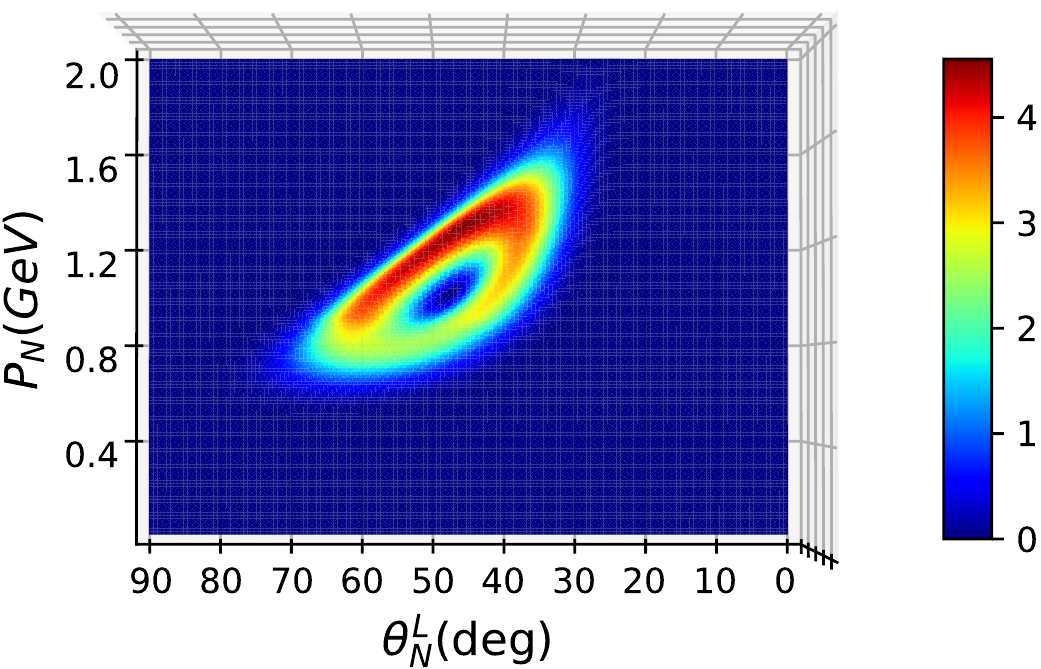}}
	\caption{\label{3drmfO16-without-1s-comparation}Semi-inclusive cross section for $^{40}$Ar and DUNE flux taking $k'$ = 1.5 GeV, $\theta_l$ = 30$\degree$ and $\phi_N^L$ = 180$\degree$ using IPSM spectral function including all shells (top panel) and removing the $s$-shells (bottom). Cross sections in $10^{-37}$cm$^2$/GeV$^{2}$.}
\end{figure}

Using the general expression for the trajectory $E_m(p_m)$ we can also analyze the origin of the high peak in the cross section shown in Fig.~\ref{16OrmfdonneDUNE} for $\phi_N^L$ = 180$\degree$. Since this result is not present in the case of the RFG (Fig.~\ref{16OfermidonneDUNE}), we assume its origin is linked to the complex, non-constant, structure shown by the global momentum distribution in the IPSM. As already shown in Fig.~\ref{maximum_trajectories}, the IPSM trajectory corresponding to $\phi_N^L=180\degree$ is consistent with significant contribution in the momentum distribution at low missing momentum values. This is the region where the various $s$-shells entering in $^{40}$Ar clearly dominate, giving rise to the maximum in the cross section observed in 
Fig.~\ref{16OrmfdonneDUNE}. In fact, if one excludes the $s$-shell contributions the semi-inclusive cross section decreases significantly.
%{\ttblue Maria: does the following mean that "the maximum located ... disappears"?} {\ttred Juan: rewritten}
%disappearing the maximum located in the vicinity of $p_N \simeq1.0$ GeV and $\theta_N^L \simeq 50\degree$. 
This is illustrated in Fig.~\ref{3drmfO16-without-1s-comparation} that shows the contour graph of the semi-inclusive cross section including all shells in $^{40}$Ar (top panel) and removing the contribution of the $s$-shells (bottom panel). Note the global reduction in the cross section, but also how importantly the strength in the cross section is modified in the ($p_N, \theta_N^L$)-plane. The peak presented in the top panel
located in the vicinity of $p_N \simeq1.0$ GeV and $\theta_N^L \simeq 50\degree$, due to the $s$-shell contributions, has completely gone in the bottom graph leaving a hole where the cross section is very small (close to zero).
 
%\begin{figure}[!htbp]
%	\centering
%	\includegraphics[width=0.48\textwidth]{pictures/Results/3d_graphs/trajec_k2_thl25-inthehole-corrected.pdf}
%	\caption{\label{hole_trajectory}Trajectory $E_m(p_m)$ for the a set of semi-inclusive variables inside the hole in Fig.~\ref{16OrmfdonneDUNE}. Final lepton variables are set to $k'$ = 2 GeV and $\theta_l$ = 25$\degree$.}
%\end{figure}

\begin{figure}[!htbp]
    \captionsetup[subfigure]{labelformat=empty}
	\centering
	\subfloat[]{\includegraphics[width=0.45\textwidth]{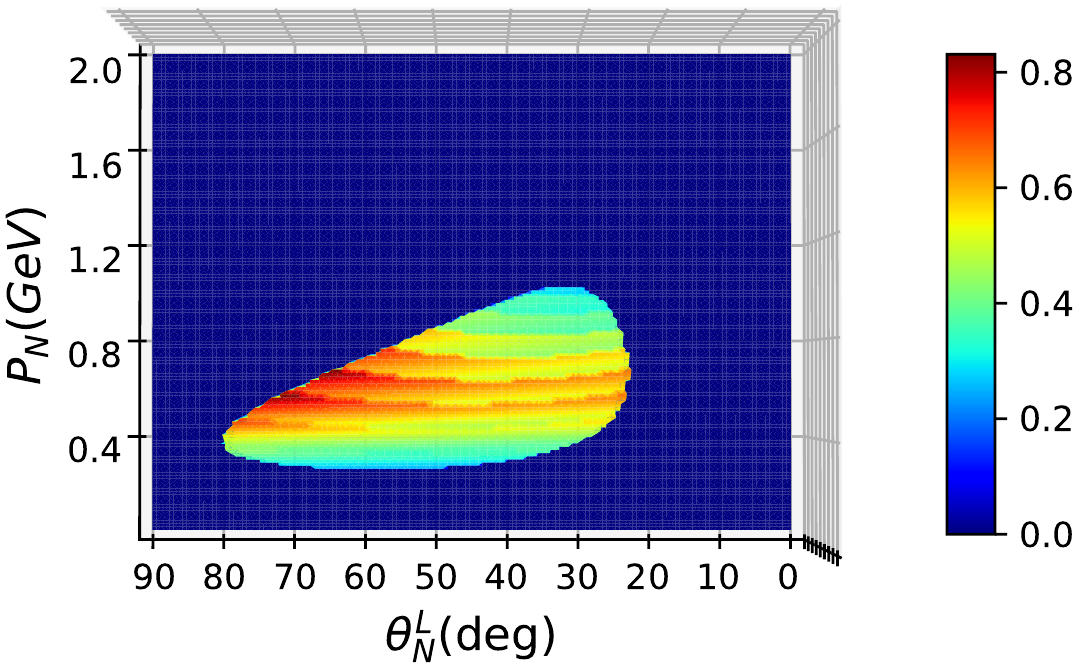}}
	\qquad
	\subfloat[]{\includegraphics[width=0.45\textwidth]{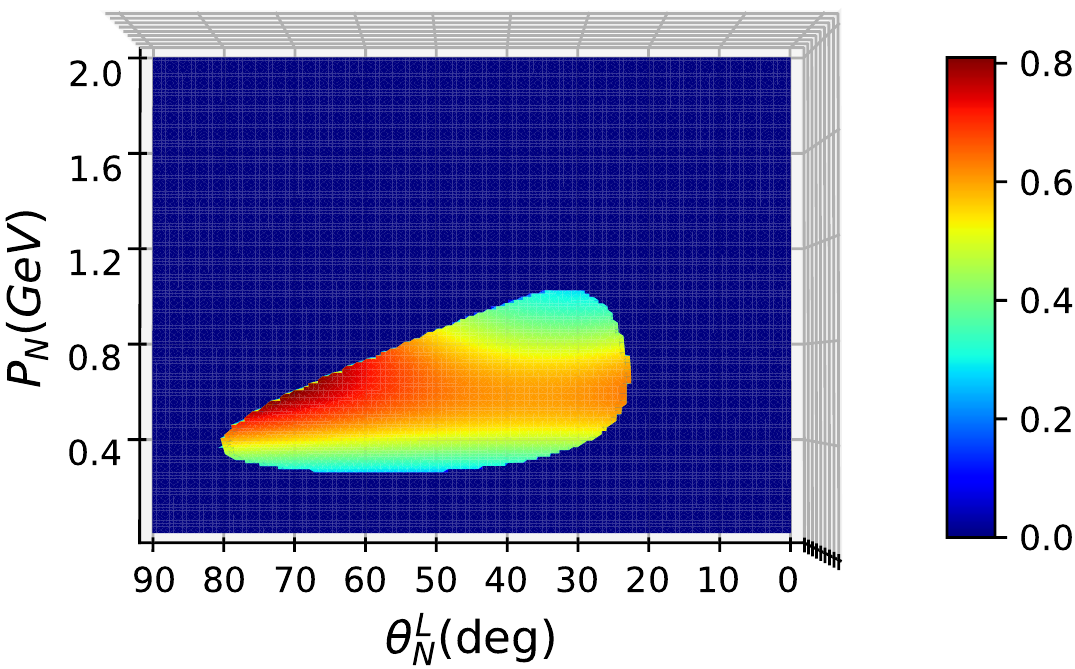}}
	\caption{\label{12Cfermi_flux_comparison} Semi-inclusive cross section for $^{12}$C with $k'$~=~0.55 GeV, $\theta_l$ = 50$\degree$ and $\phi_N^L$ = 180$\degree$ using the experimental T2K flux shown in Fig.~\ref{fluxes} (top panel) and using a Gaussian fit (bottom). Cross sections are in $10^{-37}$cm$^2$/GeV$^{2}$ and both results correspond to the RFG model.}
\end{figure}

All previous results correspond to the case of $^{40}$Ar, the target that will be used in DUNE  detector. In what follows we extend our study to the case of $^{12}$C, used in past and on-going experiments. 
%In a forthcoming publication we will consider the case of $^{16}$O. 
%{\ttblue Maria: I am not sure that saying that is safe, they could ask us why we don't do it here.} {\ttred Juan: removed.}
We present semi-inclusive results for muon neutrinos on $^{12}$C with muon variables fixed to $k'$ = 0.55 GeV and $\theta_l= 50\degree$ for $\phi_N^L= 180\degree$ using the T2K flux. In addition to the RFG and IPSM nuclear models already used in the case of DUNE ($^{40}$Ar), here we also provide predictions for NO.
%{\ttblue Maria: maybe we should explain at some point why we do the NO only for carbon (argon SF not available)?}. {\ttred Juan: you are right. We do not have the NO Argon wave functions. I asked Martin, but it is not trivial. If you think of a sentence that can be introduced somewhere would be fine, but I do not know what to say.} 
The kinematics is fixed in order to explore the impact of the neutrino flux on the shape of the semi-inclusive cross section. 
%{\ttblue Maria: I don't understand the previous sentence: do you mean that the kinematics is chosen in such a way to give the maximum criss section using the T2K flux? Or that the discontinuities due to the bins in the flux are more visible in this kinematics? The other question I have is why this discussion was not done for the Dune flux in the results aready shown. Did you use a fit of the flux in this case?} {\ttred Juan: see my comment at the end of the paragraph} 
More specifically, we analyze how the shape of the semi-inclusive cross sections changes with the experimental neutrino flux that is given in bins as shown in Fig.~\ref{fluxes}. Results for the RFG (projected cross section in the ($p_N, \theta_N^L$) plane) are presented in Fig.~\ref{12Cfermi_flux_comparison} using the experimental flux (top panel) and making use of a Gaussian fit of the flux (bottom panel). As shown, the use of the experimental flux (with the bins) leads to the appearance of some discontinuities or jumps 
in the cross section that are distributed along the $p_N$ axis as the value of $\theta_N^L$ changes. This occurs because the neutrino energy is also a function of $p_N$ and $\theta_N^L$ and it increases when we move to higher values of $p_N$. Hence, it is simply a direct consequence of the change of bin in the experimental neutrino flux. This is clearly seen in the bottom panel of Fig.~\ref{12Cfermi_flux_comparison} where we present again the semi-inclusive cross section for the same kinematics but using a continuous function fitted to the neutrino flux. As observed, the discontinuities are not present and the colors in the cross section present a smooth and gradual change~\footnote{A similar study could also be applied to the DUNE flux, although here being aware of the larger size of the bins.}. 
%{\ttred Juan: I included this footnote to justify why we restrict our discussion to T2K.}

\begin{figure*}[!htbp] %16O rmf phin:180/165 DUNE
    \captionsetup[subfigure]{labelformat=empty}
	\centering
	\subfloat[]{{\includegraphics[width=0.48\textwidth]{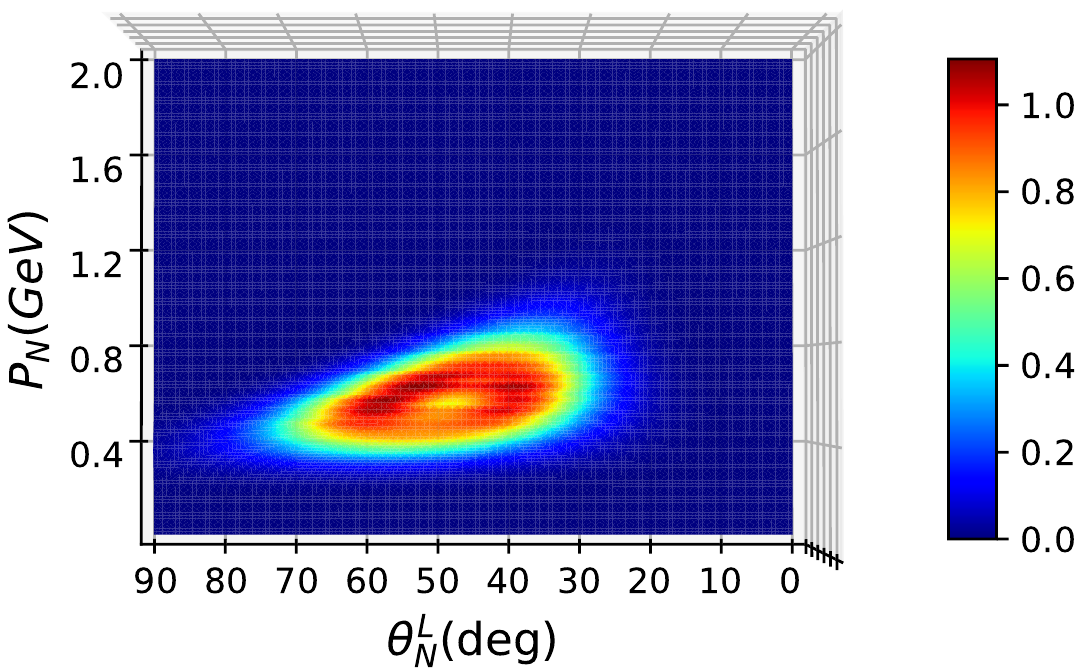} }}%
	\hspace{3.5mm}%
	%\qquad
	\subfloat[]{{\includegraphics[width=0.48\textwidth]{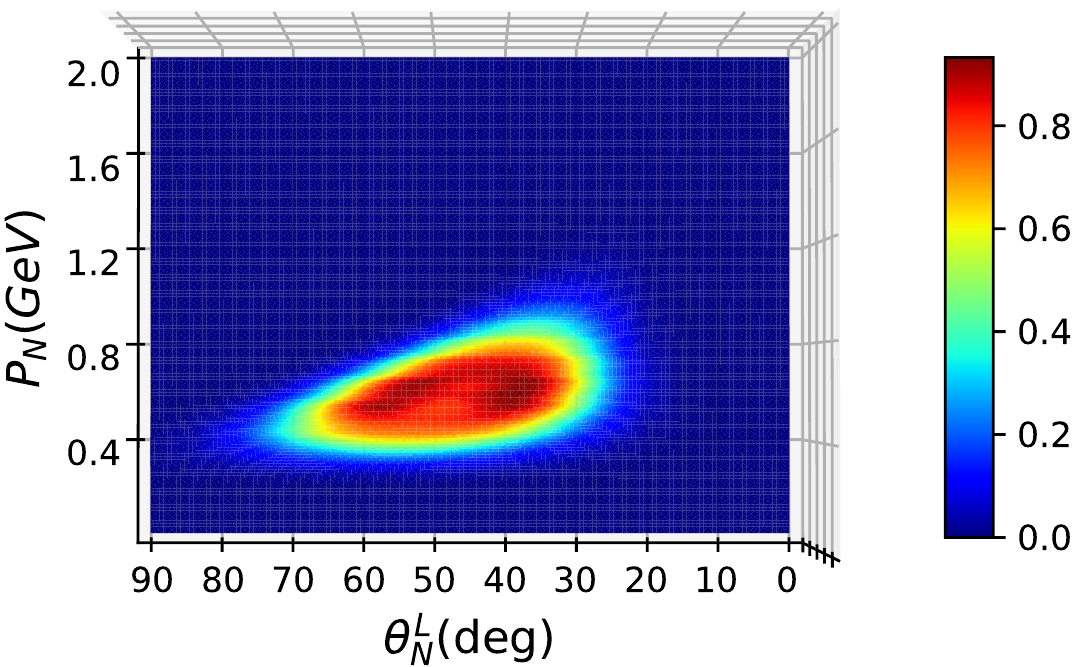} }}%
	\caption{\label{12C_ipsm_and_sofia}Semi-inclusive cross section for $^{12}$C and T2K flux setting $k'$ = 0.55 GeV, $\theta_l$ = 50$\degree$ and $\phi_N^L$ = 180$\degree$ for the IPSM (left panel) and the NO model (right). Cross sections in $10^{-37}$cm$^2$/GeV$^{2}$.}
\end{figure*}

For completeness, we show in Fig.~\ref{12C_ipsm_and_sofia} the semi-inclusive cross section for the kinematics defined above and the two remaining nuclear models: IPSM (left panel) and NO (right panel). In both cases we have used the T2K flux presented in 
Fig.~\ref{fluxes}. The shapes of the semi-inclusive cross sections for both models are highly correlated with the shapes of the momentum distributions (see bottom panel in Fig.~\ref{mom_dis_o16}). Notice that the cross section for the IPSM in Fig.~\ref{12C_ipsm_and_sofia} (left) shows a small hole in the center of the region that is not present in the case of the NO model (right). An analysis of the trajectory curves shows that this particular region in the $(p_N, \theta_N^L$)-plane corresponds to very small values of the missing momentum $p_m$. Note that the behavior of the momentum distribution for the two models, IPSM and NO, differs at low-$p_m$: the former decreases as $p_m$ approaches zero whereas the latter does not. 

\subsection{\label{Semi-inclusive experimental} Comparison with semi-inclusive experimental data. }

In this section we compare our predictions with some recent data from the T2K collaboration~\cite{PhysRevD.98.032003} corresponding to neutrino scattering on $^{12}$C with a muon and an ejected proton measured in the final state. We only present some preliminary results based on the PWIA and the three nuclear models considered in this work. A more complete analysis including a systematic comparison with all available semi-inclusive data will be presented in a forthcoming publication. The role played by the final state interactions (FSI) will be also considered in the future within the general scheme of the Distorted Wave Impulse Approximation (DWIA). 
 
\begin{figure}[!htbp]
    \captionsetup[subfigure]{labelformat=empty}
	\centering
	\subfloat[]{\includegraphics[width=0.45\textwidth]{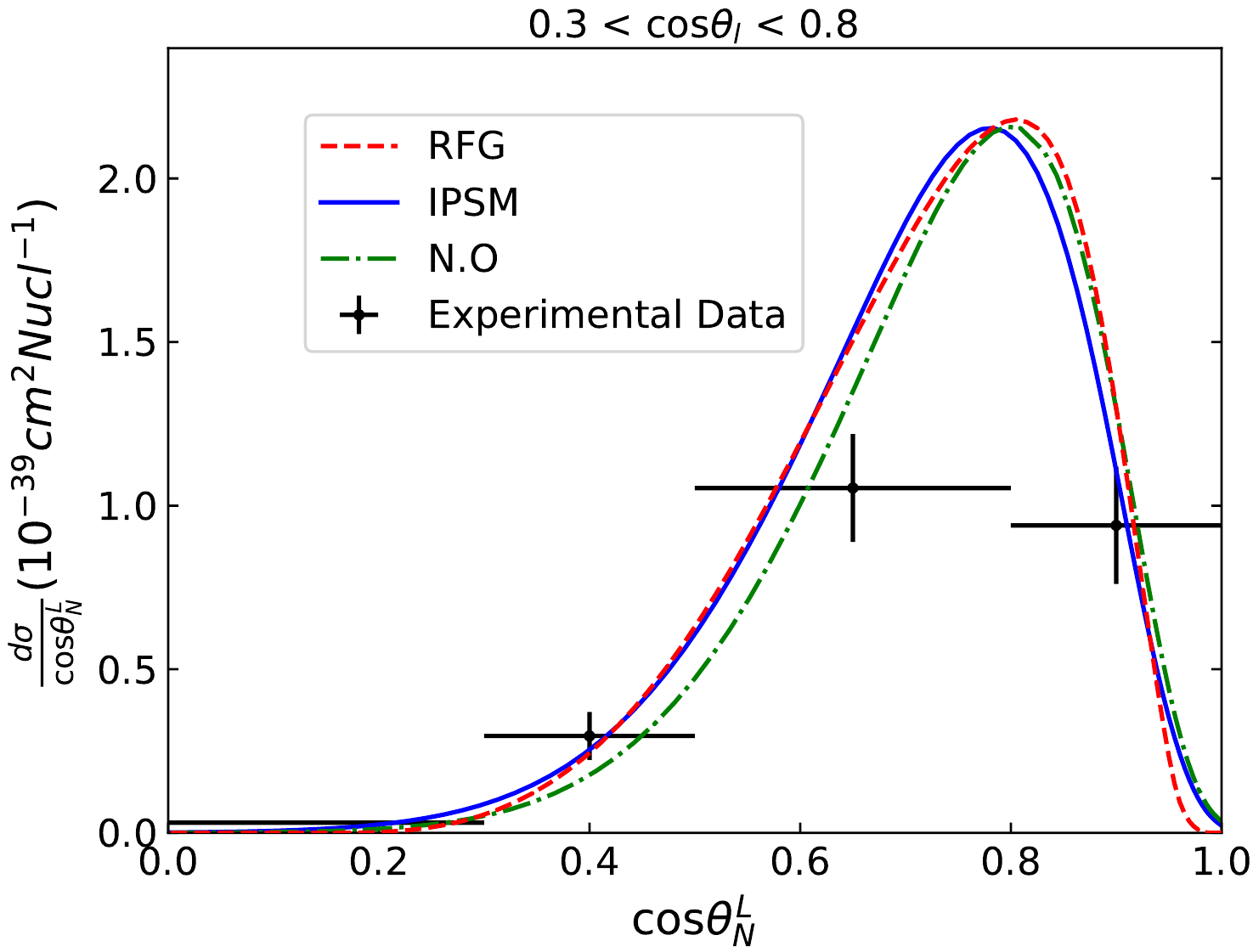}}
	\qquad
	\subfloat[]{\includegraphics[width=0.45\textwidth]{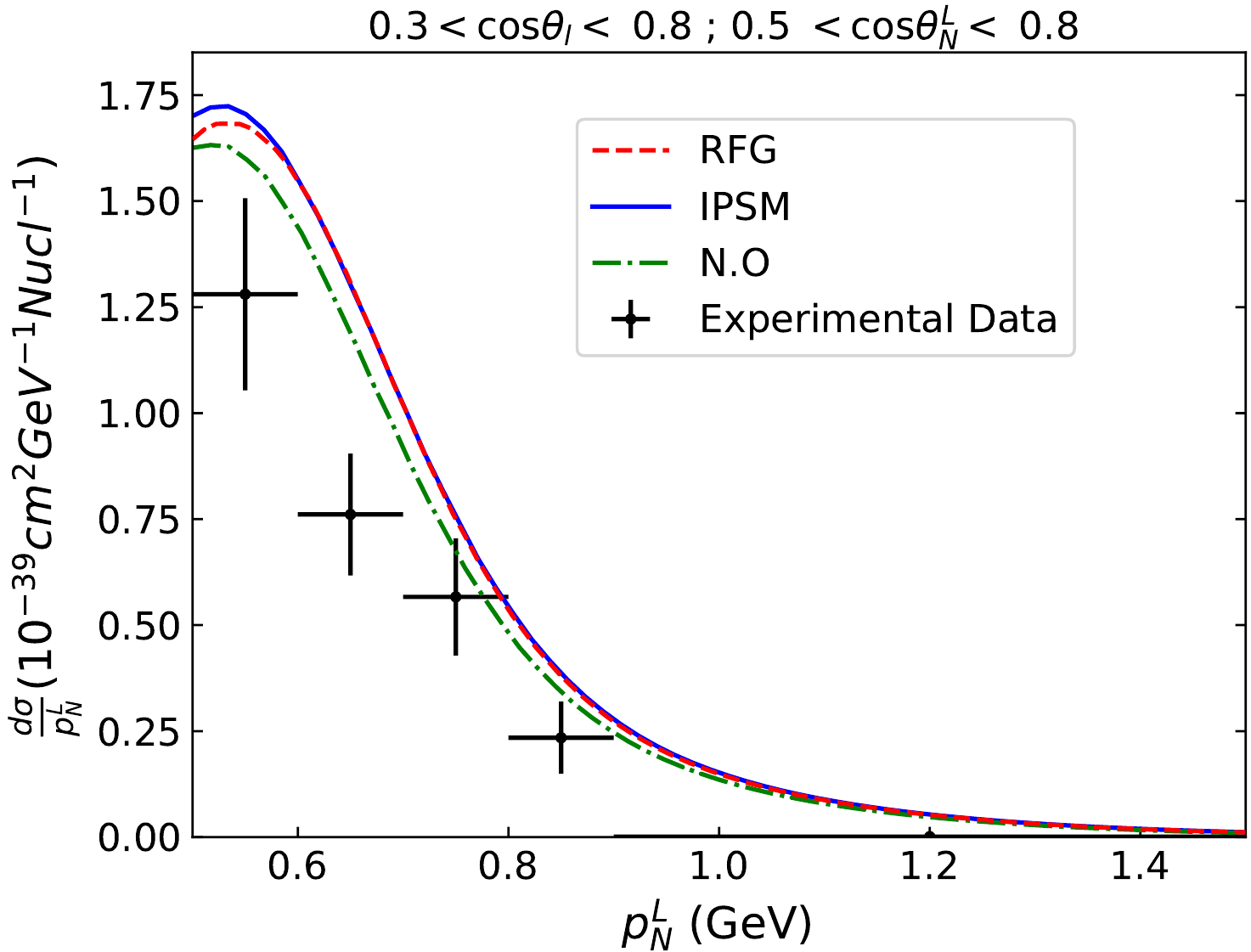}}
	\caption{\label{semi-inclusive experimental comparison}Flux-averaged single-differential cross section per target nucleon for muon neutrinos on $^{12}$C as function of $\cos{\theta_N^L}$ in a bin of $\cos{\theta_l}$ (top panel) and as function of $p_N^L$ in bins of $\cos{\theta_l}$ and $\cos{\theta_N^L}$ (bottom). Data taken from~\cite{PhysRevD.98.032003}.}
\end{figure}

Starting from the general semi-inclusive cross section as function of the final lepton and nucleon variables, we can integrate over different variables in order to get the cross section to be compared with the experiment. In the case of the T2K semi-inclusive data we define two different cross sections, namely, 

\begin{align}\label{proton1}
\left <\frac{d\sigma}{d\cos{\theta_N^L}}\right > =& \frac{2\pi}{A} \int_0^\infty dk'\int_0^{2\pi}d\phi_N^L\int_{p_N^{min}}^\infty dp_N \nonumber\\
&\times\left <\frac{d\sigma}{dk'd\Omega_{k'}dp_Nd\cos{\theta_N^L}d\phi_N^L}\right >\Delta\cos{\theta_l}
\end{align}
and
\begin{align}\label{proton2}
\left <\frac{d\sigma}{dp_N}\right > =& \frac{2\pi}{A} \int_0^\infty dk'\int_0^{2\pi}d\phi_N^L\Delta\cos{\theta_l}\Delta\cos{\theta_N^L} \nonumber\\
&\times\left <\frac{d\sigma}{dk'd\Omega_{k'}dp_Nd\cos{\theta_N^L}d\phi_N^L}\right >\, ,
\end{align}
where $\Delta\cos{\theta_i}$ is the experimental bin length. Note that in Eq.~(\ref{proton1}) the integral over $p_N$ is performed from a minimum value $p_N^{min}$ up to infinity. Since in the T2K experiment only ejected protons with momentum greater than $p_N^{min}=0.5$ GeV were detected, in the theoretical calculation we apply the same cut $p_N \ge 0.5$ GeV. T2K data compared with theoretical predictions using the three nuclear models are presented in Fig~\ref{semi-inclusive experimental comparison}. As shown, the uncertainty connected with the nuclear model is tiny. Only the NO prediction departs slightly from the IPSM and RFG results.
%{\ttblue Maria: rather than showing a shift, I would say it is lower.} {\ttred Juan: I modified the sentence.} 
Although the theoretical predictions overestimate the data by some amount, great caution should be drawn on this analysis. The present model is entirely focused on the quasielastic regime and based on the Plane Wave Impulse Approximation (PWIA). This is obviously an oversimplified description of both the reaction mechanism and the final state dynamics. FSI and ingredients beyond the IA like Meson Exchange Currents (MEC) can play a significant role in describing the data. This is consistent with the analysis presented in \cite{PhysRevD.98.032003} based on results obtained using different event generators. However, it is not yet entirely clear how precisely the event generator transport mechanisms can reproduce the effects ascribed to the final state interactions. 
%{\ttblue Maria: what is the meaning of the last sentence?} {\ttred Juan: This has already been claimed in previous works by Bill et al., I think the idea is just to state clearly that nobody knows for sure how FSI effects are incorporated in event generator transports. I wrote it to soften the general conclusion of the analysis in \cite{PhysRevD.98.032003} and justify our present approach to the problem (based on RPWIA). If it is not clear, we can simply remove it.}

%{\ttred Juan: Anything else to include in order to complete the discussion?}
 % and models shows that 2p-2h contribution and final state interactions are essential to interpret correctly the available data.

\subsection{\label{inclusive cross sections} Inclusive cross sections.}

Although the main objective of this work is the analysis of semi-inclusive CC$\nu$ reactions, in what follows we consider the case of  inclusive reactions where only the final lepton is detected. This topic has been studied in detail by several groups using very different models that incorporate not only diverse descriptions of the nuclear dynamics but also FSI, two-particle two-hole (2p-2h) contributions, nucleon resonances and deep inelastic scattering~\cite{Alvarez-Ruso17}. Moreover, extensive studies of inclusive neutrino scattering processes based on scaling arguments have been developed by our group in the 
past~\cite{Caballero:2007tz,Amaro:2005dn,Amaro05,Gonzalez-Jimenez:2014eqa,electron-vs-neutrino,Megias:2017,Megias:2016fjk}. In all the cases a systematic comparison with data has been provided. Hence, in this section our aim is simply to prove the consistency of the present calculations, originally developed for semi-inclusive processes, when applied to inclusive reactions. 

\begin{figure*}[!htbp]
	\centering
	\includegraphics[width=0.98\textwidth]{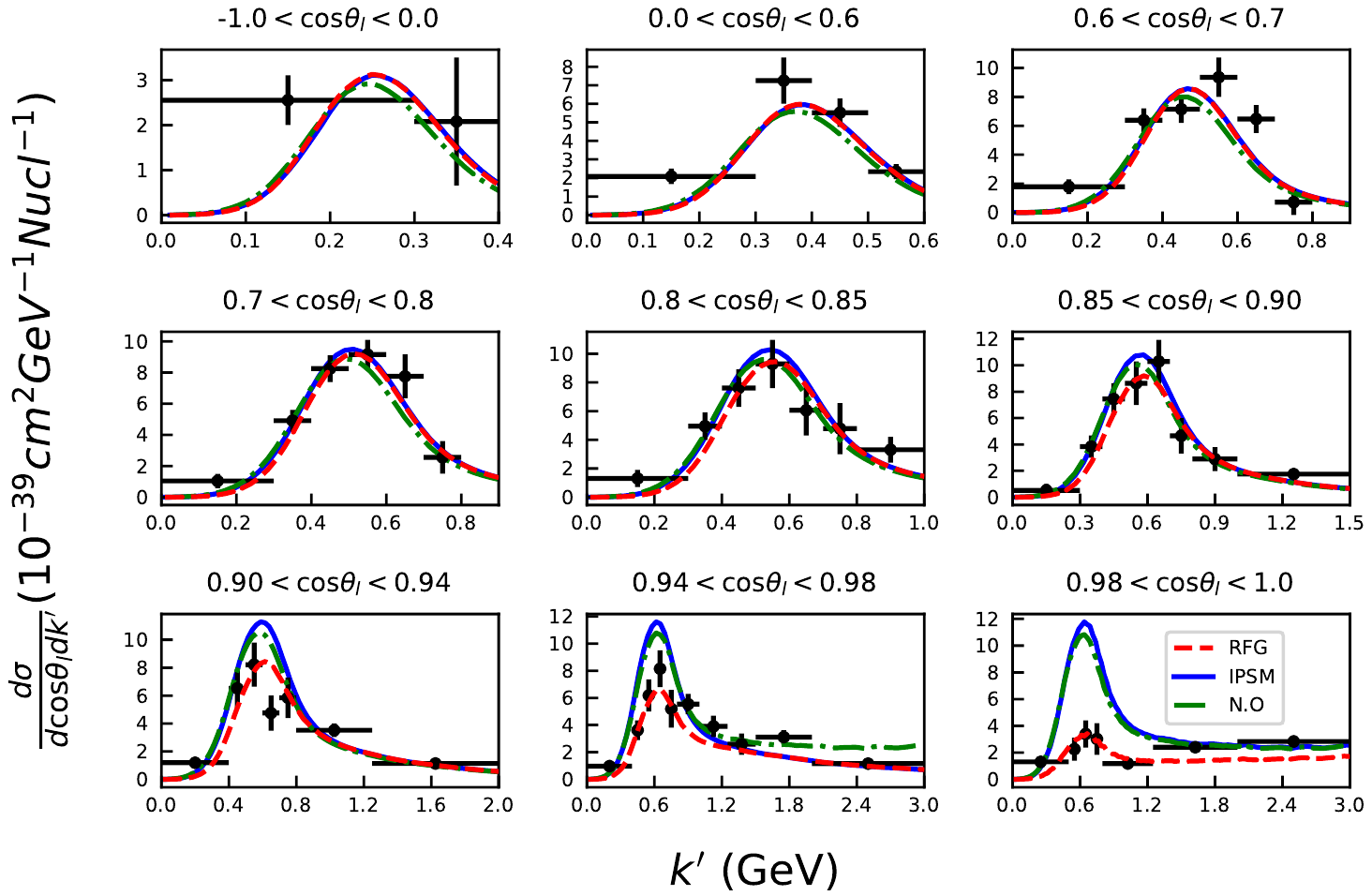}
	\caption{\label{PRD84-miniboone_inclusive}Flux-averaged double-differential inclusive cross section per target nucleon for $^{12}$C in bins of $\cos{\theta_{l}}$ as function of the momentum of the muon $k'$ for RFG model (Eq.~\eqref{inclusive-fermi}), IPSM (Eq.~\eqref{inclusive-IPSM}) and NO model  (Eq.~\eqref{inclusive-sofia}). Data taken from~\cite{PhysRevD.93.112012}.}
\end{figure*}

As we discussed in Section~\ref{From semi-inclusive to inclusive}, the inclusive results can be recovered from the semi-inclusive ones by integrating over the kinematical variables corresponding to the nucleon detected in coincidence with the lepton in the final state.  
% starting from the general expression of the semi-inclusive cross section given by Eq.~\eqref{general-semics}, we derived an expression for the inclusive cross section, Eq.~\eqref{general-inclusive-cross-section}, after integrating over the variables of the detected nucleon.
In Fig.~\ref{PRD84-miniboone_inclusive} we show the flux-averaged double-differential inclusive cross sections for $^{12}$C evaluated for the three nuclear models, RFG (red dashed line), IPSM (blue solid)  and NO (green dot-dashed). Comparison with T2K experimental 
data~\cite{PhysRevD.93.112012} is also provided. In spite of the very different momentum distributions for the three nuclear models, particularly in the case of the RFG, the inclusive cross sections are very similar, except for $\theta_\mu$ angles close to zero (i.e., small energy transfer) where the IPSM and the NO results deviate very significantly from the RFG ones being much higher than data. As discussed 
in \cite{Megias:2018oxygen,Gonzalez-Jimenez:2019ejf}, the PWIA approach fails in describing lepton-nucleus scattering reactions at low values of the momentum and energy transfers. This is a consequence of the lack of orthogonality between the bound and free nucleon wave functions, and the extremely large effects associated to the overlap between the non-orthogonal initial and final states in the near-threshold region. In the case of the RFG, the cross section at very forward scattering angles is significantly reduced and more in accordance with data. This result is largely due to the Pauli blocking effects included in the model. Notice that IPSM and NO lead to similar semi-inclusive responses (see Figs.~\ref{16OfermidonneDUNE} and~\ref{16OrmfdonneDUNE}), being very different from the predictions provided by RFG (Fig.~\ref{12C_ipsm_and_sofia}).

In the previous sections we have worked in the $q$-system in order to get the inclusive responses. A similar analysis can be performed working in the $k$-system. In this case, the neutrino-averaged inclusive cross section can be written as
\begin{align}\label{inclusive-integrating-lab}
\left <\frac{d\sigma}{dk'd\Omega_{k'}}\right > =&  \int_0^{2\pi}d\phi_N^L\int_{-1}^{+1} d\cos{\theta_N^L}\int_0^\infty dp_N \nonumber\\
&\times\left <\frac{d\sigma}{dk'd\Omega_{k'}dp_Nd\cos{\theta_N^L}d\phi_N^L}\right >\, ,
\end{align}
{\it i.e.} integrating over the outgoing nucleon variables in the $k$-system. Note that the integral over $\phi_N^L$ is not as trivial as in the $q$-system because the rotation that relates the two systems introduces extra terms in $\mathcal{F}_{\chi}^2$ that do not vanish after performing the integral. The special symmetry shown by the responses in the $q$-system is lost when expressed with respect to the $k$-system  . Although this introduces additional complexities in the problem, one can test the consistency of the calculations by solving numerically the integrals in Eq.~\eqref{inclusive-integrating-lab}. 

\begin{figure}[!htbp]
	\centering
	\includegraphics[width=0.48\textwidth]{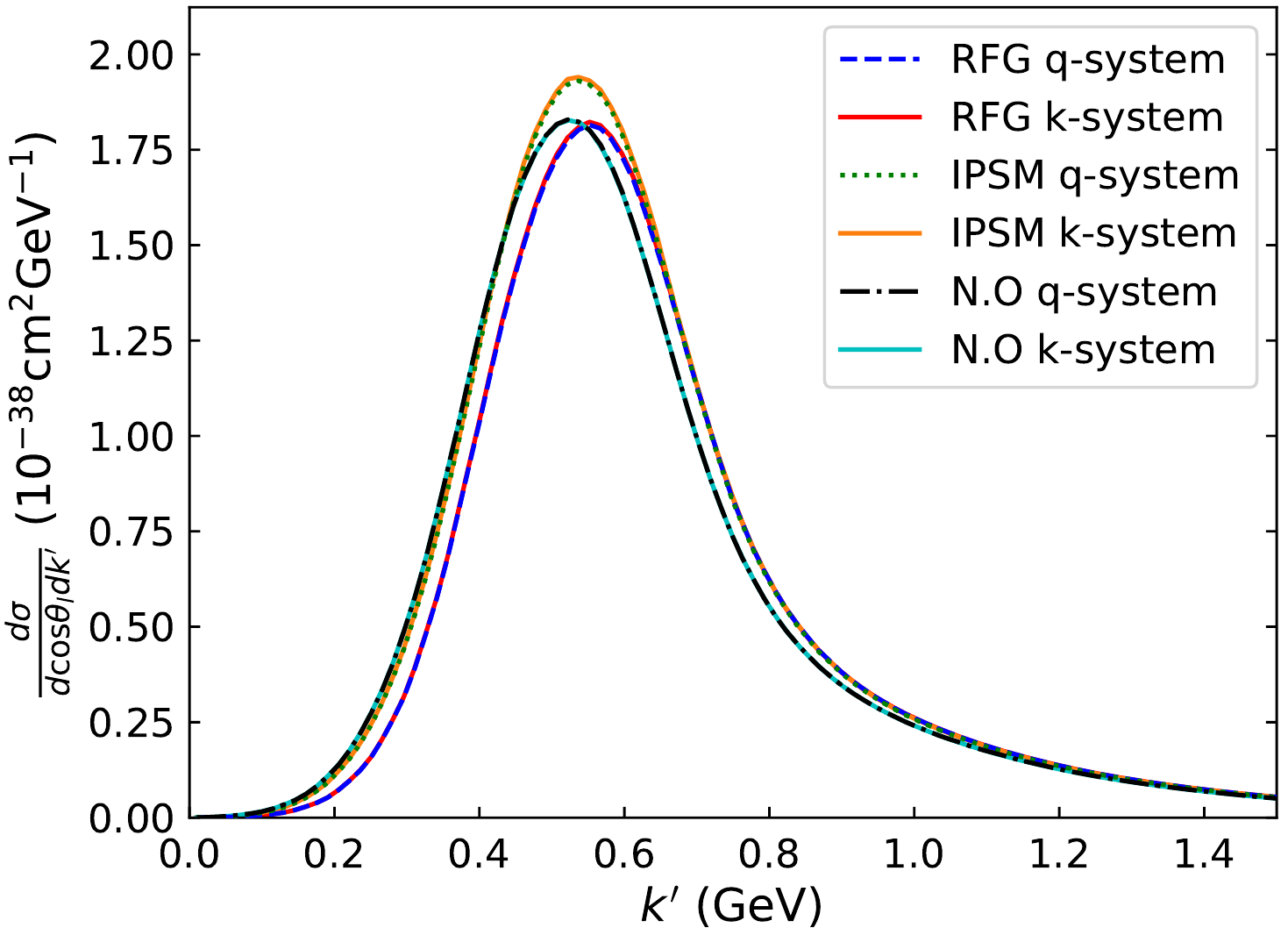}
	\caption{\label{recover}Inclusive cross section integrated over the neutrino energy weighted by the T2K flux for $^{12}$C as function of the muon momentum for $\theta_l$ = 35$\degree$ for the three models using Eq.~\eqref{inclusive-fermi} for the RFG model, Eq.~\eqref{inclusive-IPSM} for the IPSM and Eq.~\eqref{inclusive-sofia} for the NO model (labeled ``q-system") and using Eq.~\eqref{inclusive-integrating-lab} (labeled ``k-system").}
\end{figure}

For the three models considered in this work the semi-inclusive cross sections in the $k$-system to be integrated are Eqs.~\eqref{semi-inclusive_RFG},~\eqref{semi-inclusive_IPSM},~\eqref{seminclusive-sofia}. The results obtained should be consistent with the ones corresponding to Eq.~\eqref{inclusive-fermi} for the RFG, Eq.~\eqref{inclusive-IPSM} for the IPSM and Eq.~\eqref{inclusive-sofia} for the NO, respectively. 
This is illustrated in Fig.~\ref{recover} where the inclusive cross sections for $^{12}$C using the T2K flux are presented. No difference is observed between the calculations performed in the two systems for the three models. 
%{\ttred Juan: I think we should change the notation used in the figure. Why not to write simply RFG ($q$-system) and RFG ($k$-system), instead of "recovered"?}.
Although not shown here for simplicity, a further test of the consistency of the calculations has been performed using the RFG model. The simplicity of this model makes it possible to solve the problem in an analytical way getting closed expressions for the observables of interest. We have checked that these analytical results coincide with the corresponding ones obtained by solving the integrals numerically in any of the two, $q$ or $k$-systems considered. 

\subsection{\label{off-shell effects} Off-shell effects.}

To conclude, in what follows we discuss the effects in the cross sections associated to the use of different descriptions of the weak current operator. As shown in Appendix~\ref{apendixa}, the semi-inclusive responses contained in $\mathcal{F}^{2}_{\chi}$ are given by specific components of the hadronic tensor, that is a bilinear combination of the current operator matrix elements between the initial and final nucleon wave functions. As known, the weak current of the nucleon consists of a vector and an axial-vector terms. Following previous studies on electron scattering reactions~\cite{Caballero:1992tt,Caballero98a,Martinez:2002yx,Martinez:2002yw}, different options can be considered for the vector term. By analogy with the electromagnetic case, these are denoted as CC1 and CC2 prescriptions (see Appendix~\ref{apendixa} for explicit expressions). 
The two operators are equivalent for free on-shell nucleons and are connected to each other by the Gordon transformation. However, the IPSM and NO models deal in general with off-shell bound nucleons, whereas ejected nucleons are on-shell in the PWIA. Hence the two operators lead to different results. The particular case of the RFG model requires some discussion. The RFG uses relativistic free wave functions, solutions of the free Dirac equation, for all nucleons. Hence, no difference between results obtained with the two prescriptions of the vector current should exist. However, the use of a value of the separation energy fitted to the experiment in addition to the Fermi kinetic energy breaks the equivalence between the two currents, introducing at some level off-shell effects. These are linked to the energy shift 
\begin{equation}\label{omegabar-definition}
\delta = \overline{\omega} - \omega  = -E_s - T_F
\end{equation}
with $\overline{\omega}$ defined in Eq.~\eqref{omegabar-definition1}.
In the pure RFG the separation energy is negative and equal to minus the Fermi kinetic energy (see Eq.~\eqref{EsRFG}), hence $\delta=0$ and no off-shell effects are present. In the general case in which $E_s$ is fitted to experiment, the value of $\delta$ differs from zero and the election of the particular version of the vector current, CC1 or CC2, leads to different results. Appendix~\ref{apendixa} contains the detailed calculation of all the weak single-nucleon responses. The case $\delta=0$ leads to the on-shell result. 

\begin{figure}[!htbp]
	\centering
	\includegraphics[width=0.5\textwidth]{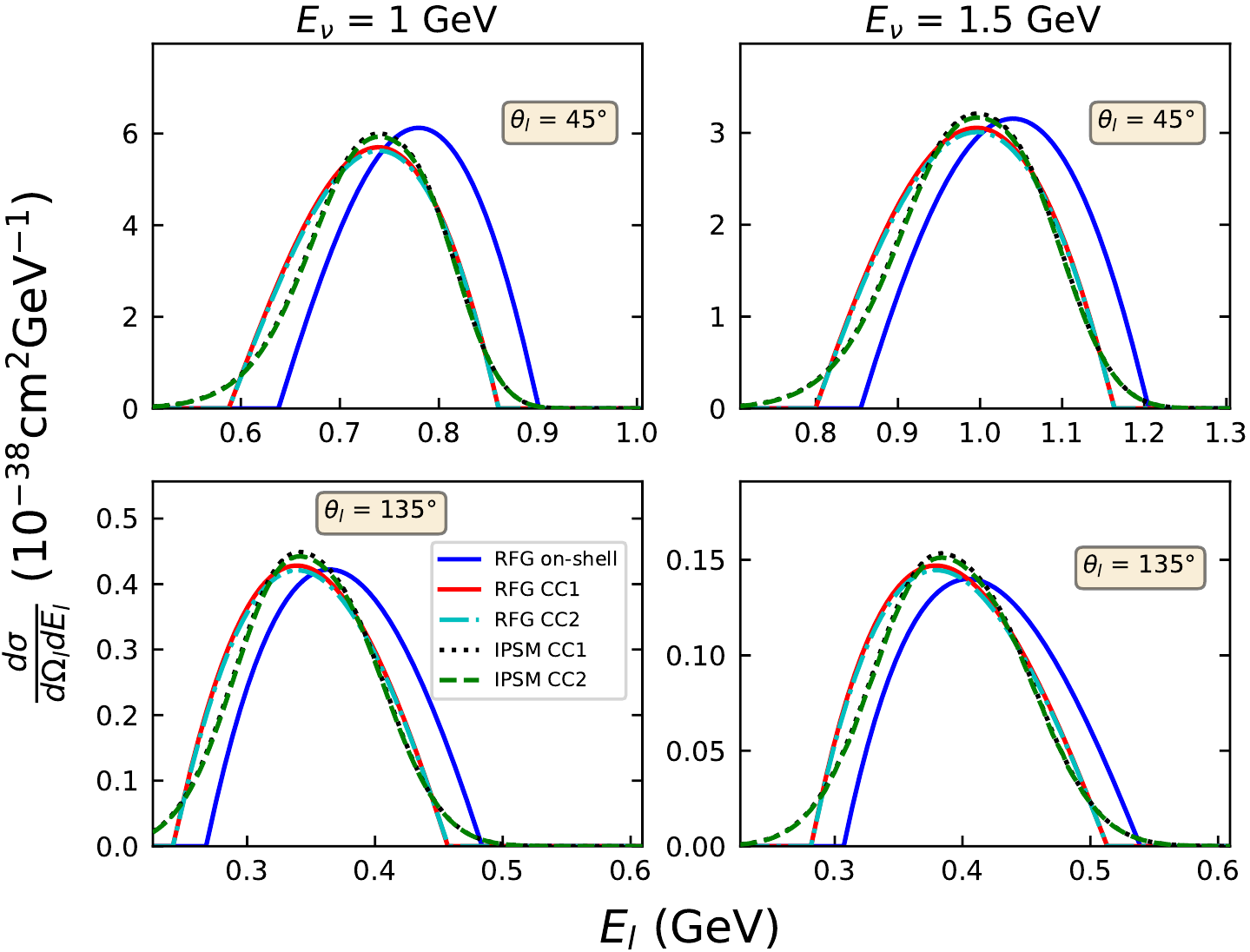}
	\caption{\label{RFG-analytic-comparation}Inclusive neutrino-$^{12}$C cross section as function of the muon energy for a fixed neutrino energy and muon scattering angle. Results are presented for the pure, on-shell, RFG model (blue solid), the off-shell ($\delta\neq 0$) RFG with the CC1 (red solid) and CC2 (cyan dot-dashed) currents. IPSM results correspond to CC1 (black dotted) and CC2 (green dashed).}
\end{figure}

The role played by off-shell effects is illustrated in Fig.~\ref{RFG-analytic-comparation}. Here we present the inclusive neutrino-$^{12}$C cross section as a function of the muon energy at different kinematical situations defined by the scattering angle and the neutrino energy. Each panel contains five curves that correspond to the RFG and IPSM models. In the former three options are considered: i) the on-shell limit, {\it i.e.,} RFG with $\delta=0$ (solid blue line), ii) RFG with off-shell effects and the CC1 prescription for the vector current (red solid) and iii) same as in the previous case but for the CC2 current (cyan dot-dashed). For the IPSM we show the results corresponding to the CC2 (green dashed) and CC1 (black dotted) currents. Comparing the results for the RFG with the two currents and $\delta$ fixed by separation energy, we observe a minor, almost negligible, discrepancy. The same comment applies to the two IPSM results. This is consistent with previous studies for electron scattering where it was shown that the use of CC1 or CC2 current operators is almost irrelevant for inclusive responses in the PWIA limit~\cite{Caballero_2006,Caballero98a,Caballero:1992tt}. On the contrary, the pure on-shell RFG result deviates significantly from the other models. As shown, the role of $\delta\neq 0$ is to shift the RFG cross section to smaller values of the muon energy by an amount that depends on the particular kinematics considered. Also the maximum in the cross section varies slightly (increasing or diminishing). Notice that the position of these maxima for the two off-shell RFG calculations coincides with the IPSM. Furthermore, the inclusive cross sections are similar except for the tails present in the IPSM due to the bound nucleon momentum distribution. 

%{\ttred Juan: I think we should change some of the curves, color....}

\begin{figure}[!htbp]
    \captionsetup[subfigure]{labelformat=empty}
	\centering
	\subfloat[]{\includegraphics[width=0.48\textwidth]{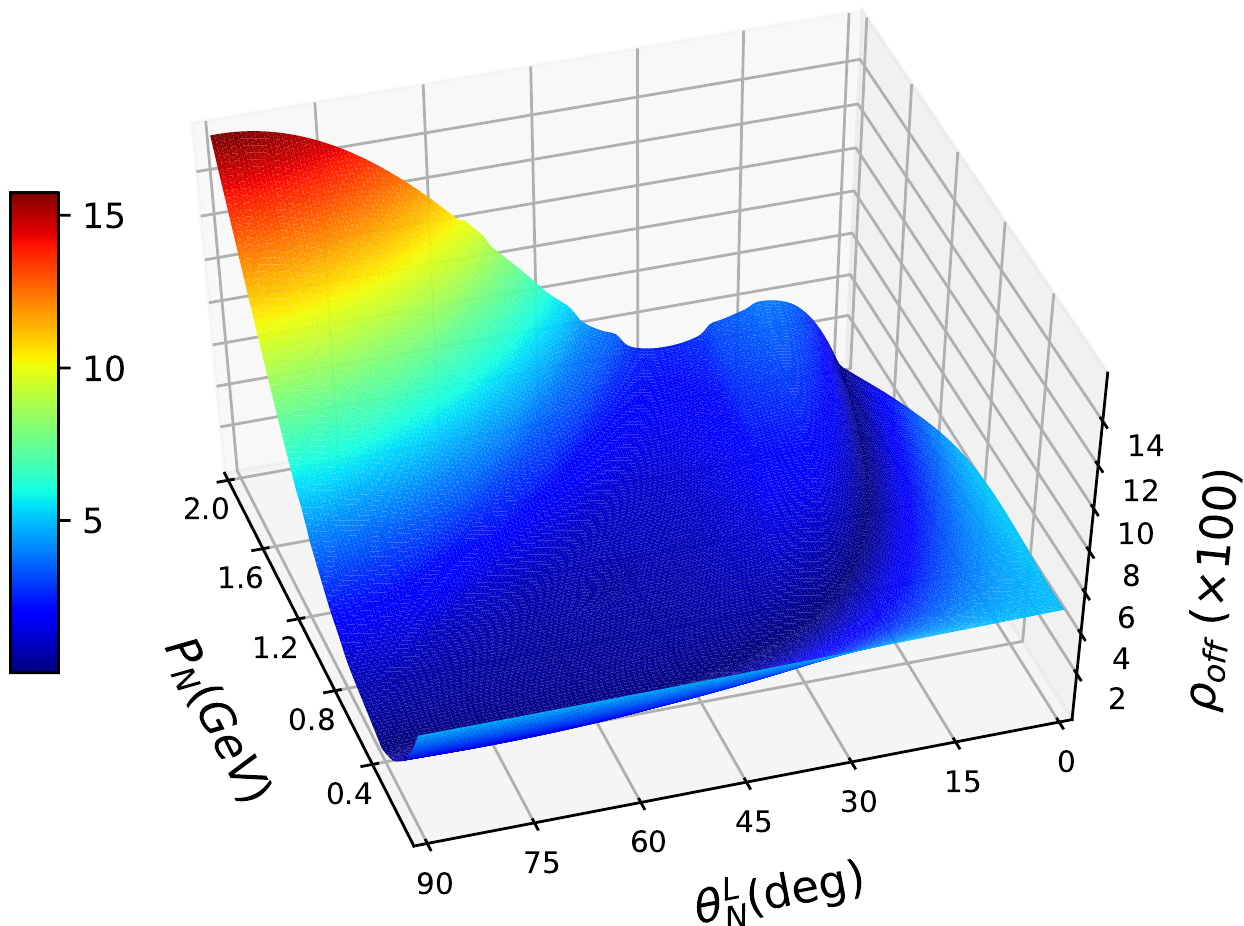}}
	\qquad
	\subfloat[]{\includegraphics[width=0.48\textwidth]{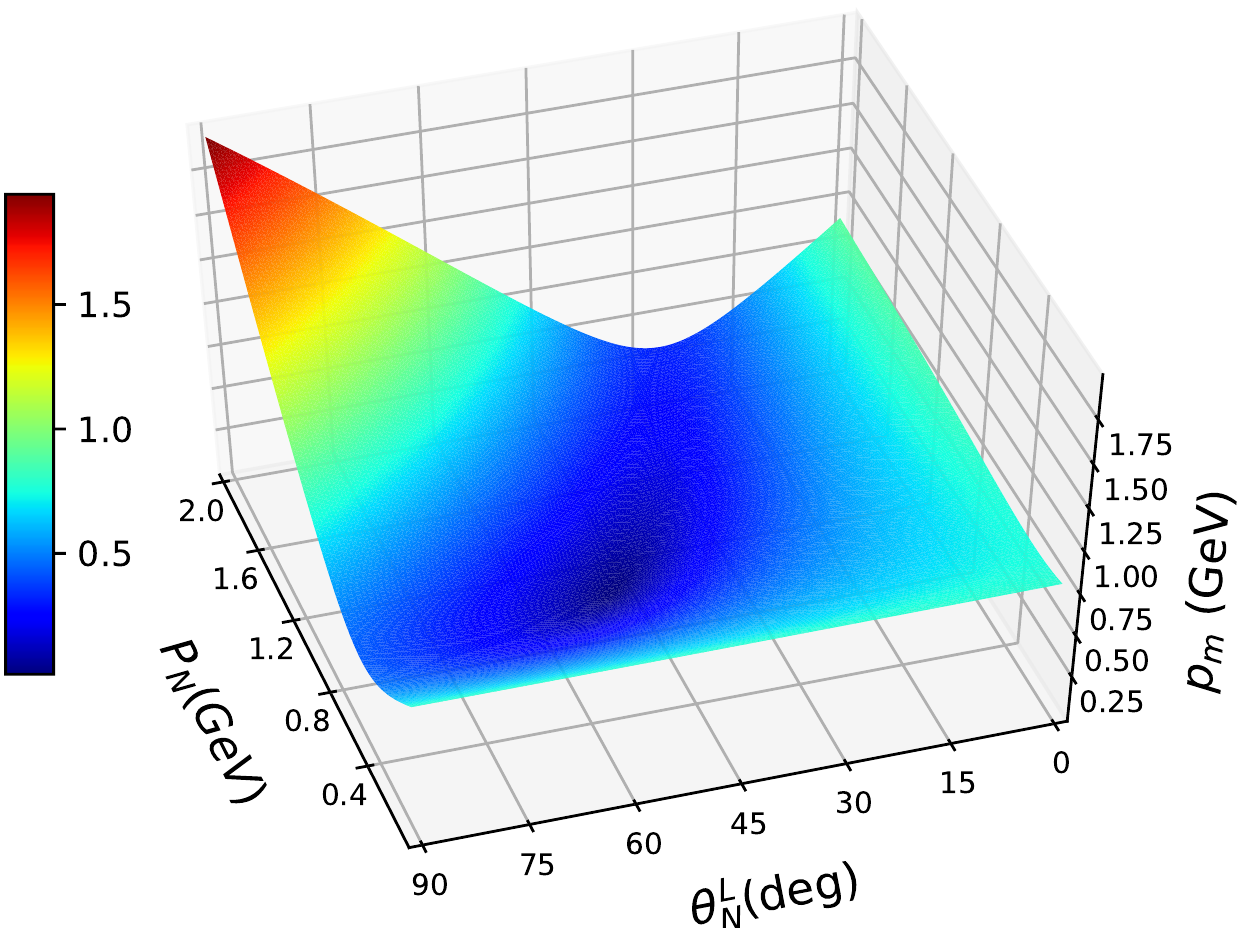}}
	\caption{\label{ratio-cc1-cc2}Top panel: the CC1/CC2 cross section ratio $\rho_{\rm off}$  defined in Eq.~\eqref{eq:rhooff}  for $^{40}$Ar in the IPSM.  Bottom panel: missing momentum $p_m$ in GeV averaged over all $^{40}$Ar shells as function of $p_N$ and $\theta_N^L$. The kinematics is  $k'$ = 1.5 GeV, $\theta_l$ = 30$\degree$ and $\phi_N^L$ = 180$\degree$.}
\end{figure}

The analysis of the off-shell effects in semi-inclusive cross sections is illustrated in Fig.~\ref{ratio-cc1-cc2}. Results in the top panel correspond to the %percent 
ratio between the difference %(in absolute value) 
and the sum of the semi-inclusive cross sections evaluated with the two current prescriptions %and 
in the IPSM applied to $^{40}$Ar: 
%{\ttblue Maria: I added the definition of this ratio, please check.} {\ttred Juan: O.K., but be careful because in the figure we show $\rho_{\rm off}$ ($\times 100$) (percent).}
\begin{equation}
\rho_{\rm off} = \left| \frac{d\sigma^{\rm CC1}-d\sigma^{\rm CC2}}{d\sigma^{\rm CC1}+d\sigma^{\rm CC2}} \right|\,.
\label{eq:rhooff}
\end{equation}
The ratio is presented as a function of the ejected nucleon variables $p_N$ and $\theta_N^L$ for the same kinematics as in previous figures and $\phi_N^L=180\degree$. As observed, off-shell effects become larger as $p_N$ and $\theta_N^L$ increase. The uncertainty introduced by the current is of the order of $\sim 12-14\%$ at $p_N\gtrsim 1.8$ GeV and $\theta_N^L\gtrsim 80\degree$. In order to understand better this result we present in the bottom panel the variation of the missing momentum $p_m$  in the ($p_N,\theta_N^L$)-plane for the same kinematics. As shown, $p_m$ is maximum in the region where the off-shell effects are the largest, {\it i.e.,} at high values of $p_N, \theta_N^L$ (left-upper corner). This is consistent with previous studies presented for semi-inclusive $(e,e'N)$ reactions in which off-shell effects were proved to be larger as the missing momentum increases. However, note that the momentum distribution drops very quickly as $p_m$ increases. As already shown in Fig.~\ref{16OrmfdonneDUNE}, the semi-inclusive cross section gets its maximum value at very low $p_m$, as illustrated in the bottom panel by the darker blue color in the center of the contour graph. Notice that this region coincides with the one where the semi-inclusive cross sections is visible (Fig.~\ref{16OrmfdonneDUNE}). The percent ratio in this region (top panel) is very small, below $\sim 2-3\%$. This implies that off-shell effects are very minor in the region where the semi-inclusive cross section gets most of its strength. However, some caution should be drawn on this general conclusion as only a specific kinematics has been explored, and more importantly, the addition of final state interactions could modify significantly these results. 
 
\section{\label{Conclusions} Conclusions}

In this paper we have presented the general formalism for semi-inclusive charged-current neutrino-nucleus reactions, {\it i.e.,} processes where an incident neutrino (antineutrino) interacts with a nucleus and a final lepton (antilepton) is detected in coincidence with some other particle. We have restricted our attention to the quasielastic kinematic regime and have assumed the impulse approximation, namely, only one-body current operators are considered. The final particle detected in coincidence with the lepton is a single  nucleon: proton (neutron) for neutrino (antineutrino) scattering. Three different models have been considered to deal with the nuclear dynamics involved in the problem: the relativistic fermi gas (RFG), the independent particle shell model (IPSM), but with fully relativistic wave functions solutions of the Dirac equation, and the natural orbitals (NO) shell model that accounts for NN correlations. 

The whole analysis has been performed assuming factorization in the cross section and the plane wave limit for the final nucleon state. Although being aware of the oversimplified description of the reaction and, particularly, the significant modifications that FSI may introduce in the analysis, we are confident that the present results help in improving our understanding on the dynamical properties of semi-inclusive cross sections. This will have an important impact in determining the neutrino beam energy with more precision, an essential requirement in the analysis of neutrino oscillation experiments. 

Flux-averaged semi-inclusive cross sections corresponding to DUNE (argon) and T2K (carbon) experiments have been presented. The results show that RFG differs completely from the two shell-based models, IPSM and NO. Not only the shape of the semi-inclusive cross section is totally different, without any sub-shell structure, but also its magnitude and behavior with the kinematic variables, particularly, with the azimuthal angle of the outgoing nucleon $\phi_N^L$. On the contrary, IPSM and NO lead to rather similar results showing only some discrepancies in the low-$p_m$ region because of the effects of NN correlations. We have checked the consistency of all calculations by recovering the inclusive observables from the semi-inclusive ones (integrating over the ejected nucleon variables) and comparing them with those already published in the literature. 
It is important to point out that the three models produce similar results for the inclusive cross sections, even being dramatically different for the semi-inclusive ones (RFG compared to IPSM and NO). The richer structure of the semi-inclusive cross section will help to better discriminate among different models, providing also a more reliable method to reconstruct the incident neutrino energy.

Theoretical predictions for the cross section against the ejected nucleon momentum have been compared with some of the recent T2K data. In spite of the above mentioned approximations considered, all the three models are capable of reproducing the data , with the theoretical curves contained within the experimental error bars. A more systematic analysis including all available data will be presented in a forthcoming work. 

Off-shell effects have been studied in the past in the case of inclusive and semi-inclusive electron scattering processes. Here we have extended this analysis to neutrino-nucleus reactions. Using for the vector part of the weak current the two usual prescriptions, CC1 and CC2, we have shown results for inclusive as well as semi-inclusive cross sections. In the latter it is shown that the region where off-shell effects are larger corresponds to high $p_m$-values, a region where the cross section is almost negligible. A similar comment applies to the inclusive cross section when comparing results for a specific model and the two prescriptions, CC1/CC2. A particular case emerges for the RFG in the on-shell limit where the inclusive cross section is shifted by a significant amount to larger values of the final lepton energy. 

To conclude,  this work represents a first step towards a more sophisticated description of the semi-inclusive reaction, where all the formalism is settled and some basic models for the nuclear initial state are applied and tested.  Some caution should be drawn on the numerical results presented, as final state interactions may introduce significant modifications. Work along this line is in progress. 

 \begin{acknowledgments}
This work was partially supported by the Spanish Ministerio de Ciencia, Innovación y Universidades and ERDF (European Regional Development Fund) under contracts FIS2017-88410-P,  by the Junta de Andalucia (grants No. FQM160 and SOMM17/6105/UGR)  (J.A.C., J.M.F.P, J.G.R.),  by the INFN under project Iniziativa Specifica NucSys and the University of Turin
under Project BARM-RILO-20 (M.B.B.). 
J.M.F.P. acknowledges support from a fellowship from the Ministerio de Ciencia, Innovación y Universidades, Program FPI (Spain). J.G.R. was supported by a Research Contract (ref. USE-19681-Q) from the University of Seville (Plan Propio de Investigación) associated to the project FIS2017-88410-P.
The authors wish to thank G.D. Megías for his helpful comments and valuable discussion. We also thank M.V. Ivanov and A.N. Antonov for providing us the code with the natural orbitals (NO) wave functions and critical comments.
\end{acknowledgments} 
  
\vspace{1cm}

\appendix

\section{\label{apendixa}The Reduced Single-Nucleon Cross Section}

This appendix contains the detailed calculation of all the reduced single-nucleon weak responses that enter in the semi-inclusive cross section introduced in previous sections. The CC neutrino (antineutrino)-nucleon scattering reactions to be considered are:
\begin{align*}
\nu_{\mu} + n \rightarrow \mu^{-} + p\, , \nonumber \\
\bar{\nu}_{\mu} + p \rightarrow \mu^{+} + n\, .
\end{align*}
%The  diagram Feynman of the neutrino process it is shown at Fig.~\ref{Fig1}, in the case of the antineutrino is similar. 

%\begin{figure}[!htbp]
%	\includegraphics{pictures/Feynman.pdf}
%\caption{\label{Fig1} Description of the kinematics involve in the CC neutrino-nucleon dispersion.}
%\end{figure}
The cross section for the previous processes can be constructed from the single-nucleon tensor that is given from the single-nucleon current matrix elements. These contain the weak charged-current operators and the wave functions for the initial (bound) and emitted nucleons. As known, the weak current operator consists of a vector and an axial-vector parts, {\it i.e.,} $\hat{J}^\mu_{weak}=\hat{J}^\mu_V-\hat{J}^\mu_A$. The axial current reads
\begin{align}
\hat{J}^\mu_A=\left[G_A\gamma^\mu + \frac{G_P}{2m_N}Q^\mu\right]\gamma^5 
\end{align}
with $G_A$ ($G_P$) the axial-vector (pseudoscalar) form factors (see~\cite{electron-vs-neutrino} and refs. therein for the specific parametrizations used).

Following the general analysis of electron scattering reactions~\cite{Caballero:1992tt,Caballero_2006,Caballero98a}, here we consider two prescriptions for the vector contribution to the weak current. These are denoted as CC1 and CC2, and are given by
\begin{eqnarray}
\left[\hat{J}^\mu_V\right]_{\rm CC2}&=& F_1\gamma^\mu +\frac{iF_2}{2m_N}\sigma^{\mu\nu}Q_\nu   \\
\left[\hat{J}^\mu_V\right]_{\rm CC1}&=& \left(F_1+F_2\right)\gamma^\mu -\frac{F_2}{2m_N}\left(\overline{P}+P_N\right)^\mu  \, ,\end{eqnarray}
where $\overline{P}^\mu$ is the on-shell four-momentum corresponding to the bound nucleon and $F_1$ ($F_2$) the isovector nucleon Dirac (Pauli) form factor. Note that the two CC1 and CC2 currents are equivalent for free on-shell nucleons. 

The above decomposition of the weak current into its vector and axial-vector parts leads to the weak hadronic tensor expressed in the form\footnote{In what follows we use the notation $W^{\mu\nu}$ for the off-shell single-nucleon tensor.}
\begin{equation}
W^{\mu\nu}=W^{\mu\nu}_{VV} + W^{\mu\nu}_{AA} + W^{\mu\nu}_{VA}\, .
\end{equation}
In what follows we present the explicit expression of the tensor obtained for the two prescriptions of the vector current. For the CC1 case we have
\begin{align}
m_{N}^2W^{\mu\nu}_{VV} = &\bigl(F_1 + F_2\bigr)^2\bigl(\overline{P}^\mu P_N^\nu + \overline{P}^\nu P_N^\mu + \frac{\Bar{Q}^2}{2}g^{\mu\nu}\bigr) - \nonumber \\
-\biggl[F_2\bigl(F_1+F_2\bigr)
&-\frac{F_2^2}{2}\bigl(1-\frac{\overline{Q}^2}{4m_{N}^2}\bigr)\biggr]\bigl(\overline{P} + P_N\bigr)^\mu\bigl(\overline{P} + P_N\bigr)^\nu \, ,
\end{align}
%and the vector-axial tensor using CC1 configuration is given by
\begin{align}
m_{N}^2W^{\mu\nu}_{VA} = -2iG_A\bigl(F_1+F_2\bigr)\epsilon^{\alpha\beta\mu\nu}P_{N \alpha}\overline{P}_\beta \, ,
\end{align}
while for the CC2 current the tensor results
\begin{widetext}
\begin{align}
m_{N}^{2}W^{\mu\nu}_{VV}&=F^{2}_{1}\biggl(\overline{P}^{\mu}P^{\nu}_{N} + \overline{P}^{\nu}P^{\mu}_{N} + \frac{\overline{Q}^{2}}{2}g^{\mu\nu}\biggr)
+F_{1}F_{2}\biggl(Q\cdot\overline{Q}g^{\mu\nu} - \frac{Q^{\mu}\overline{Q}^{\nu} + Q^{\nu}\overline{Q}^{\mu}}{2}\biggr)\nonumber \\ &
+\frac{F^{2}_{2}}{4m_{N}^{2}} 
\biggl[ P_{N}\cdot Q(\overline{P}^{\mu}Q^{\nu}  +  \overline{P}^{\nu}Q^{\mu}) +  \overline{P} \cdot Q(P^{\mu}_{N}Q^{\nu} + P^{\nu}_{N}Q^{\mu})  -Q^{2}(P^{\nu}_{N}\overline{P}^{\mu}  \nonumber \\
&+P^{\mu}_{N}\overline{P}^{\nu})-\biggl(2m_{N}^{2} - \frac{\overline{Q}^{2}}{2}\biggr)Q^{\mu}Q^{\nu} 
+g^{\mu\nu}\biggl(2m_{N}^{2}Q^{2} - \frac{Q^{2}\overline{Q}^{2}}{2} - 2(P_{N} \cdot Q)(\overline{P} \cdot Q)\biggr)\biggr]  \, ,
\label{VVtensor}
\end{align}
\end{widetext}
%and the vector-axial contribution is
\begin{widetext}
\begin{align}
m^{2}_{N}W^{\mu\nu}_{VA}=i\biggl[ G_{A}\epsilon^{\mu\nu\alpha\beta} \left( -2F_{1}P_{n\alpha}\overline{P}_{\beta} +F_{2}(P_{N} + \overline{P})_{\alpha}Q_{\beta} \right)  
+\frac{G_{p}F_{2}}{4m^{2}_{N}}\left(Q^{\mu}\epsilon^{\nu\alpha\beta\sigma} - Q^{\nu}\epsilon^{\mu\alpha\beta\sigma} \right)P_{n\alpha}Q_{\beta}\overline{P}_{\sigma}\biggr] \, . \label{VAtensor}
\end{align}
\end{widetext}

Finally, the axial-axial tensor, common to the two prescriptions, is given by
\begin{align}
m_{N}^{2}W^{\mu\nu}_{AA}&=G^{2}_{A}\left[P^{\mu}_{N}\overline{P}^{\nu} + P^{\nu}_{N}\overline{P}^{\mu} -g^{\mu\nu}\biggl(2m_{N}^{2} -\frac{\overline{Q}^{2}}{2}\biggr) \right] \nonumber \\
&-\frac{G^{2}_{p}}{8m^{2}_{N}}\overline{Q}^{2}Q^{\mu}Q^{\nu} - \frac{G_{A}G_{P}}{2}\left(\overline{Q}^{\mu}Q^{\nu} + \overline{Q}^{\nu}Q^{\mu}\right)\, . \nonumber \\ \label{AAtensor}
\end{align}
%Unless we specify, we are using the CC2 configuration.
\begin{figure}[!htbp]
		\centering
		\includegraphics{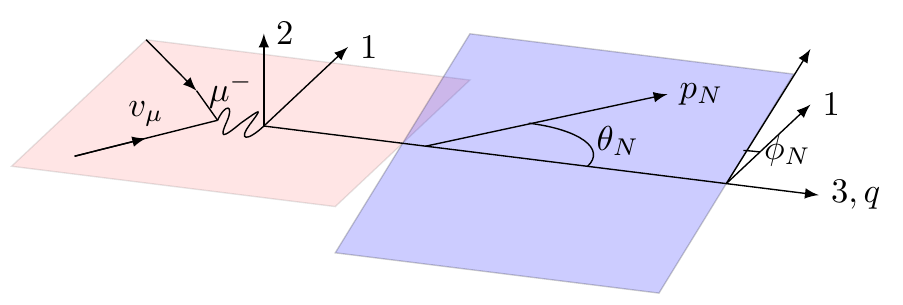}
		\caption{\label{qframe}Kinematic variables in the $q$-system where the $q$ direction is chosen as the $z$-axis (here denoted as 3-axis). The plane of the outgoing nucleon (reaction plane) is represented in blue and the scattering plane in pink.}
\end{figure}

Fig.~\ref{qframe} shows the scattering and reaction planes in the $q$-system, that is, with the $z$-axis (here denoted as $3$) chosen along the momentum transfer $\nq$. Using the reference frame given by the three orthogonal axes, 1, 2 and 3, with $(1,2)$ defining the scattering plane, the different kinematical variables introduced in the previous expressions of the tensor are given by
\begin{align} 
P^{\mu}_{N}=&(E_{N}, p_{N}\sin{\theta_{N}}\textbf{e}_{1} + p_{N}\cos{\theta_{N}}\textbf{e}_{3}), \\
\overline{Q}^{\mu}=&(E_{N} - \overline{E}, q\textbf{e}_{3}), \\
Q^{\mu}=&(\omega, q\textbf{e}_{3}), \\
\overline{P}^{\mu}=&(\overline{E}, p_{\perp}\textbf{e}_{1} + p_{\parallel}\textbf{e}_{3}) \, ,
\end{align}
where
\begin{align}
p_{\perp}&=\frac{\left|\textbf{p}_{m} \times \textbf{q}\right|}{q}=p_{N}\sin{\theta_{N}}\, , \\
p_{\parallel}&=\frac{\textbf{p}_{m} \cdot \textbf{q}}{q}=p_{N}\cos{\theta_{N}} - q\,, \\
\overline{E} &=\sqrt{p^{2}_{m} + m^{2}_{N}} \label{redefinitions}\,, \\
\overline{P} \cdot Q &= -\frac{Q^{2}}{2} - \frac{\delta^{2}}{2} - \delta(\overline{E} + \omega)\, , \\	
P_{N} \cdot Q&= \frac{Q^2}{2}-\frac{\delta^2}{2}-\delta \overline{E} \, , \\	
\overline{Q}^{2}&=Q^{2} + \delta^{2} + 2\omega\delta\, , \\
\delta&=\sqrt{p_{N}^{2} + m^{2}_{N}} -\sqrt{p^{2}_{m} + m^{2}_{N}}-\omega\nonumber\\
&=E_{N} - \overline{E} - \omega\, .
\end{align}

The different weak hadronic responses are given by taking the appropriate components of the single-nucleon tensor 
\begin{align}
R^{CC}_{VV}=W^{00}_{VV}, \\
R^{CC}_{AA}=W^{00}_{AA}, \\
R^{CL}_{VV}=W^{03}_{VV}, \\
R^{CL}_{AA}=W^{03}_{AA}, \\
R^{LL}_{VV}=W^{33}_{VV}, \\
R^{LL}_{AA}=W^{33}_{AA}, \\
R^{T}_{VV}=W^{11}_{VV} + W^{22}_{VV}, \\
R^{T}_{AA}=W^{11}_{AA} + W^{22}_{AA}, \\
R^{TT}_{VV}=W^{22}_{VV} - W^{11}_{VV},\\
R^{TT}_{AA}=W^{22}_{AA} - W^{11}_{AA}, \\
R^{TC}_{VV}=2\sqrt{2}W^{01}_{VV}, \\
R^{TC}_{AA}=2\sqrt{2}W^{01}_{AA}, \\
R^{TL}_{VV}=2\sqrt{2}W^{31}_{VV}, \\
R^{TL}_{AA}=2\sqrt{2}W^{31}_{AA}, \\
R^{T'}_{VA}=2W^{12}_{VA},\\
R^{TC'}_{VA}=2\sqrt{2}W^{02}_{VA},\\
R^{TL'}_{VA}=2\sqrt{2}W^{32}_{VA}.
\end{align}

Concerning the leptonic tensor, it is given by
\begin{equation}
L_{\mu\nu}=K_{\mu}K'_{\nu} + K_{\nu}K'_{\mu} - K \cdot K' g_{\mu\nu} - i\chi\epsilon_{\mu\nu\alpha\beta}K^{\alpha}K'^{\beta}\, ,
\end{equation}
where $\chi=1$ ($\chi =-1$) for neutrinos (antineutrinos). The neutrino and muon four-momenta in the $q$-system are written as
%{\ttred Juan: Are the covariant ones?}
\begin{align}
K_{\mu}=\biggl(k,\frac{-kk'\sin{\theta_{l}}}{q}\textbf{e}_{1} - \frac{k(k -k'\cos{\theta_{l}})}{q} \textbf{e}_{3} \biggr), \\
K'_{\mu}=\biggl(\varepsilon',-\frac{kk'\sin{\theta_{l}}}{q}\textbf{e}_{1} + \frac{k'(k' - k\cos{\theta_{l}})}{q} \textbf{e}_{3} \biggr)
\end{align} 
and  the kinematic leptonic factors are defined as~\cite{Donnelly:2014}\begin{align}
V_{CC}=\frac{2}{\upsilon_0}L_{00}, \\
V_{CL}=\frac{2}{\upsilon_0}L_{03}, \\
V_{LL}=\frac{2}{\upsilon_0}L_{33}, \\
V_{T}=\frac{L_{11} + L_{22}}{\upsilon_0}, \\
V_{TT}=\frac{L_{22} - L_{11}}{\upsilon_0}, \\
V_{TC}=\frac{2}{\sqrt{2}\upsilon_0}L_{01}, \\
V_{TL}=\frac{2}{\sqrt{2}\upsilon_0}L_{31}, \\
V_{T'}=\frac{2}{\upsilon_0}L_{12}, \\
V_{TC'}=\frac{2}{\sqrt{2}\upsilon_0}L_{02}, \\
V_{TL'}=\frac{2}{\sqrt{2}\upsilon_0}L_{32}.
\end{align}

Finally, the reduced single-nucleon cross section that enters in the general expressions for the semi-inclusive cross sections given in previous sections is defined as 
% {\ttblue Maria: actually the definition of $\mathcal{F}^{2}_{\chi}$ has not yet been given. Can you write its exact relation with the contraction $L_{\mu\nu} W^{\mu\nu}$ and explain where the terms $\cos{\phi_{N}}$, $\cos{2\phi_{N}}$ come from? From the definition of the tensors, (A21-A37) and (A41-A50) this is not clear. If you contract the leptonic and hadronic tensors defined above you don't get these terms.} {\ttred Juan: check that everything is right}
\begin{align}
&\mathcal{F}^{2}_{\chi}=\frac{2}{v_0}L_{\mu\nu}W^{\mu\nu}\nonumber \\ 
&=V_{CC}(R^{CC}_{VV} + R^{CC}_{AA}) + 2V_{CL}(R^{CL}_{VV} + R^{CL}_{AA})\nonumber \\  
&+V_{LL}(R^{LL}_{VV} + R^{LL}_{AA})
+ V_{T}(R^{T}_{VV} + R^{T}_{AA})\nonumber \\
&+V_{TT}(R^{TT}_{VV} + R^{TT}_{AA})%\cos{2\phi_{N}}
+ V_{TC}(R^{TC}_{VV} + R^{TC}_{AA})%\cos{\phi_{N}}
\nonumber \\
&+V_{TL}(R^{TL}_{VV} + R^{TL}_{AA})%\cos{\phi_{N}}
\nonumber \\
&-\chi\left( V_{T'}R^{T'}_{VA} +V_{TC'}R^{TC'}_{VA}%\cos{\phi_{N}} 
+ V_{TL'}R^{TL'}_{VA}%\cos{\phi_{N}} 
\right)\, .
\end{align}
The different single-nucleon responses given above in the $q$-system present a special symmetry with respect to the relative orientation between the scattering and reaction planes.
The whole dependence with $\phi_N$ only enters through $\cos\phi_N$ for the interference responses: $R^{TC}_{VV}$,
$R^{TC}_{AA}$, $R^{TL}_{VV}$, $R^{TL}_{AA}$, $R^{TC'}_{VA}$ and $R^{TL'}_{VA}$, and through $\cos 2 \phi_N$ in $R^{TT}_{VV}$ and $R^{TT}_{AA}$. Then, after integration over the azimuthal angle, $\phi_N$, one gets
\begin{align}
\overline{\mathcal{F}^{2}_{\chi}}&= V_{CC}(R^{CC}_{VV} + R^{CC}_{AA}) + 2V_{CL}(R^{CL}_{VV} + R^{CL}_{AA})+\nonumber \\
&V_{LL}(R^{LL}_{VV} + R^{LL}_{AA})
+ V_{T}(R^{T}_{VV} + R^{T}_{AA})- \chi V_{T'}R^{T'}_{VA}  
\end{align}
for the function defined in Eq.~\eqref{eq:Fchi2-av} entering in the inclusive cross section.

After some algebra, the leptonic kinematic factors can be written in the form \cite{Donnelly:2017}:
\begin{align}
V_{CC}=1- \frac{\Delta_{1}}{\upsilon_{0}},
\end{align}
\begin{align}
V_{CL}=-\frac{1}{q}\left(\omega + \frac{\Delta_{4}\kappa}{\upsilon_{0}}\right),
\end{align}
\begin{align}
V_{LL}&=\frac{\omega^{2}}{q^{2}} + \frac{\Delta_{1}}{\upsilon_{0}} + \frac{\Delta^{2}_{4}}{\upsilon_{0}q^{2}} + \frac{2\Delta_{4}\kappa \omega}{q^{2}\upsilon_{0}},
\end{align}
\begin{align}
V_{T} &= |Q^{2}|\left(\frac{1}{2q^{2}} + \frac{1}{\upsilon_{0}}\right) + \Delta_{1}\left(\frac{1}{2q^{2}} + \frac{1}{\upsilon_{0}} \right) - \nonumber \\
&\frac{\Delta^{2}_{1} - \Delta_{3} + \Delta_{1}|Q^{2}|}{2\upsilon_{0}q^{2}},
\end{align}
\begin{align}
V_{TT}&=- \left[\frac{\Delta_{1} + |Q^{2}|}{2q^{2}}\left(1 - \frac{\Delta_{1}}{\upsilon_{0}}\right) + \frac{\Delta_{3}}{2q^{2}\upsilon_{0}}\right],
\end{align}
\begin{align}
V_{TC}&=-\frac{1}{\sqrt{2}\upsilon_{0}}  \sqrt{1 + \frac{\upsilon_{0}}{q^{2}}}\sqrt{\Delta_{3} + (\Delta_{1} + |Q^{2}|)(\upsilon_{0} - \Delta_{1})},
\end{align}
\begin{align}
V_{TL}&=\frac{1}{\sqrt{2}q^{2}\upsilon_{0}}  \sqrt{\Delta_{3} + (\Delta_{1} + |Q^{2}|)(\upsilon_{0} - \Delta_{1})}(\Delta_{4} + \omega\kappa),
\end{align}
\begin{align}
V_{T'}&=\frac{1}{\upsilon_{0}}\left(|Q^{2}|\sqrt{1 + \frac{\upsilon_{0}}{q^{2}}} - \frac{\Delta_{4}\omega}{q}\right),
\end{align}
\begin{align}
V_{TC'}&=-\frac{1}{\sqrt{2}\upsilon_{0}}\sqrt{\Delta_{3} + (\Delta_{1} + |Q^{2}|)(\upsilon_{0} - \Delta_{1})} ,
\end{align}
\begin{align}
V_{TL'}&=\frac{\omega}{\sqrt{2}q\upsilon_{0}}\sqrt{\Delta_{3} + (\Delta_{1} + |Q^{2}|)(\upsilon_{0} - \Delta_{1})}
\end{align}
with
\begin{align}
\Delta_{1}&=m^{2}_{\nu} + m^{2}_{l}=m^{2}_{l}, \\
\Delta_{2}&=2k\varepsilon' -2kk', \\
\Delta_{3}&=4k^{2}k'^{2} - 4k^{2}\varepsilon'^{2}, \\
\Delta_{4}&=m_{l}^{2} - m^{2}_{v}=m_{l}^{2}, \\
\kappa&=k + \varepsilon', \\
\upsilon_{0}=(k + \varepsilon')^{2} - q&^{2}=\Delta_{2} + \Delta_{1} + 4kk'\cos^{2}{\frac{\theta_{l}}{2}}.
\end{align}

As already mentioned, the $q$- and $k$-systems are related by a rotation in the scattering plane of an angle $\theta_q$ that defines the direction of the neutrino momentum, $\nk$, with respect to the momentum transfer, $\nq$. Thus, we can write
\begin{equation}
\cos{\theta_{q}}=\frac{k-k'\cos{\theta_{l}}}{q}\, .
\end{equation}

The connection between the angular variables of the ejected nucleon momentum in the $k$-system, $\theta_N^L, \phi_N^L$, and the corresponding ones in the $q$-system, $\theta_N, \phi_N$, is as follows:
\begin{align}
\cos{\theta_{N}}=\cos{\theta^{L}_{N}}\cos{\theta_{q}} - \cos{\phi^{L}_{N}}\sin{\theta^{L}_{N}}\sin{\theta_{q}}\, , \\
\sin{\theta_{N}}=\sqrt{1 - \cos^{2}{\theta_{N}}}\, , \\
\cos{\phi_{N}}=\frac{\cos{\phi^{L}_{N}}\sin{\theta^{L}_{N}}\cos{\theta_{q}} + \cos{\theta^{L}_{N}}\sin{\theta_{q}}}{\sin{\theta_{N}}}\, , \\
\sin{\phi_{N}}=\frac{\sin{\phi^{L}_{N}}\sin{\theta^{L}_{N}}}{\sin{\theta_{N}}}\, .
\end{align}

Using these general results to evaluate the VV, AV and AA components of the single-nucleon tensor (\ref{VVtensor}, \ref{VAtensor}, \ref{AAtensor}), and taking the appropriate components, the single-nucleon responses in terms of the several kinematical variables and the off-shell term $\delta$, can be written as (for simplicity, we only show the responses for the CC2 prescription)
\begin{widetext}
\begin{align}
8m^{4}_{N}R^{CC}_{VV}&=4\overline{E}^{2}(4F^{2}_{1}m^{2}_{N}+ F^{2}_{2}|Q^{2}|) + 4\overline{E}\omega(4F^{2}_{1}m^{2}_{N} + F^{2}_{2}|Q^{2}|) -4F^{2}_{1}m^{2}_{N}|Q^{2}| \nonumber \\
&-8F_{1}F_{2}m^{2}_{N}(\omega^{2} + |Q^{2}|) + F^{2}_{2}(\omega^{2}|Q^{2}|  - 4m^{2}_{N}(\omega^{2} + |Q^{2}|)) \nonumber \\ 
&- 2\delta(2\overline{E} + \omega)(F^{2}_{2}(2\overline{E}\omega + \omega^{2} - |Q^{2}|)  -4F^{2}_{1}m^{2}_{N}) + \delta^{2}(-4F^{2}_{2}\overline{E}^{2} \nonumber \\ &-12F^{2}_{2}\omega\overline{E} +4F^{2}_{1}m^{2}_{N} + F^{2}_{2}(|Q^{2}| - 5\omega^{2}) )  -4\delta^{3}F^{2}_{2}(\overline{E} + \omega) - \delta^{4}F^{2}_{2}\, , 
\end{align}
\begin{align}
8m^{4}_{N}R^{CC}_{AA}&= 16\overline{E}^{2}G^{2}_{A}m^{2}_{N}  + 16\overline{E}G^{2}_{A}m^{2}_{N}\omega -8G_{A}G_{p}\omega^{2}m^{2}_{N} \nonumber \\
&-4G^{2}_{A}m^{2}_{N}(4m^{2}_{N} + |Q^{2}|)  + G^{2}_{p}\omega^{2}|Q^{2}| + \delta(16\overline{E}G^{2}_{A}m^{2}_{N} \nonumber \\ 
&+ 8 G^{2}_{A}m^{2}_{N}\omega -8G_{A}G_{p}m^{2}_{N}\omega  -2G^{2}_{p}\omega^{3}) + \delta^{2}(4G^{2}_{A}m^{2}_{N} - G^{2}_{p}\omega^{2})\, , 
\end{align}
\begin{align}
8m^{4}_{N}R^{CL}_{VV}&=2\overline{E}(2p_{\parallel} + q)(4F^{2}_{1}m^{2}_{N}  + F^{2}_{2}|Q^{2}|) 
+\omega(8F^{2}_{1}m^{2}_{N}p_{\parallel} \nonumber \\
&-8F_{1}F_{2}qm^{2}_{N} + F^{2}_{2}(-4m^{2}_{N}q   +2p_{\parallel}|Q^{2}| + q|Q^{2}|)) \nonumber \\
&+\delta(F^{2}_{2}(-(4\overline{E}^{2}q + \overline{E}\omega(4p_{\parallel} + 6q)  
+ 2\omega^{2}(p_{\parallel} +q) -|Q^{2}|(2p_{\parallel} + q))) \nonumber \\ &+8F^{2}_{1}m^{2}_{N}p_{\parallel} -4F_{1}F_{2}qm^{2}_{N}  )  -\delta^{2}F^{2}_{2}(4\overline{E}q + 2\omega p_{\parallel} + 3\omega q) - \delta^{3}F^{2}_{2}q\, ,
\end{align}
\begin{align}
8m^{4}_{N}R^{CL}_{AA}&=8\overline{E}G^{2}_{A}m^{2}_{N}(2p_{\parallel} + q) + \omega(8G_{A}^{2}m^{2}_{N}p_{\parallel}-8G_{A}G_{p}qm^{2}_{N} + G^{2}_{p}q|Q^{2}|)\nonumber \\  
&+ \delta(8G^{2}_{A}m^{2}_{N}p_{\parallel} -4G_{A}G_{p}m^{2}_{N}q  - 2G^{2}_{p}\omega^{2}q) - \delta^{2}G^{2}_{p}\omega q\, ,
\end{align}
\begin{align}
8m^{4}_{N}R^{LL}_{VV}&=F^{2}_{1}(16m^{2}_{N}p_{\parallel}(p_{\parallel} + q)  + 4m^{2}_{N}|Q^{2}|) - 8F_{1}F_{2}m^{2}_{N}\omega^{2} + F^{2}_{2}\bigl[4p_{\parallel}q(\omega(\overline{E} + \omega) \nonumber \\  
&-q(p_{\parallel} +q ))+ |Q^{2}|((2p_{\parallel} + q)^{2} + 4m^{2}_{N} - |Q^{2}| + 2\overline{E}\omega  ) - 4m^{2}_{N}q^{2}\bigr] \nonumber \\ 
&+\delta \omega(-8F^{2}_{1}m^{2}_{N}  -8F_{1}F_{2}m^{2}_{N} + F^{2}_{2}\left[2q(2p_{\parallel} + q) -4\omega^{2} -8\omega\overline{E} -4\overline{E}^{2}\right]) \nonumber  \\ 
&-\delta^{2}(4F^{2}_{1}m^{2}_{N} + F^{2}_{2}\left[6\omega(\overline{E} + \omega)-\omega^{2}\right]) - 2F^{2}_{2}\delta^{3}\omega\, , \nonumber \\
\end{align}
\begin{align}
8m^{4}_{N}R^{LL}_{AA}&=G^{2}_{A}(16m^{2}_{N}p_{\parallel}(p_{\parallel} + q) +16m^{4}_{N}  + 4m^{2}_{N}|Q^{2}|) + G^{2}_{p}q^{2}|Q^{2}| - \delta(2G_{p}^{2}q^{2}\omega \nonumber \\ 
&+ G^{2}_{A}8m^{2}_{N}\omega) -\delta^{2}(4G^{2}_{A}m^{2}_{N} +G^{2}_{p}q^{2})  -8m^{2}_{N}G_{A}G_{p}q^{2}\, ,
\end{align}
\begin{align}
8m^{4}_{N}R^{T}_{VV}&=16F_{1}F_{2}m^{2}_{N}|Q^{2}| + 4F^{2}_{2}(2m^{2}_{N} + p^{2}_{\perp})|Q^{2}| 
+8F^{2}_{1}m^{2}_{N}(2p^{2}_{\perp} + |Q^{2}|) \nonumber \\ 
&-16\delta F_1 (F_{1} + F_{2})m^{2}_{N}\omega + \delta^{2}(8F^{2}_{2}\overline{E}^{2}  
+8F^{2}_{2}\omega\overline{E} \nonumber \\
&-8F^{2}_{1}m^{2}_{N} -2F^{2}_{2}|Q^{2}|)  + 4\delta^{3}F^{2}_{2}(2\overline{E} + \omega) + 2\delta^{4}F^{2}_{2}\, ,
\end{align}
\begin{align}
8m^{4}_{N}R^{T}_{AA}=8G^{2}_{A}m^{2}_{N}(4m^{2}_{N} + 2p^{2}_{\perp} + |Q^{2}|)  -16\delta G^{2}_{A}m^{2}_{N}\omega - 8\delta^{2}G^{2}_{A}m^{2}_{N}\, ,
\end{align}
\begin{align}
8m^{4}_{N}R^{TT}_{VV}=-4p^{2}_{\perp}(4F^{2}_{1}m^{2}_{N} + F^{2}_{2}|Q^{2}|)\, ,
\end{align}
\begin{align}
8m^{4}_{N}R^{TT}_{AA}=-16G^{2}_{A}m^{2}_{N}p^{2}_{\perp}\, ,
\end{align}
\begin{align}
8m^{4}_{N}R^{TC}_{VV}&=4\sqrt{2}p_{\perp}(2\overline{E} + \omega)(4F^{2}_{1}m^{2}_{N} + F^{2}_{2}|Q^{2}|)+4\sqrt{2}\delta p_{\perp}(F^{2}_{2}(-2\overline{E}\omega  \nonumber \\
&-\omega^{2} + |Q^{2}|) +4F^{2}_{1}m^{2}_{N})  -4\delta^{2}\sqrt{2}F^{2}_{2}\omega p_{\perp}\, ,
\end{align}
\begin{align}
8m^{4}_{N}R^{TC}_{AA}=16\sqrt{2}G^{2}_{A}m^{2}_{N}p_{\perp}(2\overline{E} + \omega)  + 16\sqrt{2}\delta G^{2}_{A}m^{2}_{N}p_{\perp}\, ,
\end{align}
\begin{align}
8m^{4}_{N}R^{TL}_{VV}&=4\sqrt{2}p_{\perp}(2p_{\parallel} +q)(4F^{2}_{1}m^{2}_{N} + F^{2}_{2}|Q^{2}|) \nonumber \\
&-4\sqrt{2}\delta F^{2}_{2}p_{\perp}q(2\overline{E} + \omega) -4\sqrt{2}\delta^{2}F^{2}_{2}p_{\perp}q\, , \nonumber \\
\end{align}
\begin{align}
8m^{4}_{N}R^{TL}_{AA}=16\sqrt{2}G^{2}_{A}m^{2}_{N}p_{\perp}(2p_{\parallel} + q)\, ,
\end{align}
\begin{align}
8m^{4}_{N}R^{T'}_{VA}=32G_{A}m^{2}_{N}(F_{1} + F_{2})(\omega p_{\parallel} - \overline{E}q)  + 16\delta G_{A}m^{2}_{N}(2F_{1}p_{\parallel} - F_{2}q)\, ,
\end{align}
\begin{align}
8m^{4}_{N}R^{TC'}_{VA}=-32\sqrt{2}G_{A}m^{2}_{N}p_{\perp}q(F_{1} + F_{2})  -4\sqrt{2}\delta F_{2}G_{p}\omega p_{\perp}q\, ,
\end{align}
\begin{align}
8m^{4}_{N}R^{TL'}_{VA}=-32\sqrt{2}G_{A}p_{\perp}\omega m^{2}_{N}(F_{1} + F_{2}) -4\sqrt{2}p_{\perp}\delta(8G_{A}F_{1}m^{2}_{N} + G_{p}F_{2}q^{2})\, .
\end{align}
\end{widetext}
The isovector nucleon form factors, $F_1$ and $F_2$, can be expressed in terms of the proton and neutron electric and 
magnetic form factors~\cite{Lomon:2002, Lomon:2006},
\begin{align} % cambio de notaciÃÂÃÂ³n a G_{Electrico/Magnetico}^{proton/neutron} como en la tesis
F_{1}&=\frac{1}{1 + \tau}\bigl[G_{E}^{p} - G_{E}^{n} + \tau(G_{M}^{p} - G_{M}^{n})\bigr]\, , \\
F_{2}&=\frac{1}{1 + \tau}(G_{M}^{p} - G_{M}^{n} - G_{E}^{p} + G_{E}^{n})
\end{align}
with $\tau\equiv |Q^{2}|/(4m_N^2)= (q^{2} - \omega^{2})/(4m_N^2)$.
 
\bibliography{referencias}

\end{document}